\documentclass[numsec,webpdf,modern,large]{oup-authoring-template}

\onecolumn

\usepackage[a4paper]{geometry}
\usepackage{graphicx}
\usepackage{amsmath}
\usepackage{natbib}
\usepackage{caption}
\usepackage{subcaption}
\usepackage{comment}
\usepackage{bbm}
\usepackage{mathtools}
\usepackage{float}

\usetikzlibrary{bayesnet,shapes,fit}

\newcommand{\bfbeta}{\boldsymbol{\beta}}

\newcommand{\bfOmega}{\boldsymbol{\Omega}}
\newcommand{\bfSigma}{\boldsymbol{\Sigma}}

\newcommand{\bfU}{ {\boldsymbol U} }
\newcommand{\bfv}{ {\boldsymbol v} }

\newcommand{\bfw}{ {\boldsymbol w} }

\newcommand{\bfZ}{ {\boldsymbol Z} }

\newcommand{\normal}{\mathsf{N}}

\numberwithin{thm}{section}
\numberwithin{defn}{section}
\numberwithin{property}{section}
\numberwithin{lemma}{section}
\numberwithin{cor}{section}

\newcommand{\judgeI}{\ensuremath{\beta_{i, t}}}

\newcommand{\real}{\ensuremath{\mathbbm{R}}}

\newcommand{\dd}{\mathrm{d}}

\DeclareMathOperator{\atantwo}{atan2}
\DeclareMathOperator{\diag}{diag}
\DeclareMathOperator{\Tr}{tr}

\begin{document}

\journaltitle{The Journal of the Royal Statistical Society, Series C (Applied Statistics)}
\DOI{DOI HERE}
\copyrightyear{2023}
\pubyear{2019}
\access{Advance Access Publication Date: Day Month Year}
\appnotes{Paper}

\firstpage{1}

\title[Circular Dynamic Factor Models]{Dynamic Factor Models for Binary Data in Circular Spaces:  An Application to the U.S.\ Supreme Court}

\author[1,$\ast$]{Rayleigh Lei}
\author[1]{Abel Rodriguez}

\authormark{Rayleigh Lei and Abel Rodriguez}

\address[1]{Department of Statistics, University of Washington, \orgaddress{4110 E Stevens Way NE, 98195, Washington, USA}}

\corresp[$\ast$]{Corresponding author. \href{email:rlei13@uw.edu}{rlei13@uw.edu}}

\received{Date}{0}{Year}
\revised{Date}{0}{Year}
\accepted{Date}{0}{Year}

\abstract{Latent factor models are widely used in the social and behavioral science as scaling tools to map discrete multivariate outcomes into low dimensional, continuous scales.  In political science, dynamic versions of classical factor models have been widely used to study the evolution of justices' preferences in multi-judge courts. In this paper, we discuss a new dynamic factor model that relies on a latent circular space that can accommodate voting behaviors in which justices commonly understood to be on opposite ends of the ideological spectrum vote together on a substantial number of otherwise closely-divided opinions.  We apply this model to data on non-unanimous decisions made by the U.S. Supreme Court between 1937 and 2021, and show that, for most of this period, voting patterns can be better described by a circular latent space.
}


\maketitle

\section{Introduction}

Latent factor models (e.g., see \citealp{rummel1988applied} or \citealp{cattell2012scientific}) are a widely-used set of techniques in the social and behavioral sciences.  Depending on the context, they are used either as a dimensionality reduction tool to model multivariate data and/or as a scaling tool that maps observed outcomes onto unobserved constructs, such as intelligence or ideology. One particularly important example is the class of Item Response Theory (IRT) models (e.g., see \citealp{fox2010bayesian} or \citealp{embretson2013item}), in which the observed outcomes are either binary or categorical.  These models have been widely used in the applications in psychometry (e.g., \citealp{reise2016item} and \citealp{paganin2022computational}) and clinical assessment (e.g., \citealp{thomas2011value}), among other disciplines.  In the context of political science applications, IRT models can be derived as special cases of the celebrated class of spatial voting models \citep{davis1970expository,enelow1984spatial,poole1985spatial,JackmanMultidimensionalAnalysisRoll2001}.

An important extension of latent factor models considers situations where the observed data consists of multivariate time series and the latent factors are allowed to evolve over time (e.g., see \citealp{barhoumi2013dynamic} and \citealp{stock2016dynamic}). Versions of these models that allow for categorical observations were originally developed in \cite{forni2000generalized} and \cite{pena2004forecasting}, and have seen applications in a wide range of disciplines (e.g., see \citealp{bellego2012macro}, \citealp{fountas2018analysis} and \citealp{angelopoulos2020commodity}). In this paper, we are interested in the application of such models to the estimation of preferences in deliberative bodies, such as legislatures and multi-judge courts.

In political science, \cite{MartinQuinnDynamicIdealPoint2002a} pioneered the use of dynamic factor models by extending the work of \cite{JackmanMultidimensionalAnalysisRoll2001} to study the evolution of preferences of justices in the U.S.\ Supreme Court.
Their highly-cited paper has been extremely impactful, both because of the innovative nature of the methodology and because of the substantive contribution to the understanding of the behavior of U.S.\ Supreme Court justices.  Indeed, estimates of justices' preferences are the foundation upon which accounts of policy-motivated behavior of Supreme Court justices are built, and the topic has received attention in the political science and legal studies literature since at least the mid 1960s (e.g., see \citealp{SchubertJudicialMindAttitudes1965}, \citealp{RohdeSpaethSupremeCourtDecision1976}, \citealp{segal1989ideological} and \citealp{EpsteinEtAlJudicialCommonSpace2007}). Accordingly, the Martin-Quinn scores have been widely used as inputs for further analyses of justices' behavior (e.g., see \citealp{black2009agenda}, \citealp{casillas2011public} and \citealp{owens2011justices}).  Furthermore, the underlying methodology has been applied in other areas of political science (e.g., see \citealp{bailey2017estimating}, \citealp{bertomeu2017estimating}, and \citealp{lo2018dynamic}) and extended by various authors (e.g., see \citealp{BaileyComparablePreferenceEstimates2007},  \citealp{linzer2015global} and \citealp{kovac2019replicating}).

Most widely-used scaling models used in political sciences (including \citealp{poole1985spatial}, \citealp{JackmanMultidimensionalAnalysisRoll2001} and \citealp{MartinQuinnDynamicIdealPoint2002a}) assume that voters' preferences (their \textit{ideal points}) lie in a low-dimensional Euclidean space.  However, while the choice of a Euclidean geometry makes intuitive sense and allows for a straightforward interpretation of the model, it can fail to properly describe situations in which actors often understood as being on opposite sides of the ideological spectrum prefer similar outcomes, but for different ideological reasons.  As an alternative, circular geometries for the latent space have been proposed, with early examples including \cite{weisberg1974dimensionland} and \cite{mokken2001circles}.  Recently, \cite{yu2021spatial} proposed a Bayesian factor model for binary data that uses a circular policy space and applied it to the study of voting behavior in the U.S.\ House of Representatives.  Their results indicate that, starting with the 112\textsuperscript{th} House (which met between January 3, 2011, and January 3, 2013), the behavior of the chamber has been better explained by a circular geometry.

Accounting for the geometry of the latent policy space is particularly important when modeling the voting behavior of the U.S.\ Supreme Court, where ``end-against-the-middle'' voting patterns are common.  Indeed, despite Justice Neil Gorsuch's reputation as an originalist and the fact that he voted with Justice Clarence Thomas over 60\% of the time in nonunanimous cases during the 2019 term, he also voted with Justices Sonya Sotomayor and Ruth Bader Ginsburg (often seen as staunch liberals) about 40\% of the time in these types of cases \citep{HarvardLawReviewSCOTUSStatisticsHarvard}.  Furthermore, while the majority of five-four decisions from 2012-2020 were decided by the five conservative justices, there are numerous instances of very conservative justices, such as Justices Antonin Scalia and Thomas, voting with very liberal justices, such as Justices Sotomayor and Ginsburg.

In this paper, we extend the framework of \cite{yu2021spatial} to construct a Bayesian dynamic factor model with a circular latent space, and apply it to the study of the evolution of U.S.\ Supreme Court justices' revealed preferences since 1937.  Like \cite{MartinQuinnDynamicIdealPoint2002a}, our contribution is twofold.  One aspect is methodological.  Unlike \cite{yu2021spatial}, the model in this paper relies on a generalization of the projected Gaussian distributions \citep{mardia2000directional,small2012statistical} to model the distribution of the justices' ideal points.  More specifically, the evolution of the ideal points for each justice is modeled using a (stationary) matrix-variate first-order autoregressive processes as the joint distribution for a set of latent Gaussian distributions, which are then used to induce an autoregressive prior distribution on circular coordinates.  The resulting model for the ideal points can be seen as a special case of the projected Gaussian process introduced in \cite{WangGelfandModelingSpaceSpaceTime2014}, leading to straightforward computational algorithms based on elliptical slice samplers \citep{MurrayEtAlEllipticalSliceSampling2010}.  The projected normal distribution is more flexible than the von Mises distribution.  For example, the projected normal distribution can be bimodal, a feature that can be expected in the distribution of justices' ideal points.  Another aspect of our contribution refers to the application that motivates our modeling approach.  The estimates of justices' preferences derived from the \cite{MartinQuinnDynamicIdealPoint2002a} model for some justices during the late 1940s and early 1950s can be seen as being at odds with the usual narrative around their ideological preferences. For example, while Justice Felix Frankfurter is often described as becoming more and more conservative over his tenure despite his liberal activism before joining the court (e.g., see \citealp{LeporeColdCaseJustice2014}), this reputation is at odds with the Martin-Quinn scores, which place him as the median Justice in the 1953 term. Our circular model provides an explanation to this conundrum and leads to estimates of Justices' preferences that seem to be more in line with the understanding of legal historians.

An alternative approach to model voting records in which ``extremes'' vote together is the unfolding model recently introduced by \cite{duck2022ends}, which is in turn based on the work of \cite{roberts2000general}.  One shortcoming of this class of unfolding models is the lack of connection to any class of spatial voting models or to rational choice theory, both of which have historically served as theoretical underpinnings of  empirical scaling tools in political sciences. Furthermore, the model in \cite{duck2022ends} assumes that ideal points are constant over time, and is therefore not appropriate for studying the evolution of justices' ideal points.  Developing a dynamic version of this unfolding model is challenging because the priors in \cite{duck2022ends} have compact support that sometimes need ad-hoc adjustments.
Moreover, any implementation would involve substantial computational challenges.  Indeed, the best computational algorithms available for (static) generalized unfolding models use an MC3 algorithm that relies on multiple chains at different temperatures to improve mixing. This algorithm requires a substantial amount of tuning.  A direct extension of such an algorithm to a dynamic setting would involve sampling the ideal points of each justice in each term conditional on all other parameters. Even with very careful tuning, such an approach is well known be very inefficient. Another alternative for handling ``end-against-the-middle'' voting patterns was presented in \cite{spirling2010identifying}, who developed a clustering algorithm based on Dirichlet process mixtures to identify voting blocks within a deliberative body. Their approach was later extended to dynamic settings
by \cite{CraneHiddenMarkovModel2017a}, who used it to study the U.S.\ Supreme Court.  The voting blocks identified by these methods can provide interesting insights on their own.  However, they do not yield the kind of scaling of justices preferences in a continuous scale that is most useful in downstream applications.

The remainder of the paper is organized in the following manner. Section \ref{se:data_overview} provides some background information about the U.S.\ Supreme Court data that motivates our approach. Section \ref{se:model} introduces our dynamic circular voting model and explores some of its properties.  Section \ref{section:computation} reviews our computational strategy, which relies on Markov chain Monte Carlo algorithms to generate approximate samples of the associated posterior distribution.  Section \ref{se:analysisSCOTUS} presents the results of our analysis of the U.S.\ Supreme Court voting data.  Finally Section \ref{se:discussion} discusses some of the limitations of the model as well as further extensions.

\section{Motivating data set: voting patterns in the U.S.\ Supreme Court}\label{se:data_overview}

The U.S.\ Supreme Court (SCOTUS) is the highest court in the federal judiciary of the United States, having ultimate appellate jurisdiction over all U.S.\ federal court cases and over state court cases that involve a point of U.S.\ constitutional or federal law. Since 1869, SCOTUS has been composed of nine justices.  In principle, all sitting judges participate in every case taken on by the court, and decisions are made by a majority vote of all the participating judges.
    
\begin{figure}[!t]
  \centering
  \begin{subfigure}[t]{0.45\textwidth}
  \centering
  \includegraphics[width = \textwidth]{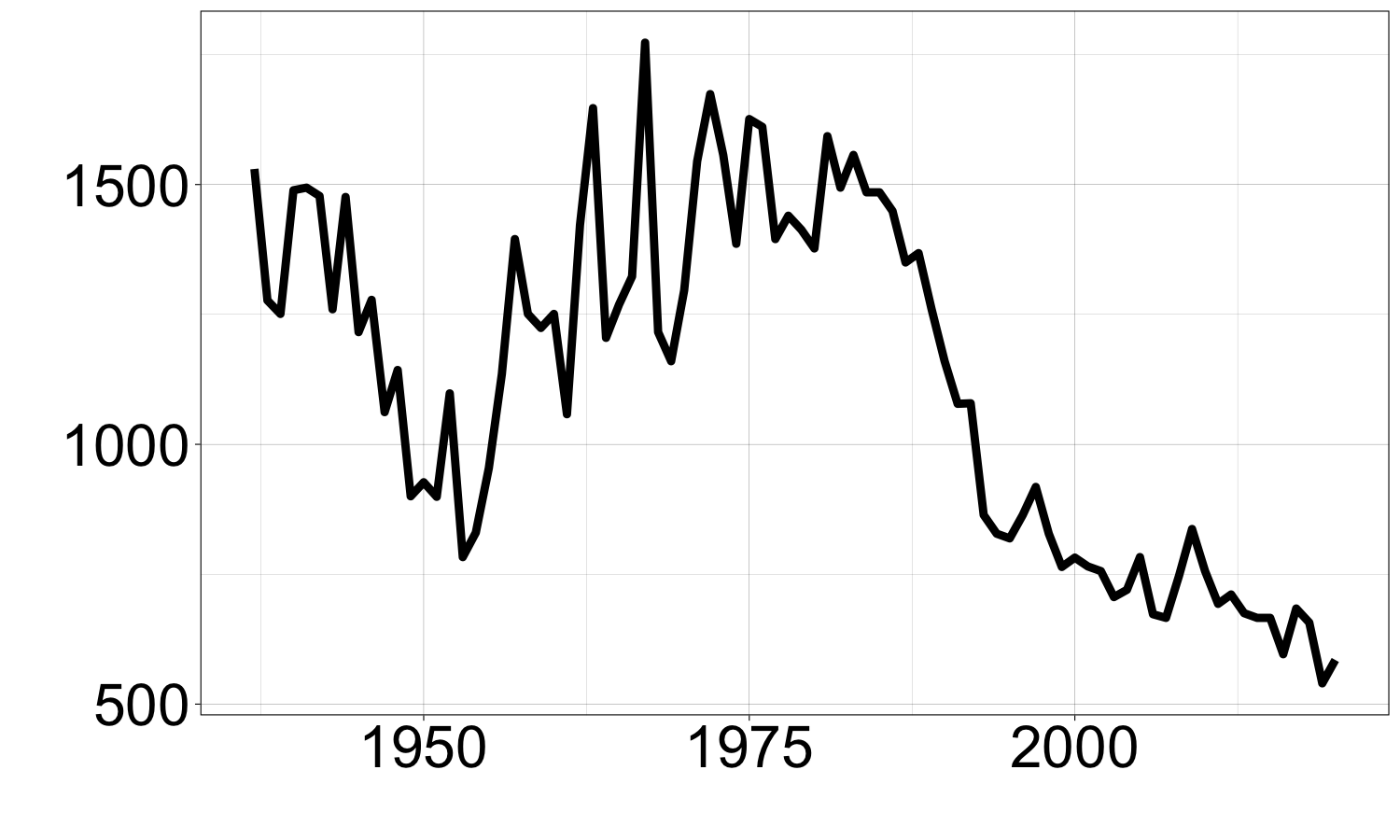}
  \subcaption{Total number of decisions}
  \label{fig:totaldecisions}
  \end{subfigure}
  \begin{subfigure}[t]{0.45\textwidth}
  \centering
  \includegraphics[width = \textwidth]{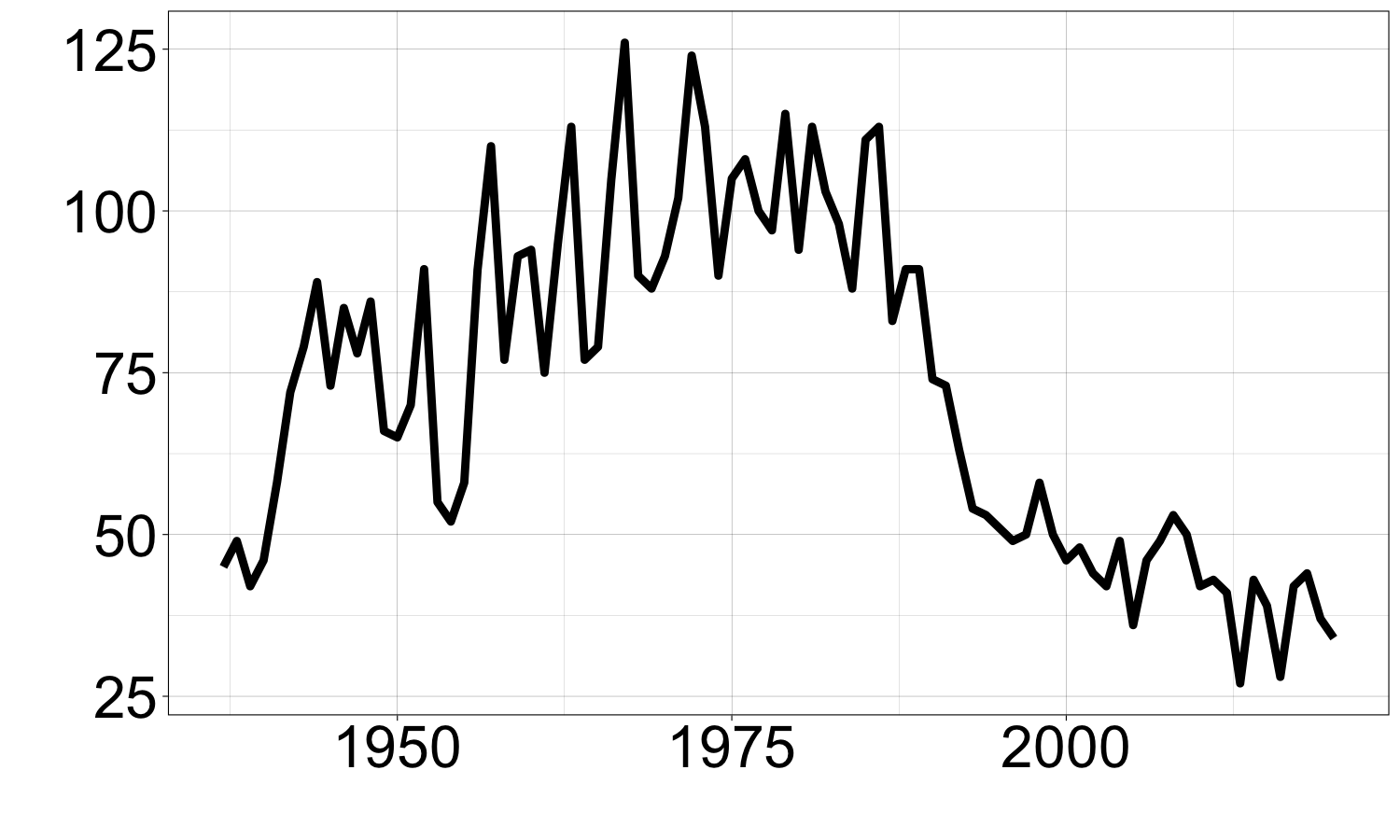}
  \subcaption{Number of nonunanimous decisions}
  \label{fig:casedecisions}
  \end{subfigure}\\
  %
  \caption{Number of  decisions made by the U.S.\ Supreme Court on each term between 1937 and 2021.}\label{fig:casenumbers}
\end{figure}

For this paper, we use the 1937-2021 data set available at \url{https://mqscores.lsa.umich.edu/}.  As in \cite{MartinQuinnDynamicIdealPoint2002a}, we work with nonunanimous decisions that are either formally decided cases with written opinions after full oral arguments, or cases decided by an equally divided vote. Ignoring unanimous votes is common practice (e.g., see \citealp{MartinQuinnParkMCMCpack}) because they provide no information about the justices' ideal points. The outcome of the justices' votes are encoded  
so that a justice's vote is set to 1 if the vote supported a reversal of a lower court's decision, and 0 otherwise.
Note that most SCOTUS decisions end up reversing lower court ones (only 98 were affirmed in this particular data set, 1.61\% of nonunanimous decisions). Equally divided votes might happen if seats are vacant or if a justice recuses themselves (a rare situation), in which case the lower court's decision stands but no precedent is set.  Figure \ref{fig:casenumbers} shows both the total number of decisions and the number of nonunanimous decisions during each term between 1937 and 2020.  These graphs show that the number of decisions made each term varies considerably. In particular, the total number of cases decided by SCOTUS started to decline in the early 1980s, with the number of nonunanimous decisions following a similar pattern.  While the number of nonunanimous decisions generally represents a relatively small percentage of the total number of decisions issued by the court, the percentage is particularly low at the very beginning of the time series.  This phenomenon was likely driven by a strong belief among the justices that the legitimacy of the court at the time heavily depended on its unanimity \citep{FeldmanScorpionsBattlesTriumphs2010}.

\begin{figure*}[!t]
  \centering
  \begin{subfigure}[t]{0.3\textwidth}
  \centering
  \includegraphics[width = .95\textwidth]{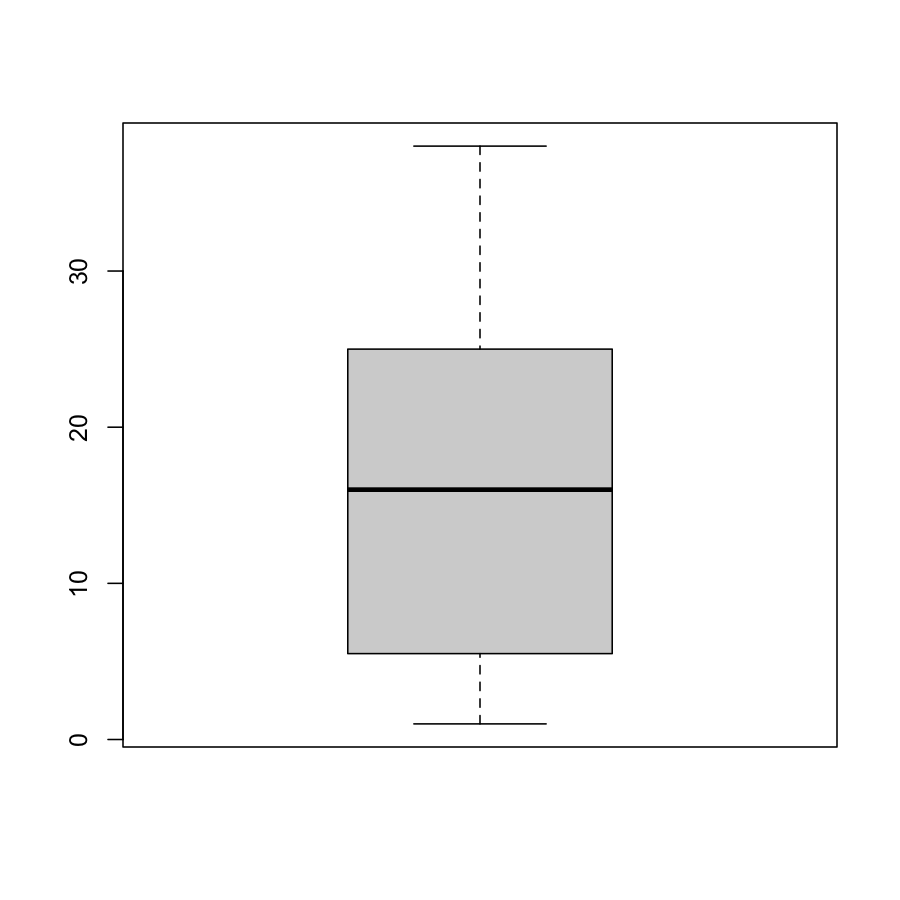}
  \caption{Length of service}\label{fig:justiceservicelength}
  \end{subfigure}
  \qquad
  \begin{subfigure}[t]{0.3\textwidth}
  \centering
  \includegraphics[width = .95\textwidth]{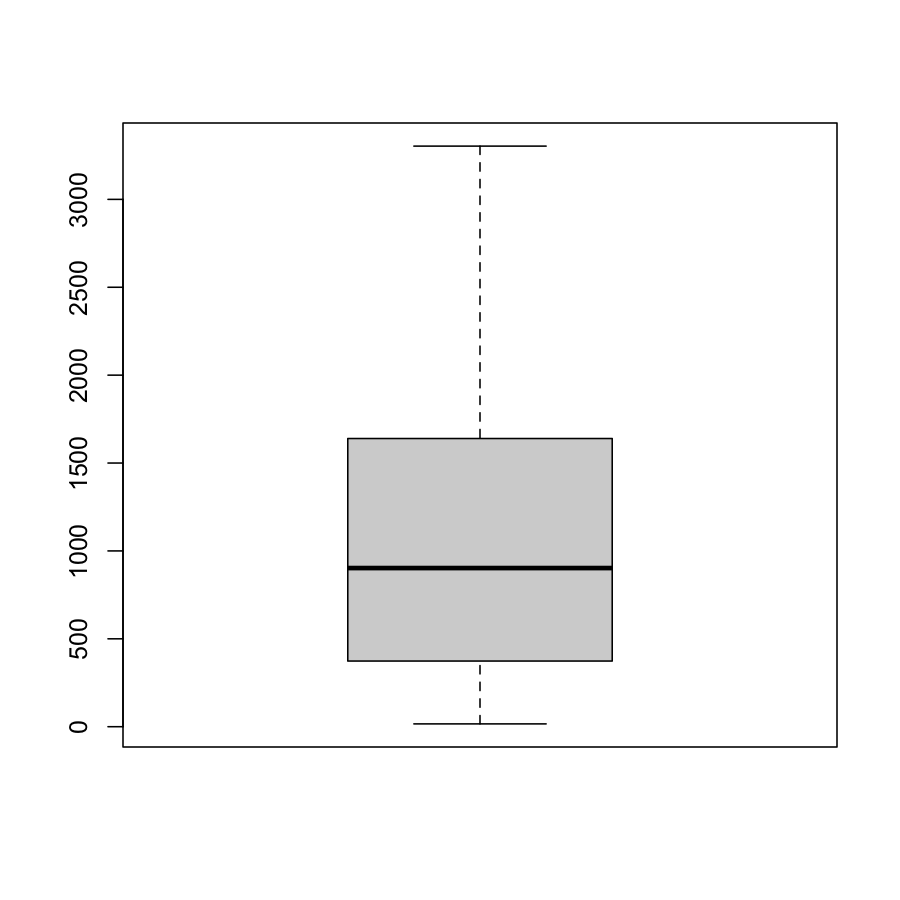}
  \caption{Number of cases that justices participated in}\label{fig:justicestotalcases}
  \end{subfigure}
  \qquad
  \begin{subfigure}[t]{0.3\textwidth}
  \centering
  \includegraphics[width = .9\textwidth]{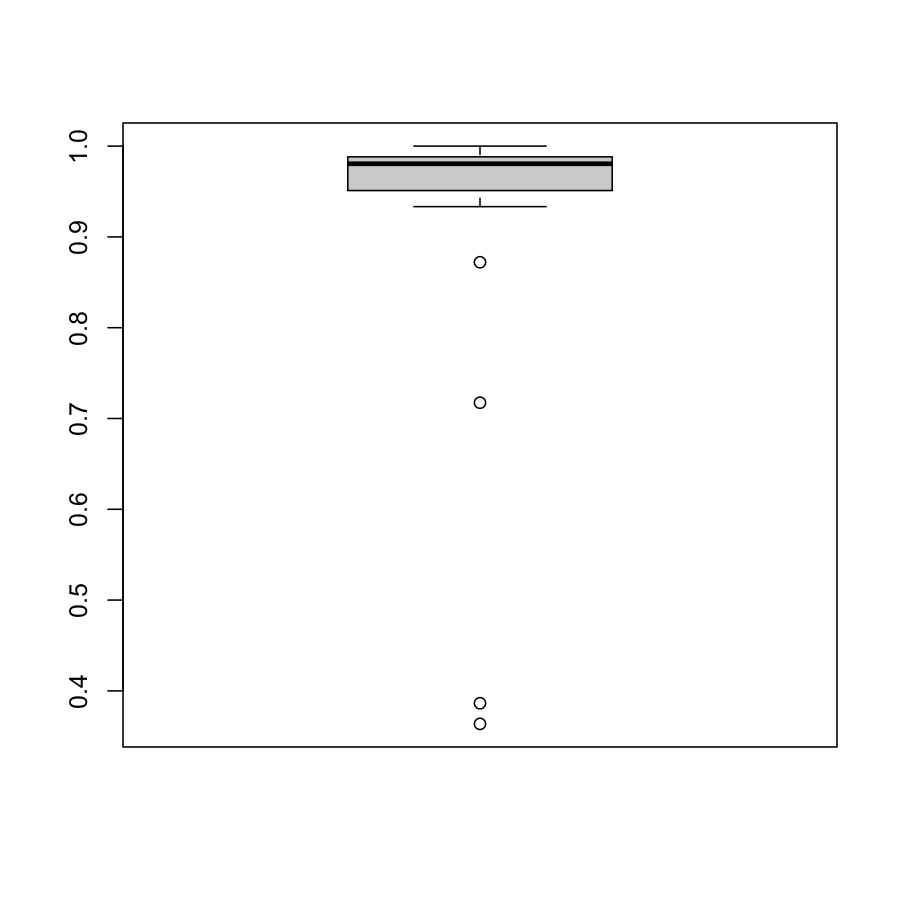}
  \caption{Average participation rate in nonunanimous cases}\label{fig:justiceaverageparticipation}
  \end{subfigure}
  %
  \caption{Length of service and participation rates for Supreme Court justices between 1937 and 2021.}\label{fig:justicessummary}
\end{figure*}

There were a total of 48 Supreme Court justices between 1937 and 2020, serving anywhere between one and 38 full or partial terms (see Figure \ref{fig:justiceservicelength}).  Virtually all these justices served their terms in consecutive years.  The one exception is Justice Robert Jackson, who took a leave of absence during the 1945 term to serve as Chief United States Prosecutor during the Nuremberg trials.  With the exception of four justices who either retired or died in the 1937 or 1938 terms, the justices on average voted in 97.5\% of the cases taken up by the court during their terms (see Figures \ref{fig:justicestotalcases} and \ref{fig:justiceaverageparticipation}).

\section{Model formulation} \label{se:model}

Let $y_{i, j, t}$ denote the vote of justice $i$ on the $j$-th decision taken in term $t$, where $i=1, \ldots, I$, $t=1, \ldots, T$, and $j=1,\ldots, J_t$.  As discussed in the previous section, we set $y_{i, j, t} = 1$ if the vote was to reverse the lower court decision, and $y_{i, j, t} = 0$ otherwise.  Because there are only up to 9 justices in the Supreme Court at any given time, a large number of the $y_{i, j, t}$ are unobserved.  Introducing this unobserved votes is not strictly necessary, but it simplifies the description of the model.  They can be treated as being missing completely at random (MCAR) and integrated out of the model when deriving our computational algorithm in Section \ref{section:computation} (see, for example, \citealp{enders2010applied,rodriguez2015measuring}).  We extend the MCAR assumption to missing values that occur when a justice is on the court, e.g., due to recusals.  This assumption, which is common in the literature, is reasonable in the context of SCOTUS because the justices have life tenure and are therefore not expected to use abstentions strategically.  Additionally, as we discussed in the previous section, the number of recusals in our data set is very smal (recall Figure \ref{fig:justiceaverageparticipation}) and we expect that, even if the MCAR assumption is incorrect, its impact will be minimal.

We construct our model using the random utility framework introduced in \cite{mcfadden1973conditional}.  We assume that each justice has two utility functions $U_A$ and $U_R$, one associated with voting to affirm the lower court decision, and one associated with reversing it.  These functions depend on the ideal point of the justice during term $t$, $\beta_{i,t}$, which lives in a latent space $\mathcal{S}$, and of two decision-specific positions ($\psi_{j,t}$ and $\zeta_{j,t}$, also in $\mathcal{S}$) that represent votes to reverse or affirm the $j$-th lower court decision during term $t$:
\begin{align*}
    U_R(\psi_{j,t}, \beta_{i,t})  & =  -d^2(\psi_{j,t}, \beta_{i,t}) + \nu_{i, j, t}, 
    &
    U_A(\zeta_{j,t}, \beta_{i,t}) & =  - d^2(\zeta_{j,t}, \beta_{i,t}) + \epsilon_{i, j, t} ,
\end{align*}
where $d(\cdot, \cdot)$ is an appropriate distance function defined on $\mathcal{S}$, and $\nu_{i, j, t}$ and $\epsilon_{i, j, t}$ are mutually independent random shocks such that $\nu_{i, j, t}-\epsilon_{i, j, t}$ is distributed with density $g_{\kappa_{j,t}}$.  A reversal vote occurs if and only if $U_R(\psi_{j,t}, \beta_{i,t}) > U_A(\zeta_{j,t}, \beta_{i,t})$, i.e., 
\begin{align*}
\Pr\left(y_{i,j,t} = 1 \mid \beta_{i,t}, \psi_{j,t}, \zeta_{j,t} \right) &= \Pr\left(U_R(\psi_{j,t}, \beta_{i,t}) > U_A(\zeta_{j,t}, \beta_{i,t}) 
\right) \\
& = \Pr \left( \nu_{i, j, t}-\epsilon_{i, j, t} > d^2(\psi_{j,t}, \beta_{i,t}) - d^2(\zeta_{j,t}, \beta_{i,t}) 
\right) \\
& = G_{\kappa_{j,t}} \left( d^2(\zeta_{j,t}, \beta_{i,t}) - d^2(\psi_{j,t}, \beta_{i,t}) \right) ,
\end{align*}
where $G_{\kappa_{j,t}}$ is the cumulative distribution function associated with $g_{\kappa_{j,t}}$.  In other words, the model assumes that justice $i$ is more likely to reverse decision $j$ at time $t$ if $\beta_{i,t}$ is closer to $\psi_{j,t}$ than it is to $\zeta_{j,t}$, with the uncertainty coming from the random $\nu_{i,j,t}$ and $\epsilon_{i,j,t}$.

\begin{figure*}[!ht]
\centering
\begin{subfigure}[t]{0.49\textwidth}
    \centering
    \includegraphics[width = .8\textwidth]{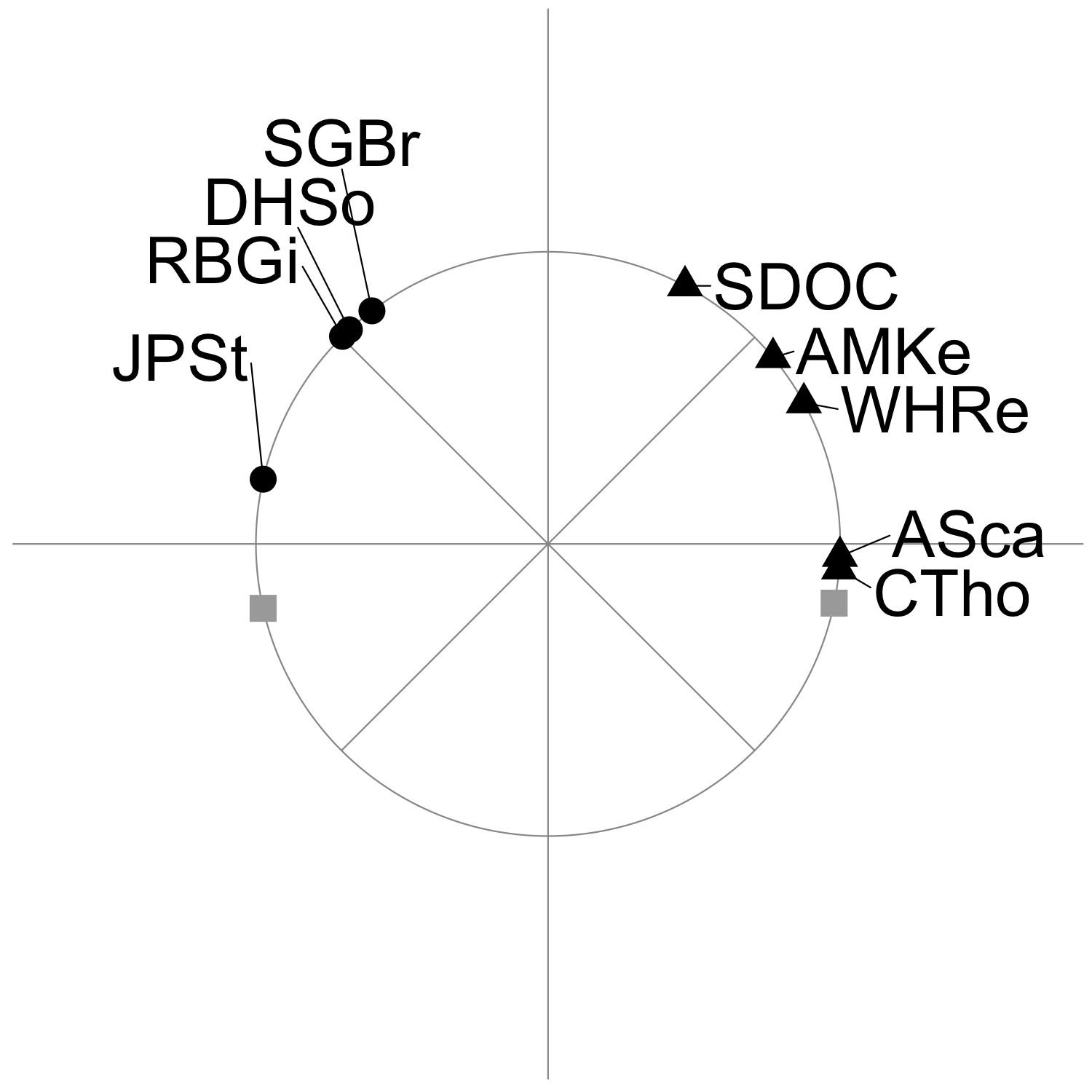}
    \caption{Circular fit to a partisan vote}
    \label{fig:circular_Euclidean}
\end{subfigure}
\begin{subfigure}[t]{0.49\textwidth}
    \centering
    \includegraphics[width = .8\textwidth]{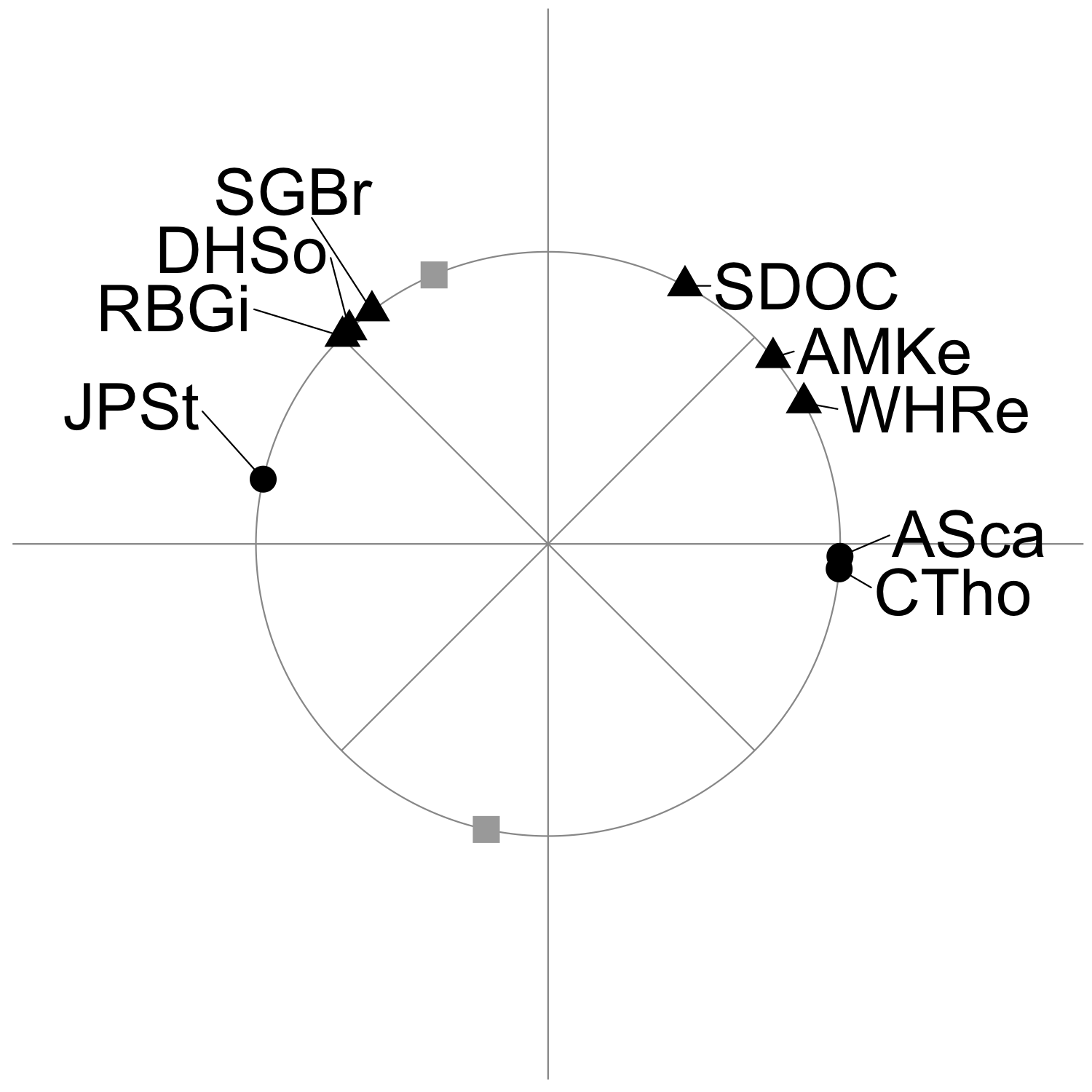}
    \caption{Circular fit to an end-against-the-middle vote}
    \label{fig:circular_circular}
\end{subfigure} \\
\centering
\begin{subfigure}[t]{0.49\textwidth}
    \centering
    \includegraphics[width = .8\textwidth]{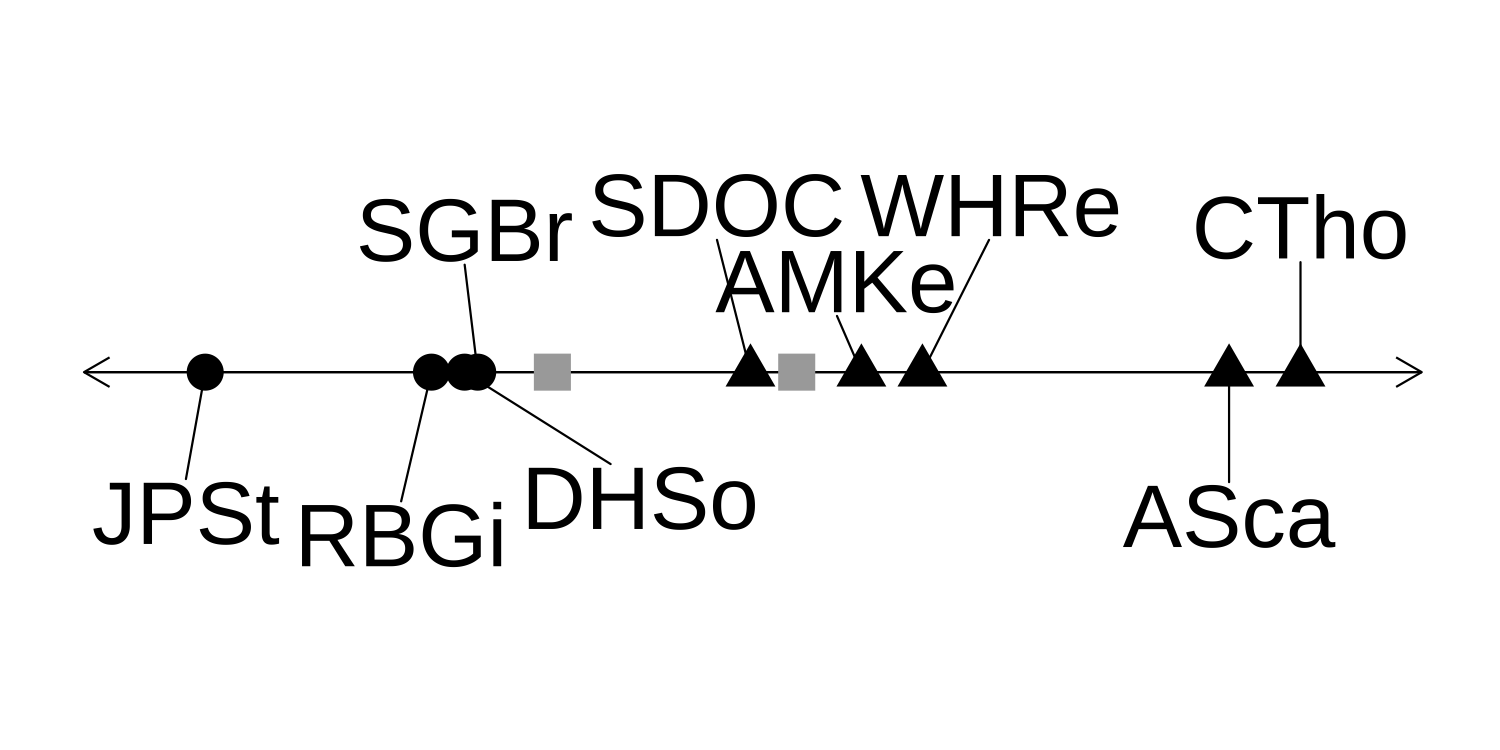}
    \caption{Euclidean fit to a partisan vote}
    \label{fig:Euclidean_Euclidean}
\end{subfigure}
\begin{subfigure}[t]{0.49\textwidth}
    \centering
    \includegraphics[width = .8\textwidth]{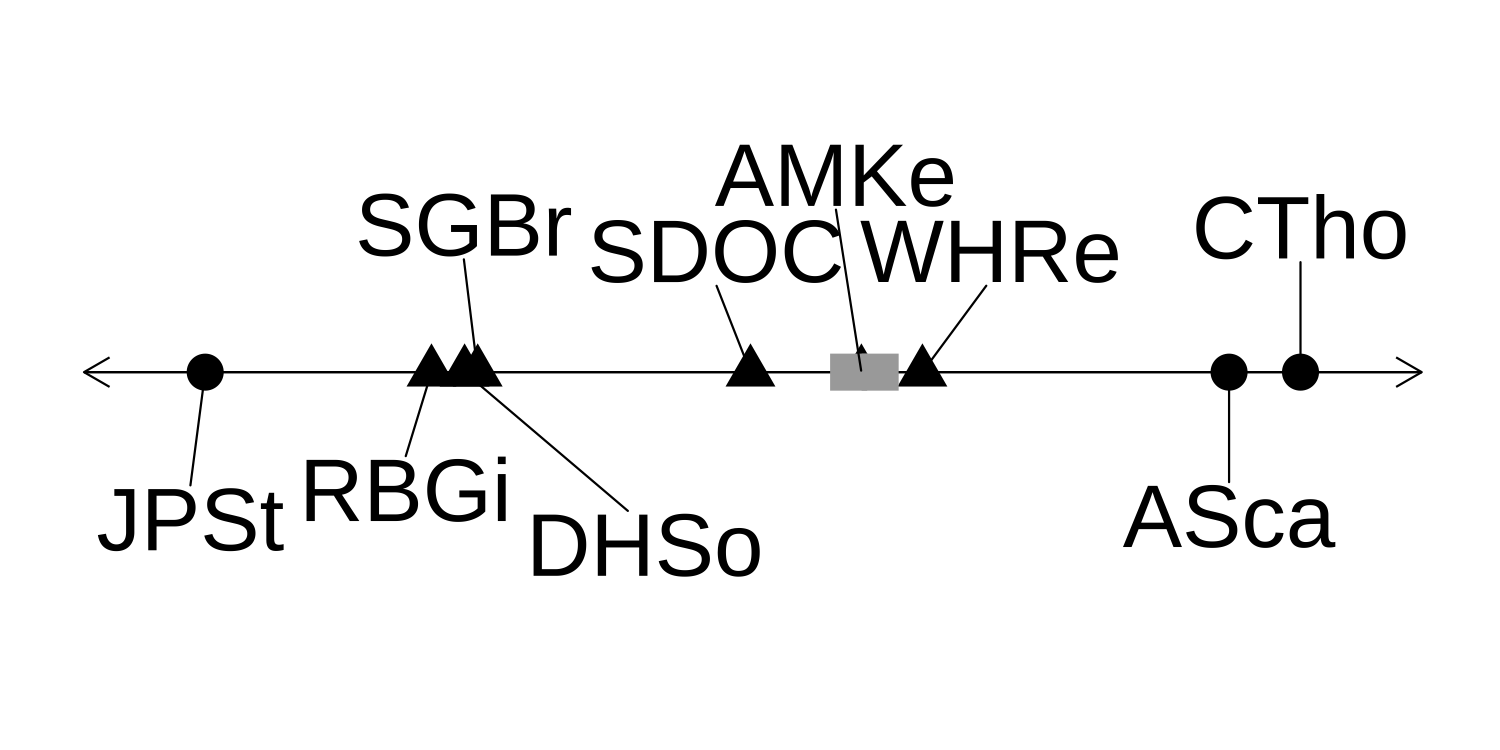}
    \caption{Euclidean fit to an end-against-the-middle vote}
    \label{fig:Euclidean_circular}
\end{subfigure}
\caption{Examples of partisan (left column) and end-against-the-middle (right column) votes in the U.S.\ Supreme Court during the 2001 term. The top row provides the estimates of the model parameters under our circular model, while the bottom row provides the estimates under the \cite{MartinQuinnDynamicIdealPoint2002a} model for the same two votes.  Triangles indicate the ideal points of justices who voted to reverse a lower court decision, whereas circles indicate those of justices who voted to affirm a lower court decision. The positions associated with votes to affirm and reverse the decisions are shown as grey squares.}\label{fig:votesexamples}
\end{figure*}

The model introduced by \cite{MartinQuinnDynamicIdealPoint2002a} assumes that $\beta_{i,t}, \psi_{j,t}, \zeta_{j,t} \in \real$,  $d(x_1,x_2) = d_E(x_1,x_2) = | x_1 - x_2 |$ is the Euclidean distance between the two points, and $\nu_{i,j,t} - \epsilon_{i,jt}$ follows a standard normal distribution for all $j$ and $t$.  Instead, in this paper we assume that $\mathcal{S}$ corresponds to the unit-radius circle and the latent positions, $\beta_{i,t}, \psi_{j,t}, \zeta_{j,t} \in [-\pi,\pi)$, are given in terms of angles (i.e., the polar representation of the points) with respect to a reference (which we select as the upper pole of the circle, e.g., see Figures 
\ref{fig:circular_Euclidean} and \ref{fig:circular_circular}).  An appropriate distance metric in this context is the geodesic distance on the unit circle, $d(x_1,x_2) = d_G(x_1,x_2) = \arccos \left( \cos \left( x_1 - x_2 \right) \right)$, the smallest angle between $x_1$ and $x_2$.  
This formulation implies that 
\begin{align}\label{eq:likel}
    \Pr \left(y_{i,j,t} = 1 \mid \beta_{i,t}, \psi_{j,t}, \zeta_{j,t} \right) &= 
    G_{\kappa_{j,t}} \left( \left\{ \arccos \left( \cos\left(\zeta_{j,t} - \beta_{i,t}\right) \right) \right\}^2 - \left\{ \arccos \left(\cos\left(\psi_{j,t} - \beta_{i,t}\right)\right) \right\}^2 \right) .
\end{align}

Unlike the Euclidean distance, the geodesic distance on the circle takes values on $[0, \pi]$.  This implies that $e_{i,j,t} = d_G^2(\zeta_{j,t}, \beta_{i,t}) - d_G^2(\psi_{j,t}, \beta_{i,t}) \in [-\pi^2, \pi^2]$ for any $\psi_{j,t}$, $\zeta_{j,t}$ and $\beta_{i,t}$.   To account for this, we assume that $\nu_{i, j, t}-\epsilon_{i, j, t}$ follows a shifted and scaled symmetric beta distribution with density
\begin{align}\label{eq:linkfunc}
g_{\kappa_{j,t}} (z) &= 
\frac{\Gamma\left( 2 \kappa_j \right)}{2 \pi^2 \Gamma\left( \kappa_j \right)\Gamma\left( \kappa_j \right)} \left( \frac{\pi^2 + z}{2 \pi^2} \right)^{\kappa_j - 1} \left( \frac{\pi^2 - z}{2 \pi^2} \right)^{\kappa_j - 1}, & \quad z \in [-\pi^2, \pi^2],
\end{align}
and cumulative distribution function $G_{\kappa_{j,t}} (z) = \int_{-\pi^2}^{z} g_{\kappa_{j,t}} (u) \dd u$.

Note that, as $\kappa_j \to \infty$, the distribution in \eqref{eq:linkfunc} converges in distribution to an appropriately scaled zero-mean Gaussian density. Together with the fact that $d_G$ is approximately equal to the $d_E$ for small distances, this means that, when the positions $\beta_{i,t}, \psi_{j,t}, \zeta_{j,t}$ are concentrated around zero and $\kappa_j$ is large, the circular model behaves as a model with a Euclidean latent space and a probit link.  On the other hand, when the latent coordinates are spread more widely over the circle, the model can accommodate voting patterns in which justices that are typically understood as having opposed preferences vote together (see Figures \ref{fig:circular_Euclidean} and \ref{fig:circular_circular} for examples that arise in our analysis of the SCOTUS data). This is a behavior that cannot be accommodated by traditional Euclidean voting models. Hence, the circular model can provide a very good approximation to the \cite{MartinQuinnDynamicIdealPoint2002a} model when the data supports it, while providing additional flexibility when the data does not support it.

It is worthwhile contrasting the behavior of Euclidean and circular models in traditional, ``partisan'' votes versus ``end-against-the-middle'' votes.  To this end, Figures \ref{fig:Euclidean_Euclidean} and \ref{fig:Euclidean_circular} show estimates of the ideal points and the affirm/reverse position  under the Euclidean model of \cite{MartinQuinnDynamicIdealPoint2002a} for two decisions, one of each kind. Note that the affirm and reverse positions in Figure \ref{fig:Euclidean_circular} are very close to each other, which implies that the justices' preferences do not explain the results of this vote. While this makes sense under the assumptions of the Euclidean model, it clearly demonstrates that information is not being effectively used.  A consequence of this is that, when end-against-the-middle votes are common, the Euclidean model can make extreme justices who participate in these votes look like centrists because that is the only explanation for their behavior that is allowed by the model.  In contrast, Figures \ref{fig:circular_Euclidean} and \ref{fig:circular_circular} show the estimated positions for the same two votes under our circular model.  In this case, rather than pushing the affirm/reverse positions for the end-against-the-middle vote close together, the circular model keeps them well separated by rotating them about 90 degrees with respect to the positions associated with a typical partisan vote.

\begin{figure*}[!ht]
\centering
\begin{subfigure}[t]{0.49\textwidth}
    \centering
    \includegraphics[width = .8\textwidth]{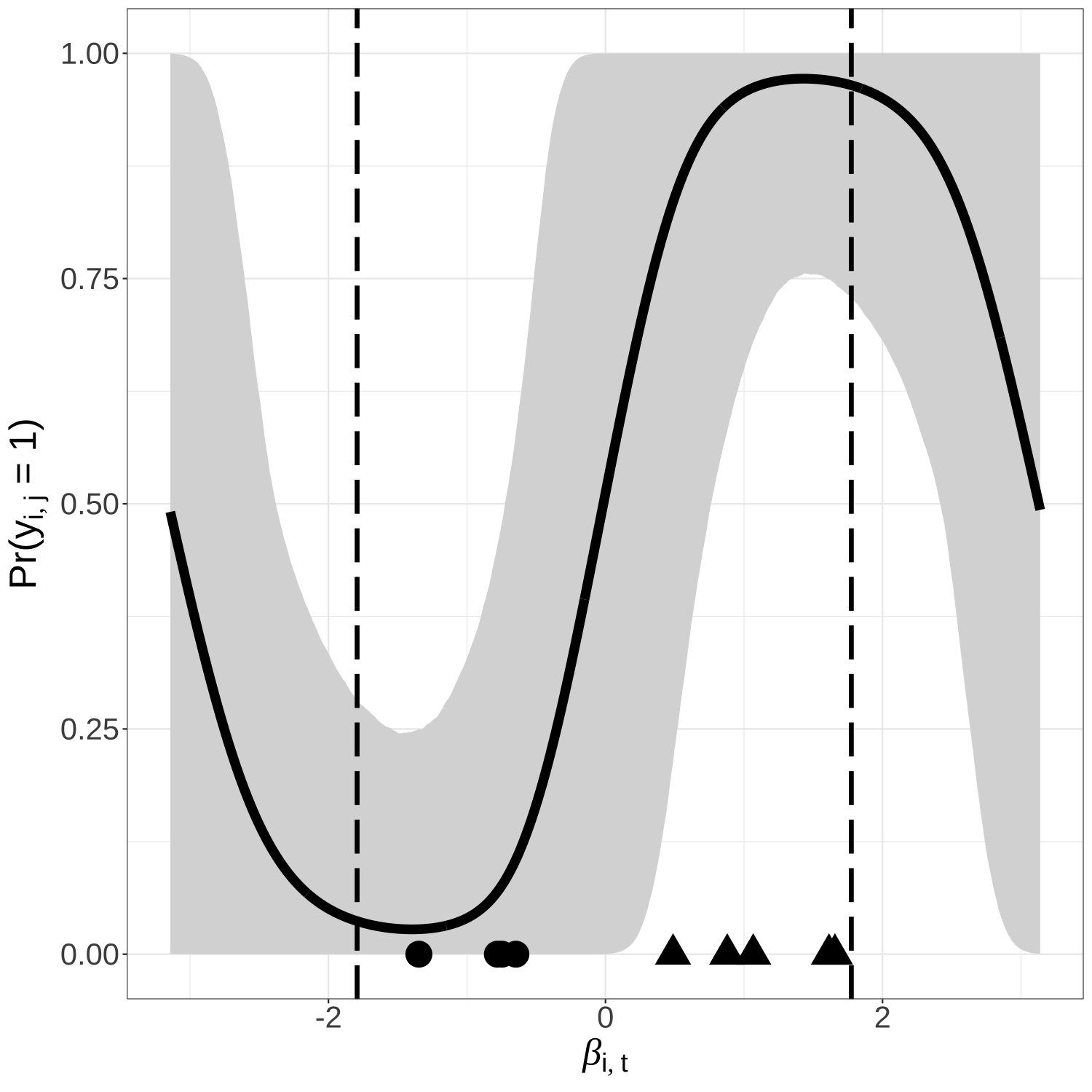}
    \caption{Response function for a partisan vote}
    \label{fig:response_partisan}
\end{subfigure}
\begin{subfigure}[t]{0.49\textwidth}
    \centering
    \includegraphics[width = .8\textwidth]{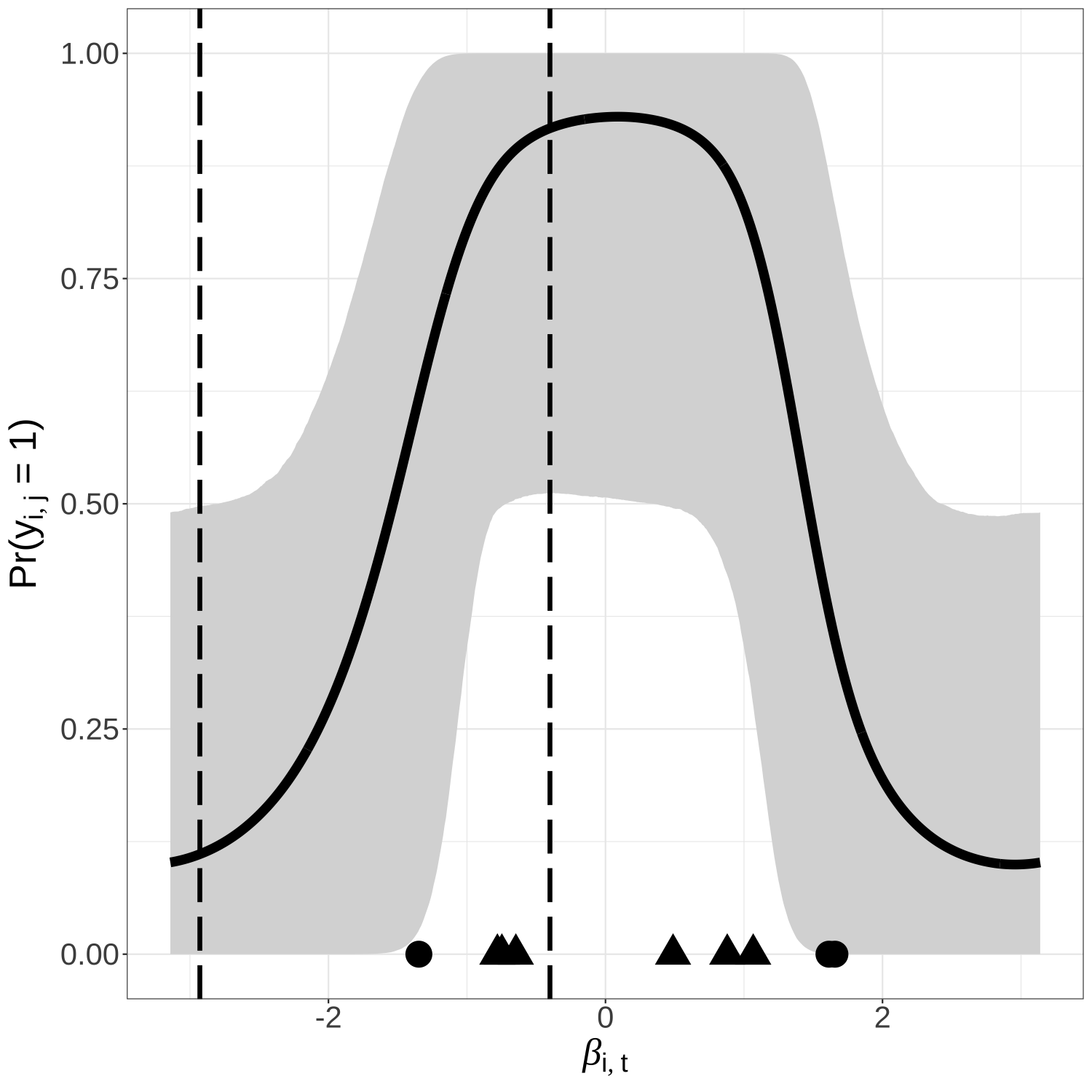}
    \caption{Response function for an end-against-the-middle vote}
    \label{fig:response_endagainstmiddle}
\end{subfigure} \\
\caption{Examples of response functions estimated by our model for partisan (left column) and end-against-the-middle (right column) votes in the U.S.\ Supreme Court during the 2001 term. Triangles indicate the ideal points of justices who voted to reverse the lower court decision whereas circles indicate justices who voted to affirm a lower court decision. The positions associated with voting to affirm and reverse the decisions are shown as dashed lines.}\label{fig:votesexamplesresponses}
\end{figure*}

Further insight can be obtained by reviewing the estimated response functions associated with each of the two votes presented in Figure \ref{fig:votesexamplesresponses}. As we would expect, the response function for the partisan vote assigns high probability for a reverse vote to justices with moderately positive ideal points and a very low probability to those with moderately negative ideal points.  On the other hand, the response function for the end-against-the-middle vote assigns high probability for a reverse vote to ``centrist'' justices with ideal points close zero, and low probability to justices with very positive or very negative ideal points.

\subsection{A hierarchical model for the evolution of revealed preferences on the circle}

One key goals for our model is to understand how the preferences of justices have changed over time.  To do this, we propose to use a joint prior for the vector of ideal points of a given judge, $\bfbeta_i = (\beta_{i,1}, \ldots,\beta_{i,T})'$, while assuming independence across justices.  To construct such prior, we introduce two independent auxiliary vectors of real-valued random variables, $\bfv_{i} = (v_{i,1}, \ldots v_{i,T})'$ and $\bfw_{i} = (w_{i,1}, \ldots w_{i,T})'$, that follow stationary first-order autoregressive processes with common autocorrelation parameters but potentially different means and evolution variances,
\begin{align}
    v_{i,t} \mid v_{i,t-1} &\sim \normal( v_{i,t} \mid \mu_1 + \rho (v_{i,t-1} - \mu_1), \tau_1^2 ) , & v_{i,1} &\sim \normal \left( \mu_1,  \frac{\tau_1^2}{1-\rho^2} \right), \label{eq:auxv}\\
    w_{i,t} \mid w_{i,t-1} &\sim \normal(w_{i,t} \mid  \mu_2 + \rho (w_{i,t-1} - \mu_2), \tau_2^2 ) , & w_{i,1} &\sim \normal \left( \mu_2,  \frac{\tau_2^2}{1-\rho^2} \right), \label{eq:auxw}
\end{align}
where $\normal(x \mid a, b)$ denotes the Gaussian distribution with mean $a$ and variance $b$, and $\rho \in (-1,1)$.  Then, the  distribution for $\bfbeta_i = (\beta_{i,1}, \ldots, \beta_{i,T})$ is obtained through a transformation of $\bfv_{i}$ and $\bfw_{i}$ such that 
\begin{align}
\beta_{i,t} &= \atantwo(v_{i, t}, w_{i, t}) = \begin{cases}
\arctan\left(\frac{w_{i, t}}{v_{i, t}}\right) & v_{i, t} \geq 0,\\
\arctan\left(\frac{w_{i, t}}{v_{i, t}}\right) + \pi & v_{i, t} < 0, w_{i, t} \geq 0,\\
\arctan\left(\frac{w_{i, t}}{v_{i, t}}\right) - \pi & v_{i, t} < 0, w_{i, t} < 0.
\end{cases}
\label{eq:angle_trans}
\end{align}

The joint density of $\bfbeta_i$ is then 
\begin{multline}\label{eq:jointdensbeta}
p(\bfbeta_i \mid \rho, \mu_1, \mu_2, \tau_1, \tau_2) =\\ 
\frac{1}{2 \pi \tau_1 \tau_2} \left| \bfOmega(\rho) \right|^{1/2} \int_{0}^{\infty} r \exp\Big\{ -\frac{1}{2}  \Tr\bfU^{T}(r,\bfbeta_i,\mu_1,\mu_2,\tau_1, \tau_2) \bfOmega(\rho) \bfU(r,\bfbeta_i,\mu_1,\mu_2,\tau_1, \tau_2)\Big\}
\dd r,
\end{multline}
where $\bfOmega(\rho)$ is a $T \times T$ tridiagonal matrix and $\bfU(\bfbeta,\mu_1,\mu_2,\tau_1, \tau_2)$ is a $T \times 2$ matrix:
\begin{align}\label{eq:OmegaU}
    \bfOmega(\rho) &= \begin{pmatrix}        
        1 & -\rho & 0  & \cdots & 0 & 0 \\
        -\rho & 1 + \rho^2 & -\rho  & \cdots & 0 & 0 \\
        0 & -\rho & 1 + \rho^2 &  \cdots & 0 & 0 \\
        \vdots & \vdots & \vdots & \ddots & \vdots & \vdots \\
        0 & 0 & 0 & \cdots & -\rho & 1
        \end{pmatrix} , 
        &
    \bfU(r,\bfbeta,\mu_1,\mu_2,\tau_1, \tau_2) &= \begin{pmatrix}        
        \frac{r \cos \beta_{1} - \mu_1}{\tau_1} & \frac{r \sin \beta_{1} - \mu_2}{\tau_2} \\
        \frac{r \cos \beta_{2} - \mu_1}{\tau_1} & \frac{r \sin \beta_{2} - \mu_2}{\tau_2} \\
        \vdots & \vdots \\
        \frac{r \cos \beta_{T} - \mu_1}{\tau_1^2} & \frac{r \sin \beta_{T} - \mu_2}{\tau_2}
    \end{pmatrix} .
\end{align}

This distribution defines an autoregressive process on the circle with stationary distribution given by a projected Gaussian distribution \citep{mardia2000directional,small2012statistical}:
\begin{align*}
p_0(\beta) = \frac{1-\rho^2}{2\pi \tau_1 \tau_2} \int_{0}^{\infty} r \exp \Bigg\{  -\frac{1-\rho^2}{2} \bigg[ \left(\frac{r\cos\beta - \mu_1}{\tau_1}\right)^2 + \left(\frac{r\sin\beta - \mu_2}{\tau_2}\right)^2 \bigg] \Bigg\} \dd r.    
\end{align*}
The prior in \eqref{eq:jointdensbeta} is a special case of the 
projected Gaussian process introduced in \cite{WangGelfandModelingSpaceSpaceTime2014} with an exponential covariance function evaluated on a uniform grid over time. The value of $\rho$ controls the level of temporal dependence (with $\rho=0$ leading to independent priors on the ideal point of each judge during each term), the ratio $\mu_1 / \mu_2$ determines the mean value of the stationary distribution, and the values of $\mu_1^2 + \mu_2^2$ together with $\tau_1^2/(1-\rho^2)$ and $\tau_2^2/(1-\rho^2)$ control the variance of the stationary distribution, as well as its kurtosis and whether it is a unimodal or bimodal distribution. 
We further discuss the interpretation of the parameters in Section \ref{se:hyper} below and in Section 1 of the supplementary materials.

For the ``reverse'' and ``affirm'' positions, $\psi_{j,t}$ and $\zeta_{j,t}$, we follow the literature and assume that they are a priori independent from each other and across all decisions. Furthermore, in order to enable the model to accommodate the full spectrum of potential voting patterns, we assign each of these parameters a uniform distribution on the circle (recall Figure \ref{fig:votesexamples}).  Assuming independence between the affirm and reverse positions for a given vote is natural, as their relative position can vary substantially depending on whether the vote is unanimous, partisan or end-against-the-middle (recall our discussion at the end of Section \ref{se:model}).  Assuming independence of the $(\psi_{j,t},\zeta_{j,t})$ pairs across votes is more debatable.  Nonetheless, it is a common assumption in the literature, perhaps because eliciting dependent priors requires substantial prior knowledge about the subject matter addressed by case.

Finally, to complete the model, we assume that the precision parameters for the link function, $\kappa_{j,t}$, are independently distributed from an exponential distribution with mean $\lambda$.  The value of $\lambda$ is assumed unknown and assigned an appropriate hyperprior (see next section).  

\subsection{Hyperprior elicitation}\label{se:hyper}

In this section, we discuss the main considerations associated with the selection of the hyperpriors for our model, and how these drive our choice in the context of the U.S.\ Supreme Court data discussed in Section \ref{se:data_overview}. Alternative values for the  hyperparameters are discussed as part of our sensitivity analysis in Section \ref{sec:sensitivity}.

\begin{figure}[!t]
\centering
\begin{subfigure}[t]{0.45\textwidth}
    \centering
    \includegraphics[width = .67\textwidth]{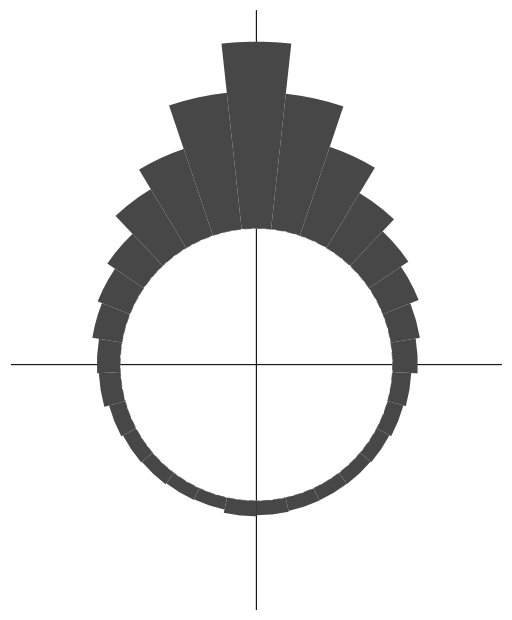}
    \caption{$\judgeI$}\label{fig:priorbeta}
\end{subfigure}
\begin{subfigure}[t]{0.45\textwidth}
    \centering
    \includegraphics[width = .8\textwidth]{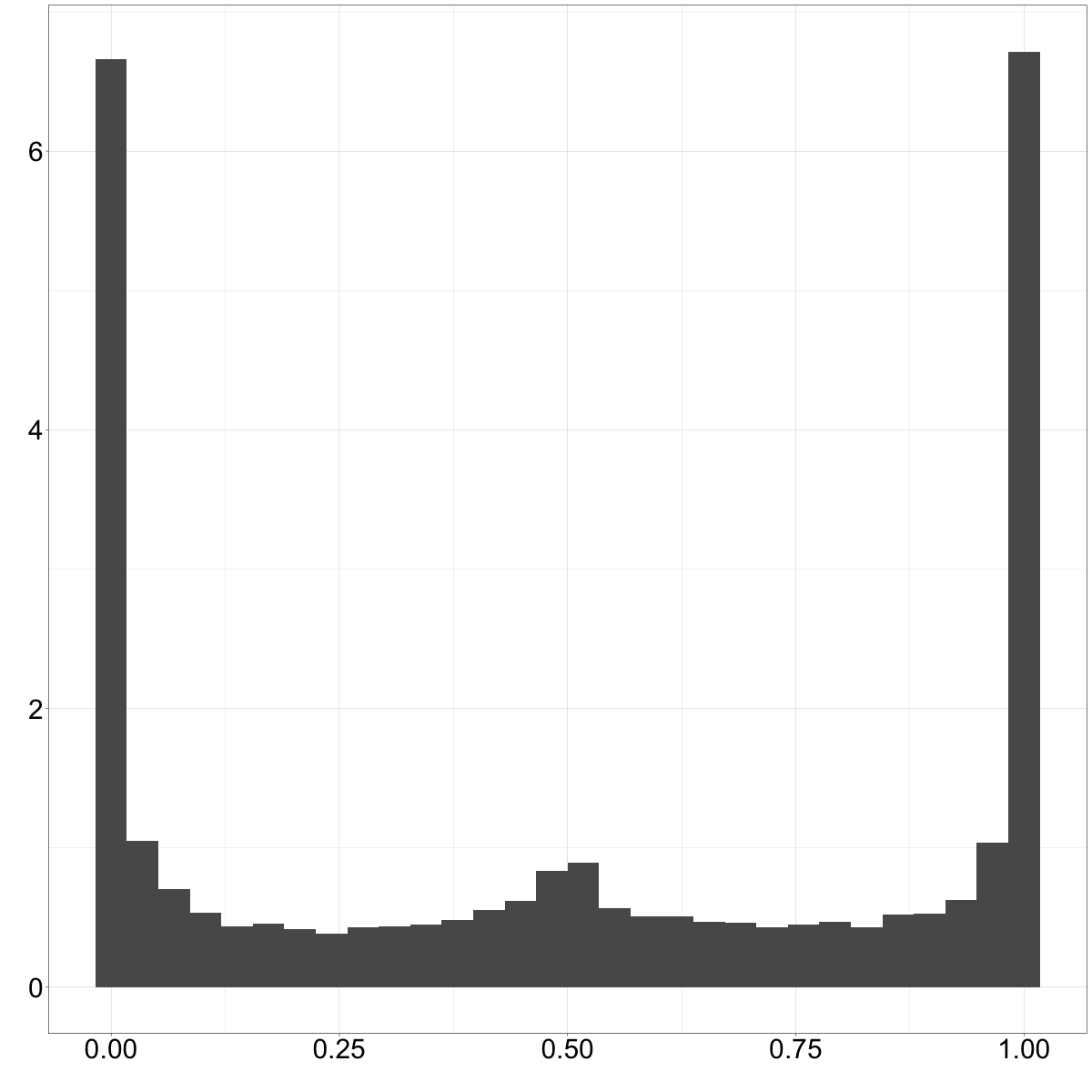}
    \caption{$\theta_{i, j, t}$}\label{fig:priortheta}
\end{subfigure}
\caption{Histograms showing 10,000 draws from the marginal prior distribution for $\judgeI$ and $\theta_{i, j, t}$. 
}
\label{fig:prior_post_comp}
\end{figure}

We consider first the hyperpriors on the parameters $\mu_1$, $\mu_2$, $\tau_1$, $\tau_2$ and $\rho$ for the projected autoregressive process.  In order to ensure that the ideal points are (weakly) identifiable, we need to fix the prior mean of the stationary distribution of the process.  We achieve this by setting $\mu_2 = 0$, which implies that the expected value of $\beta_{i,t}$ is zero.  A second consideration has to do with our assumptions about the evolution of the justice's preferences, which we expect, for the most part, to drift only slowly over time.  Based on this, we prefer priors that put high probability on values of $\rho$ that are close to 1.  More specifically, in our analyses we assigned $\rho$ a Gaussian distribution with mean 0.9 and standard deviation 0.03, truncated to the $[0,1]$ interval.  This prior implies that, a priori, $\Pr(0.85 \le \rho \le 1) \approx 0.95$. Then, the priors on $\mu_1 = \mu$, $\tau^2_1 = \tau^2$ and $\tau_2^2 = \varsigma \tau^2$ are selected so that the implied stationary distribution on $\beta_{i,t}$ is unimodal and places the majority of its mass on the $[-\pi/2, \pi/2]$ interval.  This choice reflects our prior belief that the Euclidean voting model is a good starting point for inference.  In the analyses we present in Section \ref{se:analysisSCOTUS}, $\mu$ follows a normal distribution with mean $3.073$ and standard deviation $1.588$, truncated to the positive numbers, $\tau^2$ follows an exponential distribution with mean $2.473$, and $\varsigma$ follows a Gamma distribution with mean 1 and variance 0.299.  A histogram created from 10,000 samples of the prior on $\beta_{i,t}$ implied by this specific choice of hyperpriors can be seen in Figure \ref{fig:priorbeta}.


To specify the prior distribution on $\lambda$ (the common mean of the precision parameters associated with the link function), we focus on the implied prior on
\begin{align}
\theta_{i,j,t} &= G_{\kappa_{j,t}} \Big( \left\{ \arccos \left( \cos\left(\zeta_{j,t} - \beta_{i,t}\right) \right) \right\}^2 - \left\{ \arccos \left(\cos\left(\psi_{j,t} - \beta_{i,t}\right)\right) \right\}^2 \Big),    
\end{align}
the probability that a justice votes to reverse a lower court's decision.  Note that, under our hyperpriors, this implied prior is the same for every $i$, $j$ and $t$. As noted in \cite{spirling2010identifying}, we typically want the prior on this parameter to favor values that are close to either 0 or 1.  With this in mind, we set the hyperprior for $1/\lambda$ to be an exponential distribution with mean 25. The implied prior on $\theta_{i,j,t}$ can be seen in Figure \ref{fig:priortheta}.

\subsection{Interpreting ideal points in circular spaces}\label{se:interpretation}

The fact that circular spaces are not endowed with a total ordering presents challenges when trying to interpret the estimates of the ideal points generated by our model.  Indeed, in traditional models with Euclidean policy spaces, it is common to speak of justices as being ``liberal'' or ``conservative'' and to report their ranking in terms of this ordered scale.

When all the estimates of the ideal points points of the justices fall within a semicircle, the ideal points can be mapped onto an appropriate tangent space using an order preserving transformation.  Hence, in that case, a ranking of the judges can still be obtained from our circular model by ``unfolding'' the circle onto the interval $[-\pi,\pi]$ and ranking justices on the implied linear scale.

For terms in which the range of ideal points span a wider range but no ideal point is close to either $\pi$ or $-\pi$, the results can also often be interpreted by first clustering them into a small, data-driven number of groups.  Our experience is that this number tends to be greater than 1 but otherwise very small (most often 2, and sometimes 3).  In this case, it is still appropriate to think about justices as belonging to either a ``conservative'', ``liberal'' or perhaps ``centrist'' block.

The most challenging situations arise when justices have ideal points that are close to either $\pi$ or $-\pi$, as small perturbations of this ideal point can have a very large impact on the ranking implied by the unfolding of the circle.  In this case, the interpretation of the voting blocks generated by the clustering procedure in terms of ``conservatives'' or ``liberals'' will not be appropriate for some of the justices.  However, rather than being seen as a shortcoming, this is in fact a key feature of the model that allows us to explain some of the historical conundrums that cannot be explained by Euclidean models.  We come back to this point in the context of our application.

\section{Computation}\label{section:computation}

The posterior distribution associated with our model is analytically intractable. Therefore, we rely on a Markov chain Monte Carlo algorithm to generate realizations from the posterior distribution.  This is accomplished by iteratively sampling from the full conditional distribution of (blocks of) parameters.  Posterior summaries of interest can then be approximated using their empirical counterparts.

To jointly sample from the posterior distribution of each of the ideal point sequence, $\bfbeta_{i}$, we employ an elliptical slice sampler \citep{MurrayEtAlEllipticalSliceSampling2010} that relies on the auxiliary variables $\bfv_{i}$ and $\bfw_{i}$ defined in \eqref{eq:auxv} and \eqref{eq:auxw}. Generally speaking, the elliptical slice sampler was developed to draw samples for random variables whose density can be decomposed as the product of the kernel of a zero-mean Gaussian kernel and an arbitrary function.  In our case, conditioning on the remaining parameters, we work with the reparameterization
$$
\bfZ_i = \begin{pmatrix}        
        z_{i,1,1} & z_{i,1,2} \\
        z_{i,2,1} & z_{i,2,2} \\
        \vdots & \vdots \\
        z_{i,T,1} & z_{i,T,2} \\
        \end{pmatrix} =
        \begin{pmatrix}        
        v_{i,1} - \mu & w_{i,1} \\
        v_{i,2} - \mu & w_{i,2} \\
        \vdots & \vdots \\
        v_{i,T_i} - \mu & w_{i,T_i} \\
        \end{pmatrix} ,
$$ 
and focus on sampling $\bfZ_i$ from its full conditional distribution
\begin{align*}
    p(\bfZ_i \mid \cdots )  \propto 
    \exp\left\{ - \frac{1}{2\tau^2} \Tr\left[ \bfSigma \bfZ_i^T \bfOmega(\rho) \bfZ_i \right] \right\}
    \prod_{t=1}^T \prod_{j=1}^{J_t} \left\{ \theta(z_{i,t,1}, z_{i,t,2}) \right\}^{y_{i,j}}
    \left\{  1 - \theta(z_{i,t,1}, z_{i,t,2})\right\}^{y_{i,j}},
\end{align*}
where $\bfSigma = \diag\{1, 1/\varsigma\}$, $\bfOmega(\rho)$ was defined in \eqref{eq:OmegaU}, and
\begin{align*}
    \theta(z_{i,t,1}, z_{i,t,2}) =& G_{\kappa_{j,t}} \left( \left\{ d^*(\zeta_{j,t}, z_{i,t,1}+\mu, z_{i,t,2}) \right\}^2 - \left\{ d^*(\psi_{j,t}, z_{i,t,1}+\mu, z_{i,t,2}) \right\}^2 \right) ,
    %
\end{align*}
with $d^*(\xi, z_1, z_2) = \arccos \left( \cos\left(\xi - \atantwo \left(  z_{1}, z_{2} \right) \right) \right)$. 
The elliptical slice sampler proceeds by first drawing an auxiliary matrix $\bfU^*$ from a matrix normal distribution,
\[
\bfU^{*} \mid \tau^2, \varsigma, \rho \sim \textrm{MN}\left( \bfU^{*} \mid \mathbf{0}, \bfOmega^{-1}(\rho), \tau^2\bfSigma^{-1}\right),
\]
along with a random threshold $c \sim \textrm{Unif}[0,1]$ and a random angle $a \sim \textrm{Unif} [0, 2\pi)$.  A proposal is then generated as $\bfZ^{*}_i = \cos(a) \bfZ_i  + \sin(a) \bfU^{*}$,
and this proposal is accepted if the ratio
$$
\frac{\prod_{t=1}^T \prod_{j=1}^{J_t} \left\{ \theta(z^*_{i,t,1}, z^*_{i,t,2}) \right\}^{y_{i,j}} \left\{  1 - \theta(z^*_{i,t,1}, z^*_{i,t,2})\right\}^{y_{i,j}}}{\prod_{t=1}^T \prod_{j=1}^{J_t} \left\{ \theta(z_{i,t,1}, z_{i,t,2}) \right\}^{y_{i,j}} \left\{  1 - \theta(z_{i,t,1}, z_{i,t,2})\right\}^{y_{i,j}}}
%
$$
is greater than $c$.  If the proposal $\bfZ_i^*$ is rejected, a new value of $a$ is sampled from a uniform distribution with a curtailed support and the corresponding $\bfZ_i^*$ is either accepted or rejected.  This process is repeated until either the proposal is accepted or the support for $a$ becomes empty.  Note that the auxiliary matrix $\bfU^*$ and the threshold $c$ remain the same for every proposed $\bfZ_i^{*}$ within a given iteration of the algorithm.

For the hyperparameters associated with the projected autoregressive process, note that the full conditional posterior distribution for $\rho$ given the random vectors $\bfv_i$ and $\bfw_i$ reduces to a distribution that is almost proportional to a truncated Gaussian distribution.  Hence, for this parameter, we develop a Metropolis-Hasting algorithm with independent proposals that has very high acceptance probabilities.  On the other hand, while we can directly sample from the individual full conditionals of $\mu$, $\tau$ and $\varsigma$, the very high autocorrelation among these three parameters leads us to prefer a trivariate random walk Metropolis Hastings algorithm to sample from their joint full conditional posterior distribution.  The variance-covariance matrix of the proposal distribution is tuned to target an approximate 30\% acceptance rate.  

The parameters associated with voting to affirm or reverse a given decision, $(\zeta_{j,t}, \psi_{j,t})$, are sampled independently using a Riemannian Manifold Hamiltonian Monte Carlo algorithm adapted to the circle \citep{GirolamiCalderheadRiemannManifoldLangevin2011}. Next, each $\kappa_{j,t}$ is sampled using univariate random walk Metropolis-Hastings algorithms with proposal distributions tuned to target a 40\% average acceptance rate on average. Finally, $\lambda$ is sampled via a Gibbs sampler because of conjugacy.  Full details of the MCMC strategy can be seen in the implementation available at \url{https://github.com/rayleigh/circ_supreme_court}.

\subsection{Parameter identifiability}

The likelihood function in \eqref{eq:likel} is invariant to shifts, reflections and wrapping of the latent space.  This means that model parameters are not identifiable from the likelihood alone.  We already mentioned in Section \ref{se:hyper} that we address invariance to shifts in the latent space by centering the prior of the ideal points at zero, leading to weak identifiability in the posterior distribution.  On the other hand, we address the invariance to wrappings and reflections by postprocessing the samples generated by the MCMC algorithm.  In particular, invariance to wrapping is addressed by mapping all angles to the $[-\pi,\pi]$ interval. While any interval of length $2\pi$ can be used, we map onto the $[-\pi,\pi]$ interval to be consistent with the fact that our marginal prior for $\beta_{i,t}$ is centered around zero. In the same spirit, invariance to reflections is addressed by fixing the sign of the ideal point of a selected number of justices.  In the case of the application to the U.S.\ Supreme Court discussed in Section \ref{se:analysisSCOTUS}, this involves making the ideal points of Justice William Douglas negative every term between 1938 to 1966, that of Justice Thurgood Marshall also negative from 1967 to 1990, and that of Justice Clarence Thomas positive from 1991 on. 

It is important to note here that, unlike models with Euclidean latent spaces, models based on circular spaces do not suffer from invariance to rescalings of the latent space.  This follows from the fact that distances in the latent circular space are bounded from above by $\pi$.  One consequence of this is that the concentration parameters $\kappa_{j,t}$'s are identifiable from the data without the need to introduce additional constraints.

\section{A case study:  The evolution of justices' preferences in the U.S.\ Supreme Court}\label{se:analysisSCOTUS}

In this section, we analyze the data introduced in Section \ref{se:data_overview} using the circular model described in this paper, and compare the results of the analysis with those obtained using the methodology of \cite{MartinQuinnDynamicIdealPoint2002a} (in the sequel, MQ) as implemented in the \texttt{R} package \texttt{mcmcPack} \citep{MartinQuinnParkMCMCpack}.  Results for our model are based on 40,000 samples of the posterior distribution obtained from the algorithm described in Section \ref{section:computation} after burning 250,000 iterations and thinning the original chain every 25 observations.  Results for the MQ model are based on 40,000 posterior samples obtained after burning 20,000 iterations.
\begin{figure}[!ht]
    \centering
    \includegraphics[width = .7\textwidth]{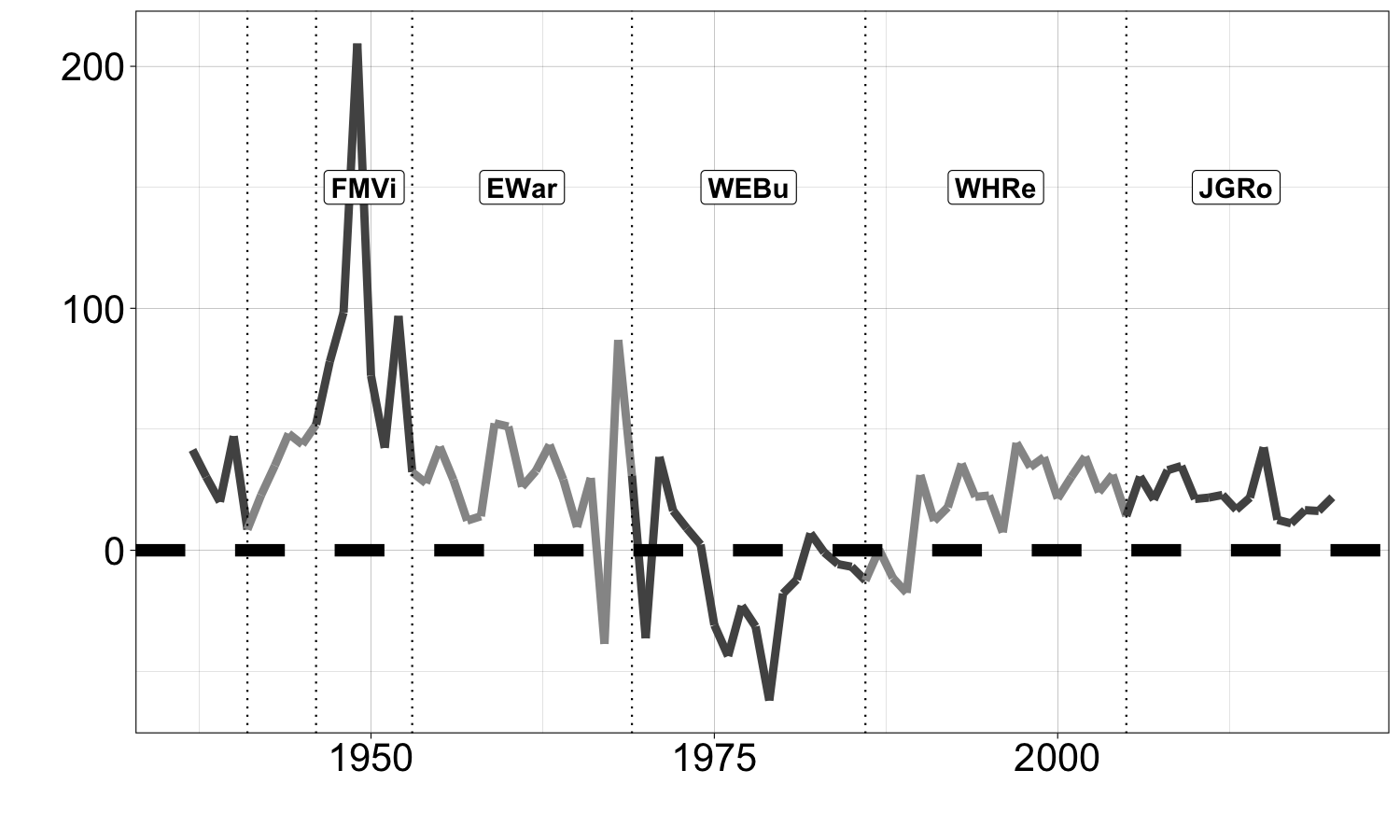}
    \caption{Difference between the WAIC computed for the MQ and circular models. Here, a negative difference indicates that MQ is preferred, whereas a positive difference indicates the circular model is preferred. 
}
\label{fig:WAIC_comparison_new}
\end{figure}

\begin{figure}[!ht]
\centering
\begin{subfigure}[t]{\columnwidth}
    \centering
    \includegraphics[width = .7\textwidth]{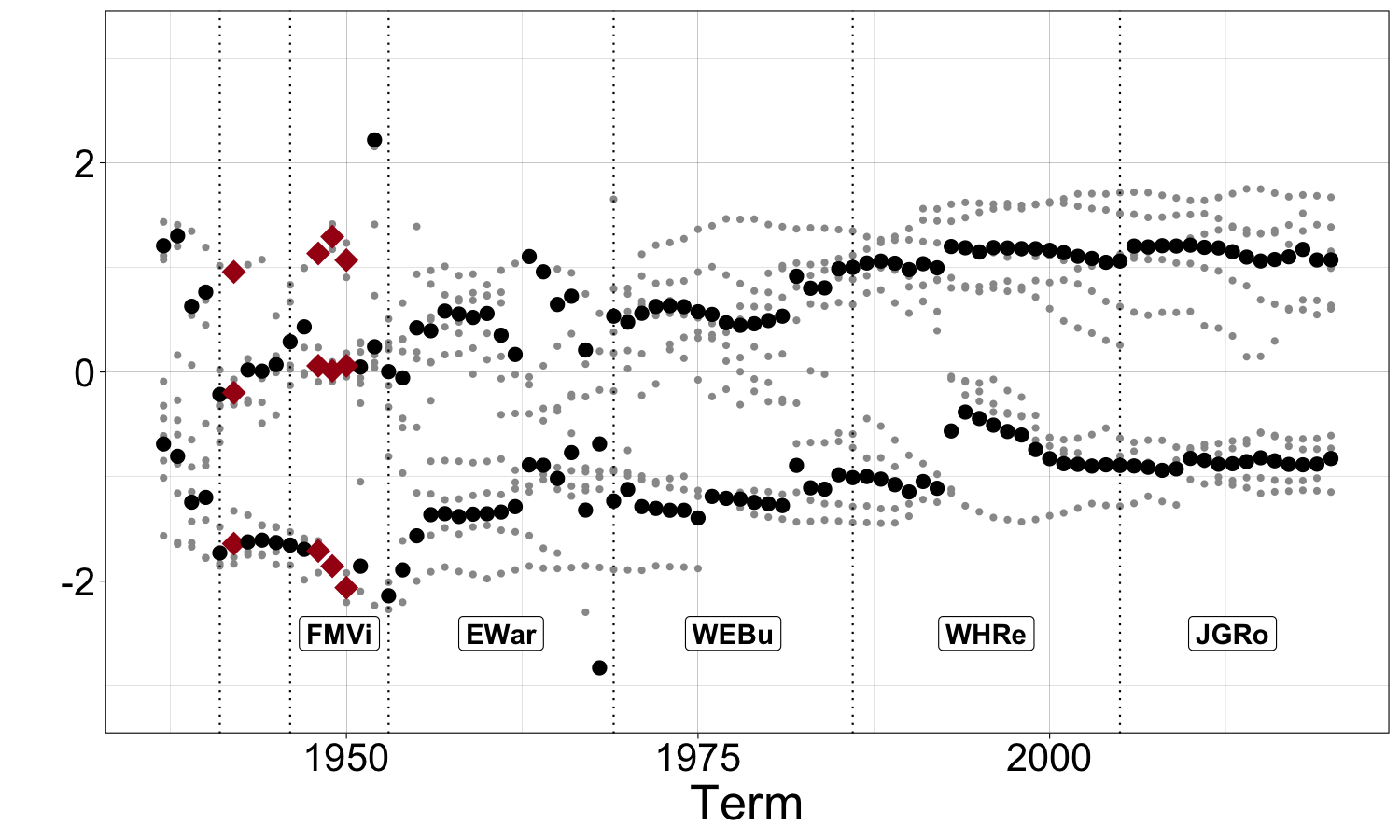}
    \caption{Posterior means of ideal points and cluster means}
    \label{fig:means_over_time}
\end{subfigure}\\
\begin{subfigure}[t]{\columnwidth}
    \centering
    \includegraphics[width = .7\textwidth]{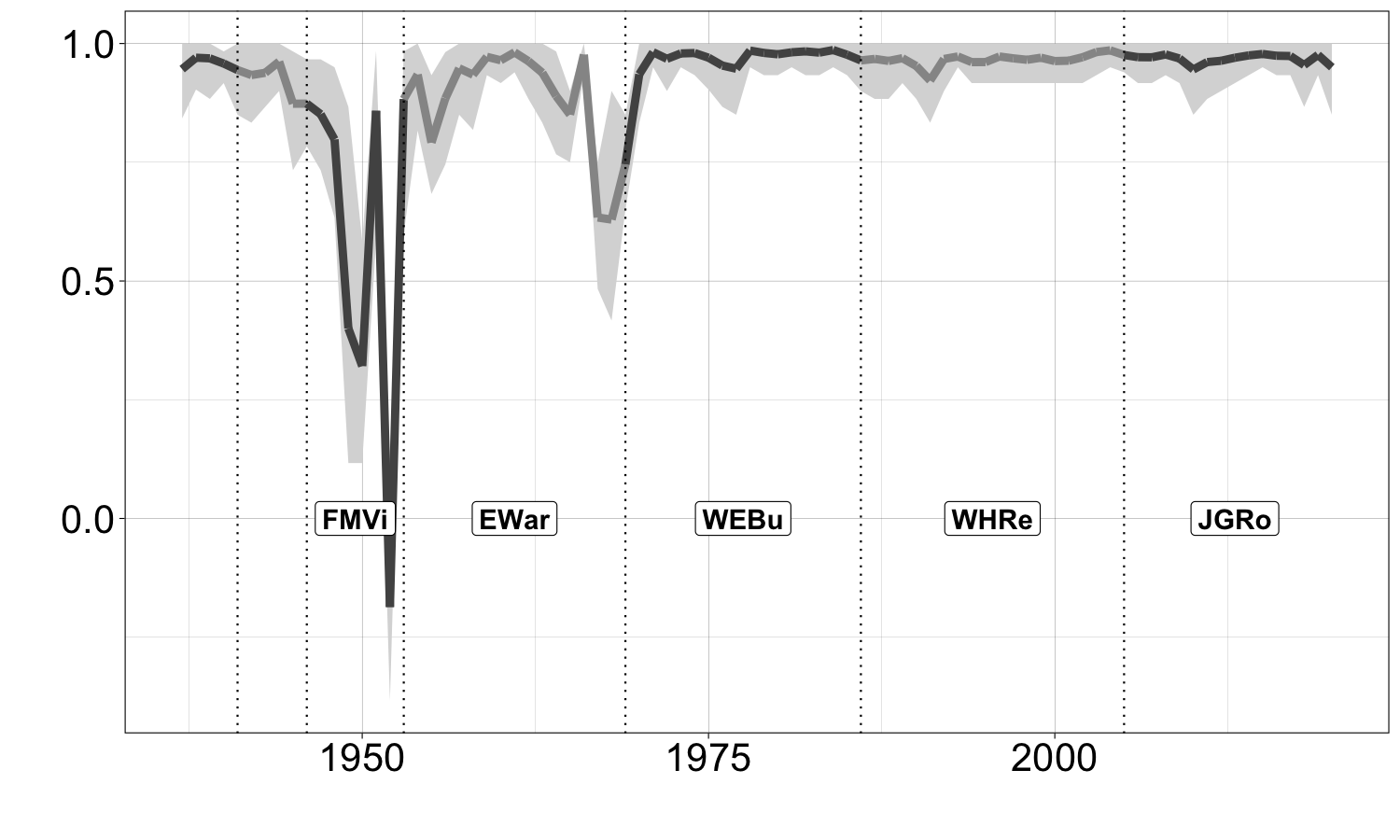}
    \caption{Spearman correlation among ranks}\label{fig:kendall_corr}
\end{subfigure}
\caption{The top plot shows the posterior mean of the ideal points of the justices as grey dots and the optimal cluster means of these ideal points as either black dots (for terms in which we identify two clusters of ideal points) or red rhombi (for the terms in which we identify 3 clusters). We use a Euclidean scale for this plot to facilitate visualization.
The bottom plot shows the posterior mean (solid line) and corresponding 95\% credible intervals (shaded region) for the Spearman correlation between the justices' rankings generated by unfolding our circular model and those recovered from the MQ model.}
\label{fig:rank_WAIC_comparison}
\end{figure}

We start our analysis by investigating model fit using the Watanabe-Akaike Information Criterion (WAIC, \citealp{watanabe2010asymptotic}, \citealp{watanabe2013widely}, \citealp{gelman2014understanding}).  Similarly to the well-known Akaike information criteria (AIC) and the Bayesian information criteria (BIC), WAIC balances goodness of fit against model complexity. However, unlike AIC and BIC, WAIC is well suited for hierarchical models where the number of effective parameters can be much smaller than the headline number. Compared to the deviance information criteria (DIC), the WAIC has the additional advantage of being invariant to reparameterizations of the model \citep{gelman2014understanding,spiegelhalter2014deviance}.  We compute the WAIC for model $m$ during term $t$ from the output of the Markov chain Monte Carlo algorithms as:
\begin{multline}\label{eq:lWAIC}
    WAIC_{t}(m) = -2\left[\sum_{i=1}^{I} \log\left( \textrm{E}_{\textrm{post}} \left\{ 
      \prod_{j}
    \theta_{i,j,t}(m)^{y_{i,j,t}} \left[1 -\theta_{i,j,t}(m)\right]^{1 - y_{i,j,t}}
    \right\} \right) \right. \\
    - \left. \sum_{i=1}^{I} \textrm{var}_{\textrm{post}}\left\{ \sum_{j}
    \left[ y_{i,j,t} \log \theta_{i,j,t}(m) + (1-y_{i,j,t}) \log(1-\theta_{i,j,t}(m)) \right] \right\}\right],
\end{multline}
where $\textrm{E}_{\textrm{post}}\{ \cdot \}$ and $\textrm{var}_{\textrm{post}}\{ \cdot \}$ denote the mean and variances computed over the posterior distribution, and $\theta_{i,j,t}(m)$ represents the probability that Justice $i$ votes to reverse the lower court decision on case $j$ during term $t$ under model $m$.  Figure \ref{fig:WAIC_comparison_new} presents the difference between the value of the WAIC for the circular model and that for the MQ model, $WAIC_t(MQ) - WAIC_t(C)$, so that positive values favor our circular model. The graph shows evidence that our model outperforms MQ over most of the period under study.  The main exceptions arise during Burger's court and the first  few terms of Rehnquist's court.

Next, we present in Figure \ref{fig:means_over_time} the posterior means of the ideal points of the justices along with cluster centers during each term, and in Figure \ref{fig:kendall_corr} the posterior mean and associated 95\% credible intervals for the Spearman correlation between the justices' ranks generated from the MQ scores and those implied by our model through the unfolding of the latent circular space (recall our discussion in Section \ref{se:interpretation}).  In Figure \ref{fig:means_over_time}, the clusters associated with  each term are identified by independently fitting a mixture of von Mises distributions with component-specific means and precisions to the posterior means of the justices' ideal points. The model is fitted with 2 to 5 components, and the optimal number is selected through a slight modification of the BIC criteria introduced in \cite{FraleyRafteryHowManyClusters1998}.
We can see that, starting in the early 70s (shortly after Warren E.\ Burger took over as Chief Justice), the ideal points consistently fall within arcs of length approximately $\pi$ and are centered roughly around 0.  Furthermore, our clustering procedure consistently indicates the presence of two clusters during this period.  As we discussed in Section \ref{se:interpretation}, both of these features mean that recovering ideological ranks from the circular scores by projecting onto a tangent space is legitimate during this period.  Interestingly, we can see from Figure \ref{fig:kendall_corr} that these ranks are, for the most part, closely aligned with those recovered from MQ. Nonetheless, the WAIC scores for our model are higher than those for MQ, which can be explained by the fact that end-against-the-middle votes, which were relatively rare during the middle years of Burger's court, became increasingly common in the Rehnquist and Roberts' courts (recall Figure \ref{fig:votesexamples} and the associated discussion in Section \ref{se:model}). 
On the other hand, before the early 70s, the ideal points of legislator under our circular model are often spread over wider arcs and sometimes show estimates that are close to either $\pi$ or $-\pi$ and/or the clustering procedure identifies 3 voting blocks.  As would be expected, there is less agreement during these periods between the rankings recovered by unfolding our circular scores and those recovered from the Martin-Quinn scores.  There is particularly wide disagreement in the late 1940s/early 1950s and during the late 1960s.  We come back to discuss those periods in more detail in Sections \ref{se:SCOTUS1949-1952} and \ref{se:SCOTUS1967-1970}.

\begin{figure}[!ht]
\begin{subfigure}{\textwidth}
    \centering
    \includegraphics[width = 0.37\textwidth]{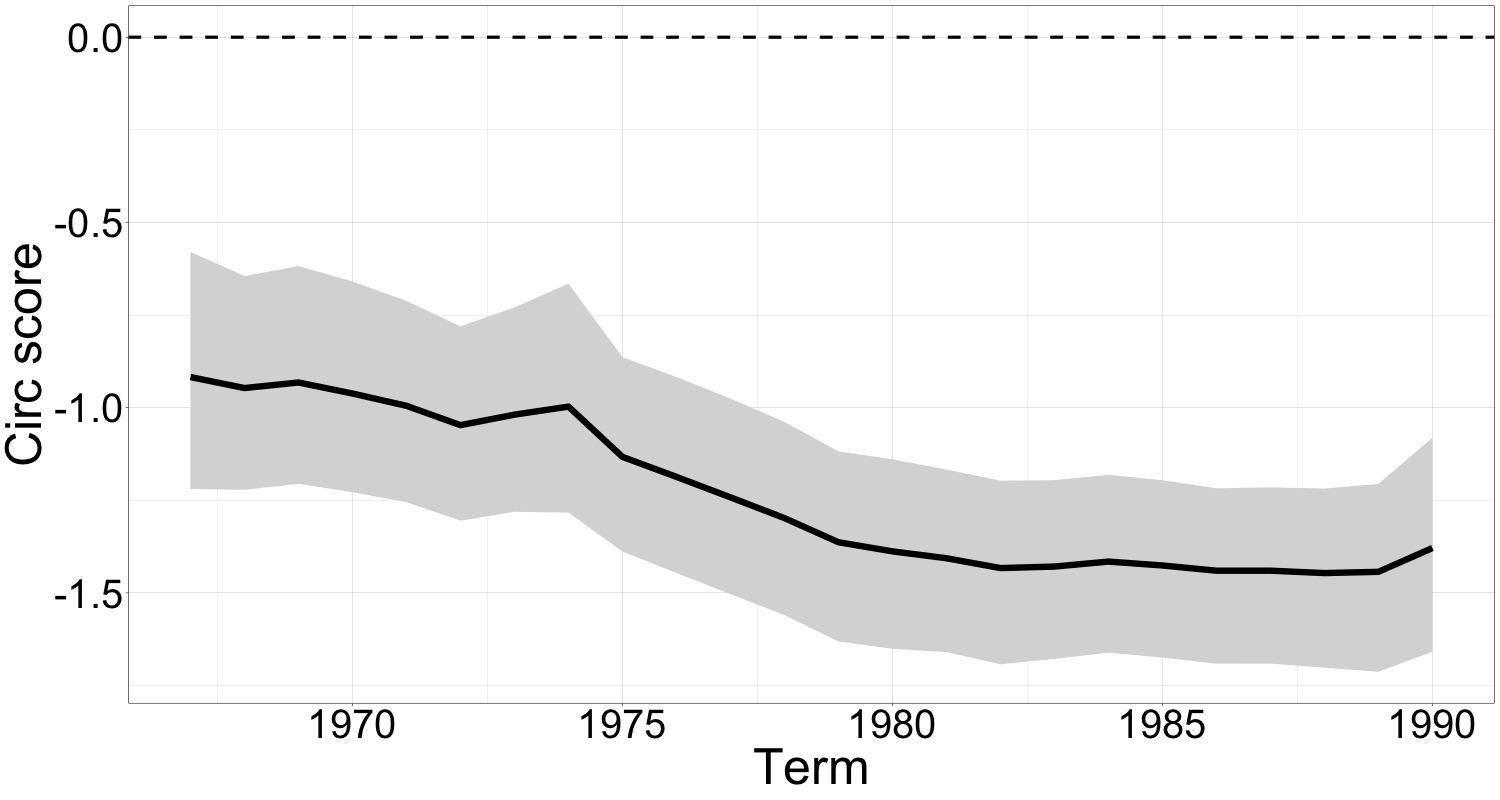}
    \includegraphics[width = 0.37\textwidth]{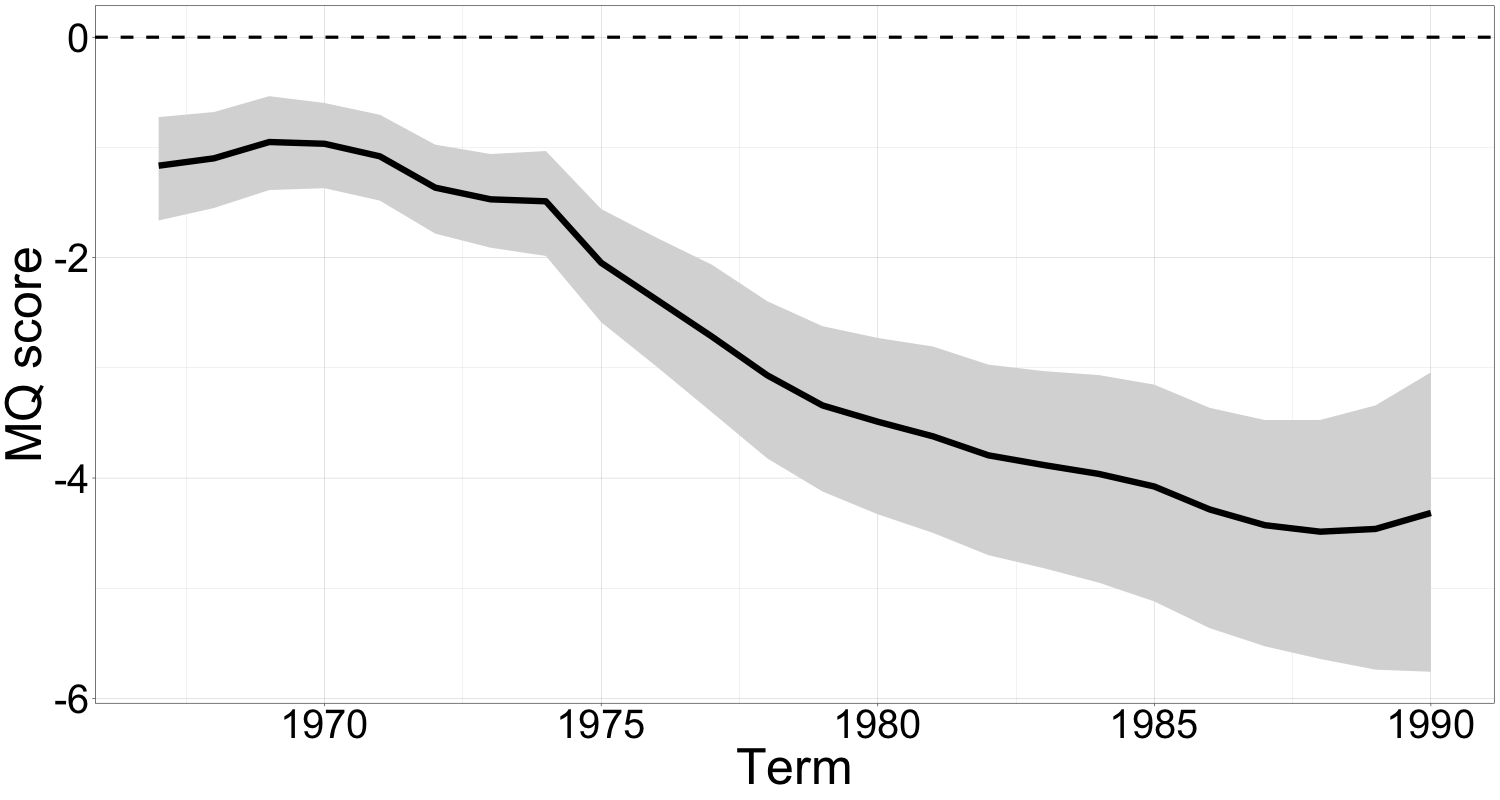}
    \subcaption{Thurgood Marshall}
\end{subfigure}

\begin{subfigure}{\textwidth}
    \centering
    \includegraphics[width = 0.37\textwidth]{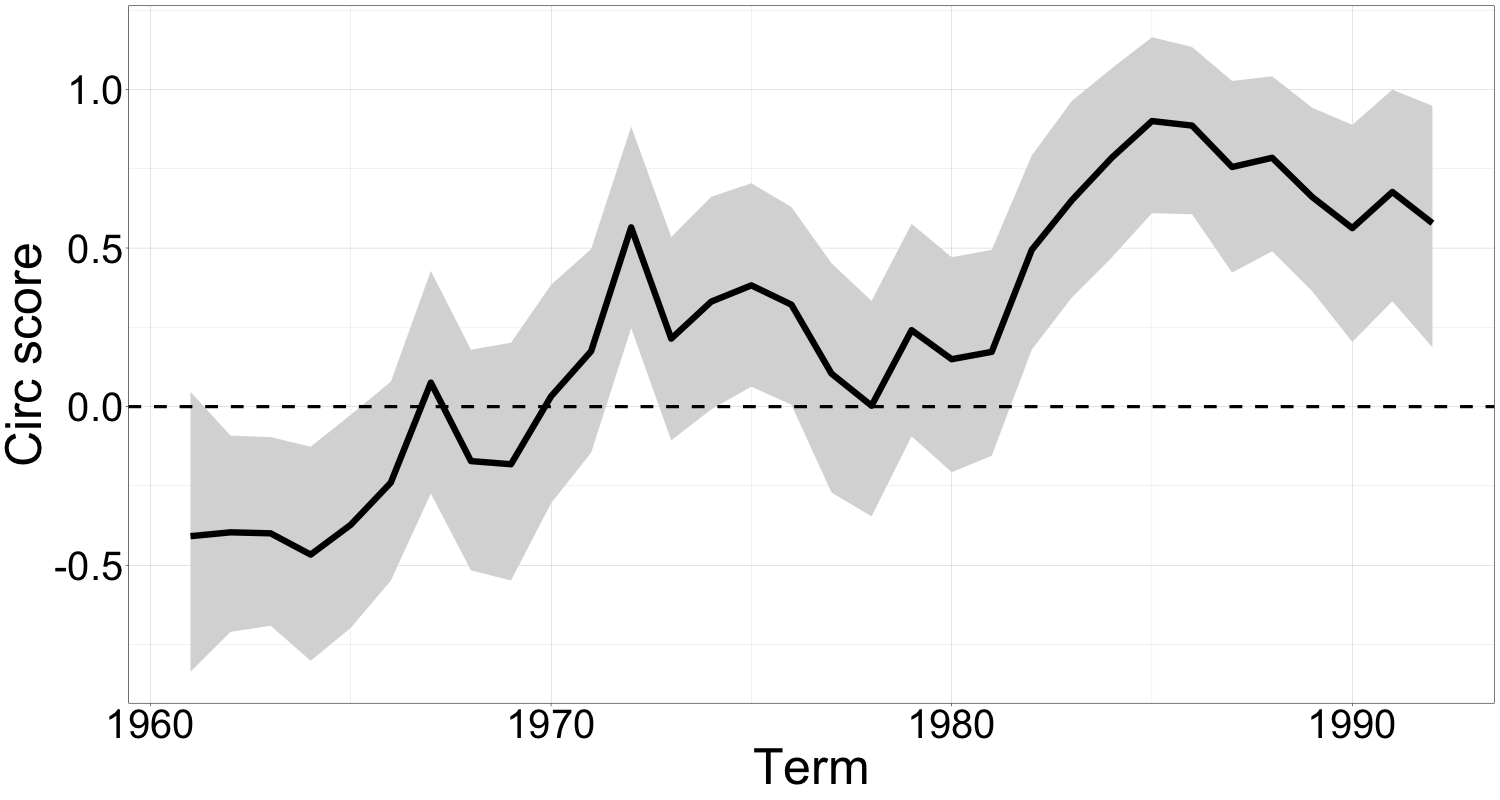}
    \includegraphics[width = 0.37\textwidth]{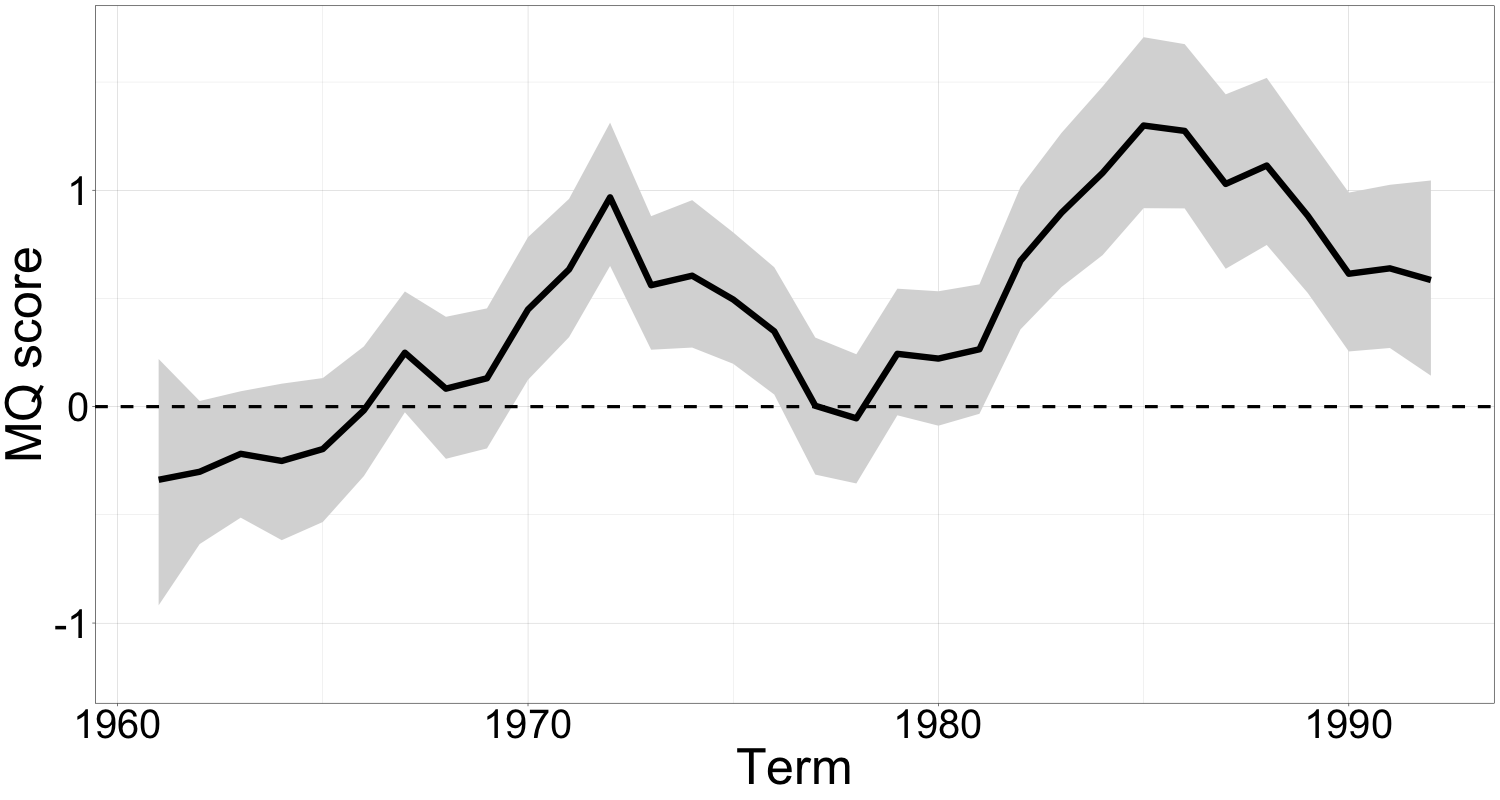}
    \subcaption{Byron White}
\end{subfigure}  

\begin{subfigure}{\textwidth}
    \centering
    \includegraphics[width = 0.37\textwidth]{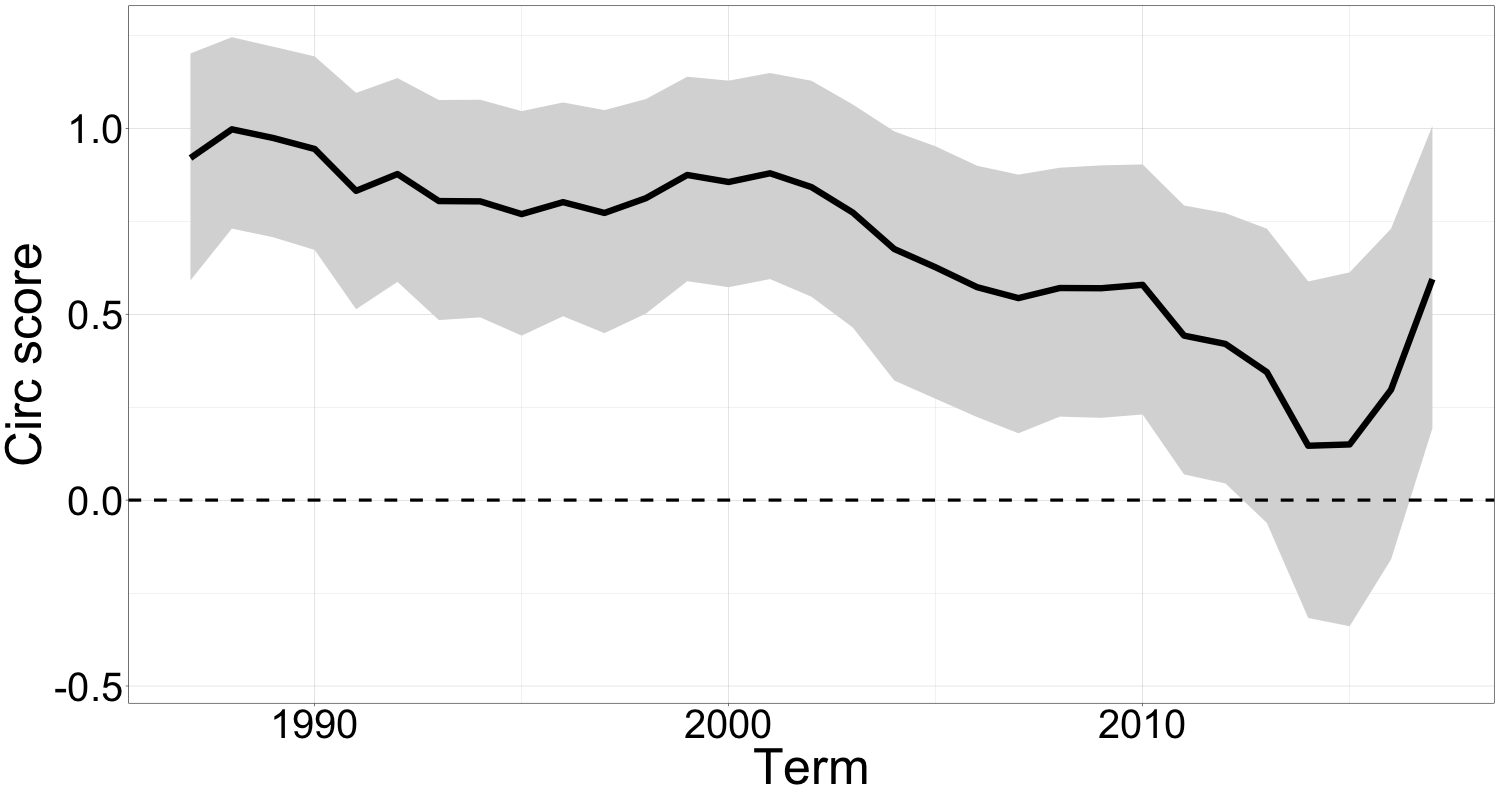}
    \includegraphics[width = 0.37\textwidth]{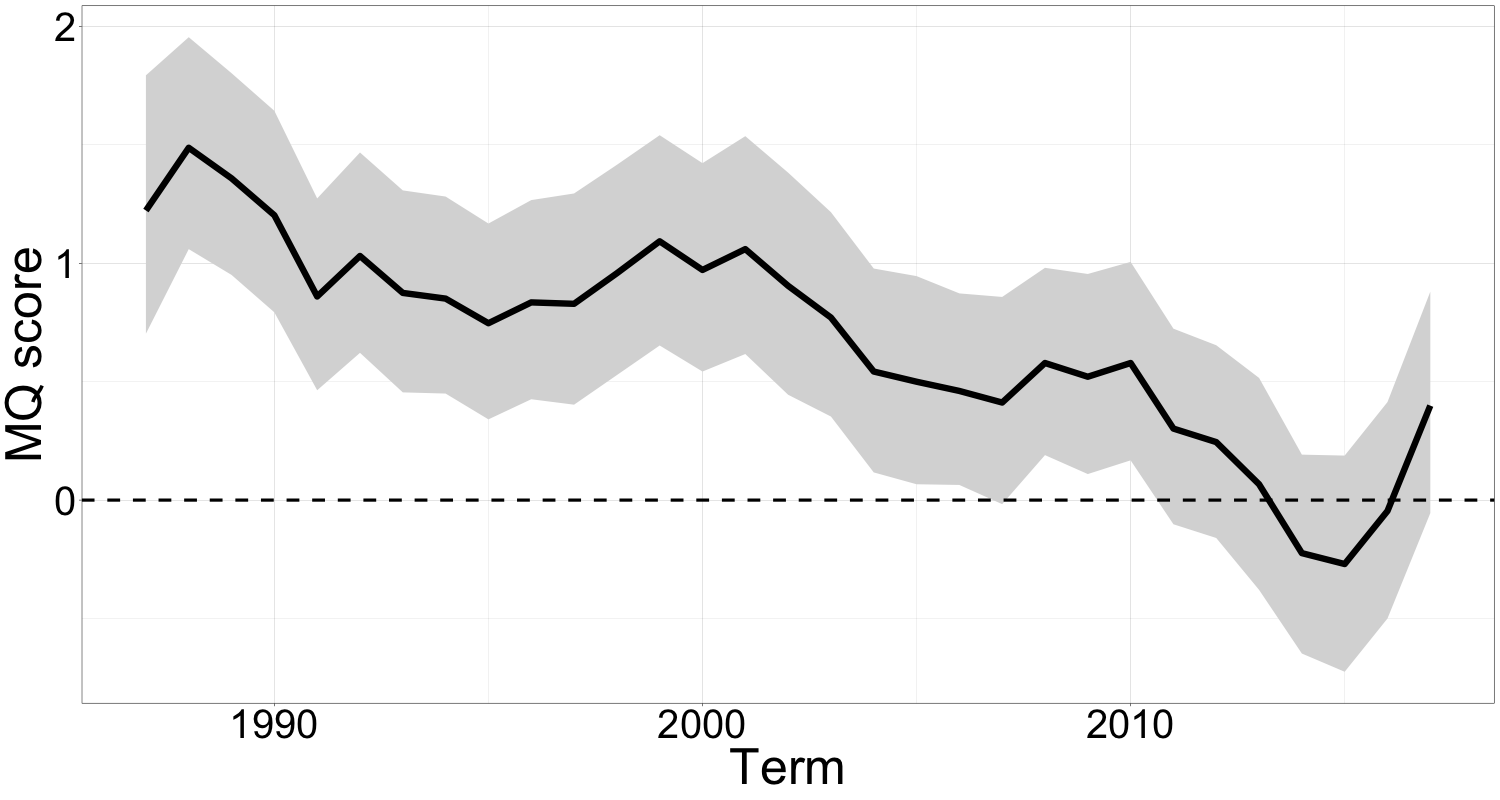}
    \subcaption{Anthony Kennedy}
\end{subfigure}    

\begin{subfigure}{\textwidth}
    \centering
    \includegraphics[width = 0.37\textwidth]{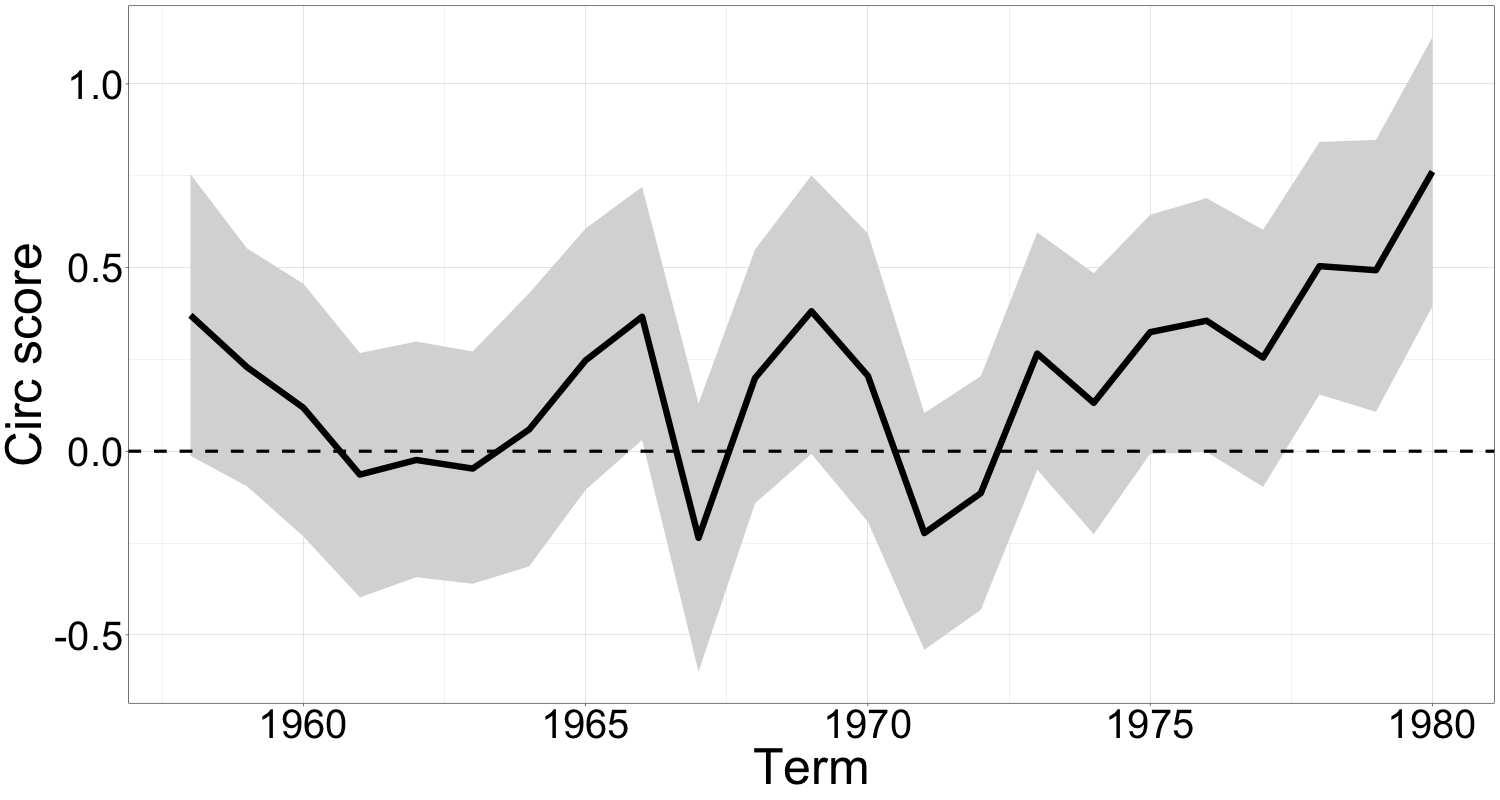}
    \includegraphics[width = 0.37\textwidth]{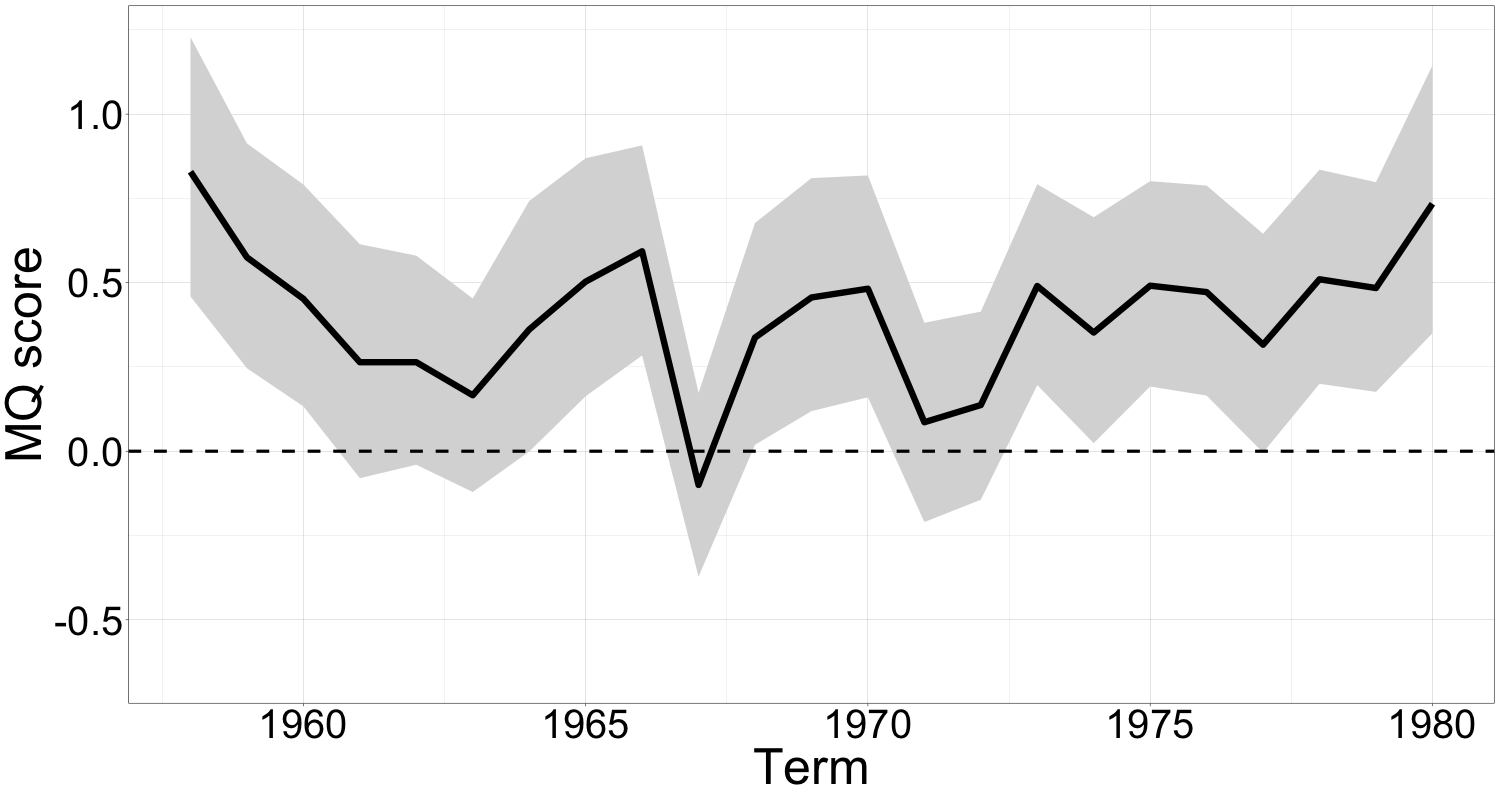}
    \subcaption{Potter Stewart}
\end{subfigure}    

\begin{subfigure}{\textwidth}
    \centering
    \includegraphics[width = 0.37\textwidth]{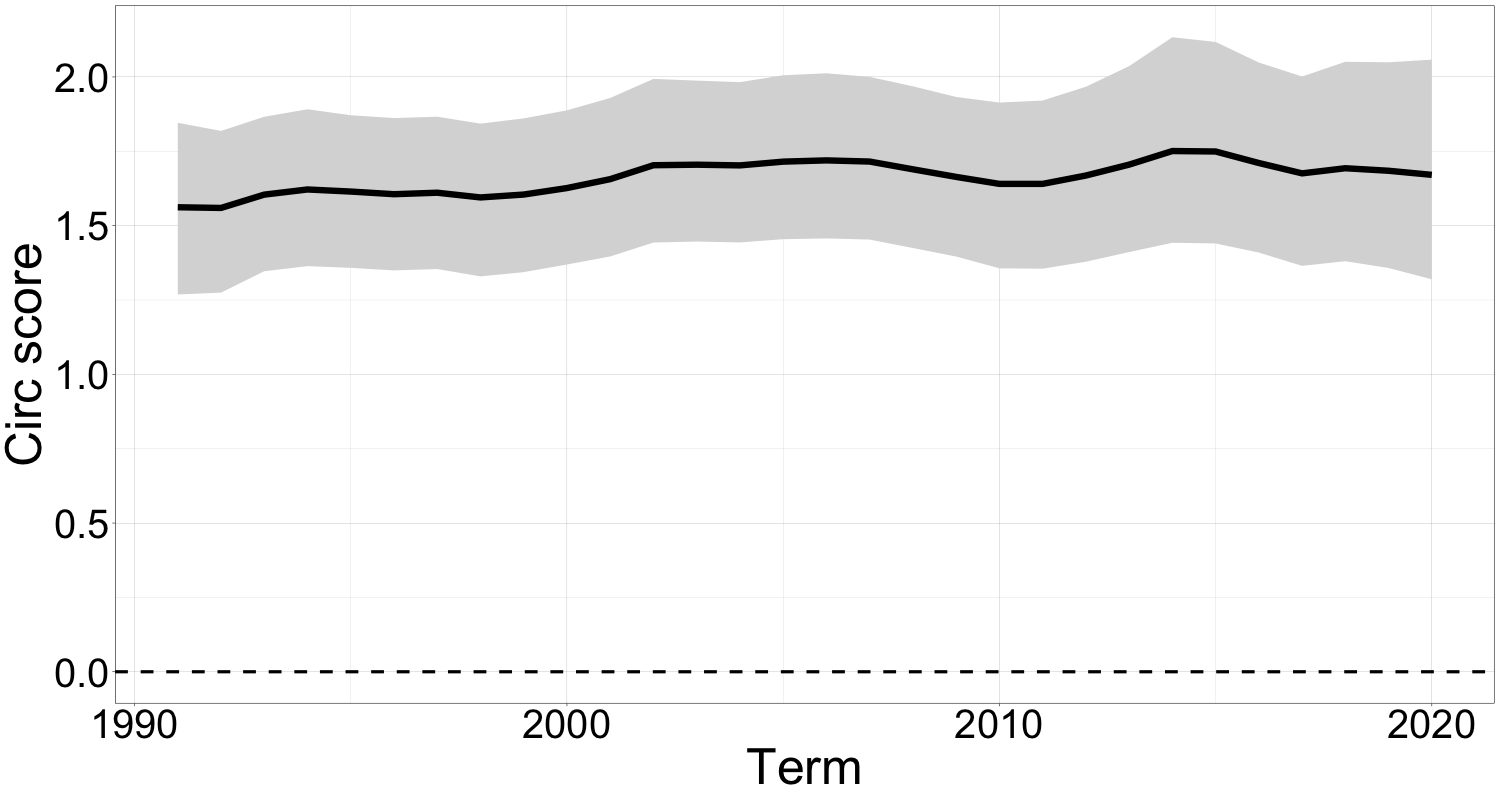}
    \includegraphics[width = 0.37\textwidth]{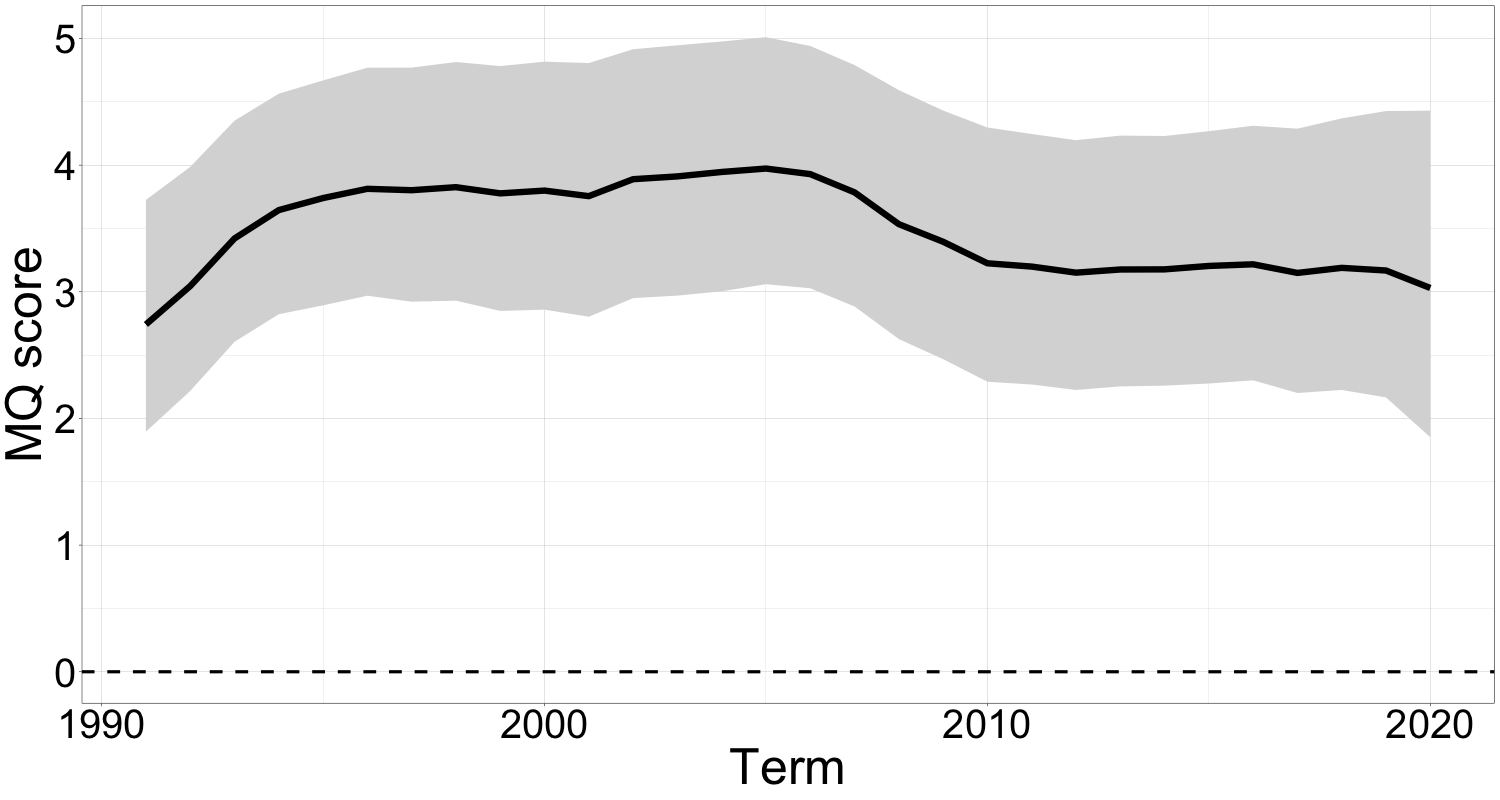}
    \subcaption{Clarence Thomas}
\end{subfigure}  
\caption{Posterior mean (solid line) and 95\% credible intervals for the trajectories of the ideal points of Justices Thurgood Marshall, Byron White, Anthony Kennedy, Potter Stewart, and Clarence Thomas. The left column displays estimates based on our circular model, whereas the right column displays the MQ scores.  Circular scores are presented using Euclidean coordinates to facilitate comparisons.
}
\label{fig:trajectories1}
\end{figure}

\begin{figure}[!ht]
\centering
\begin{subfigure}{\textwidth}
    \centering
    \includegraphics[width = 0.37\textwidth]{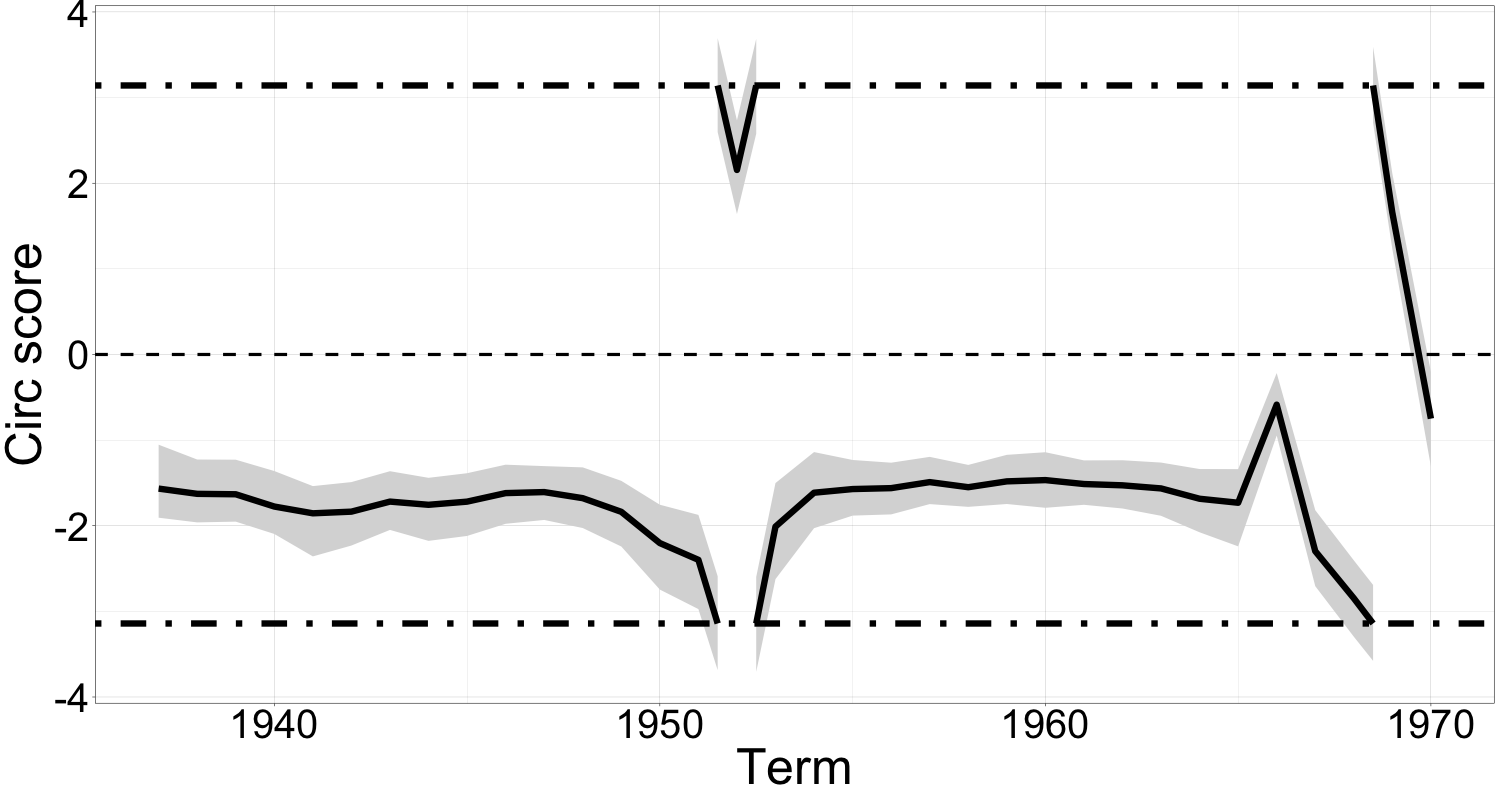}
    \includegraphics[width = 0.37\textwidth]{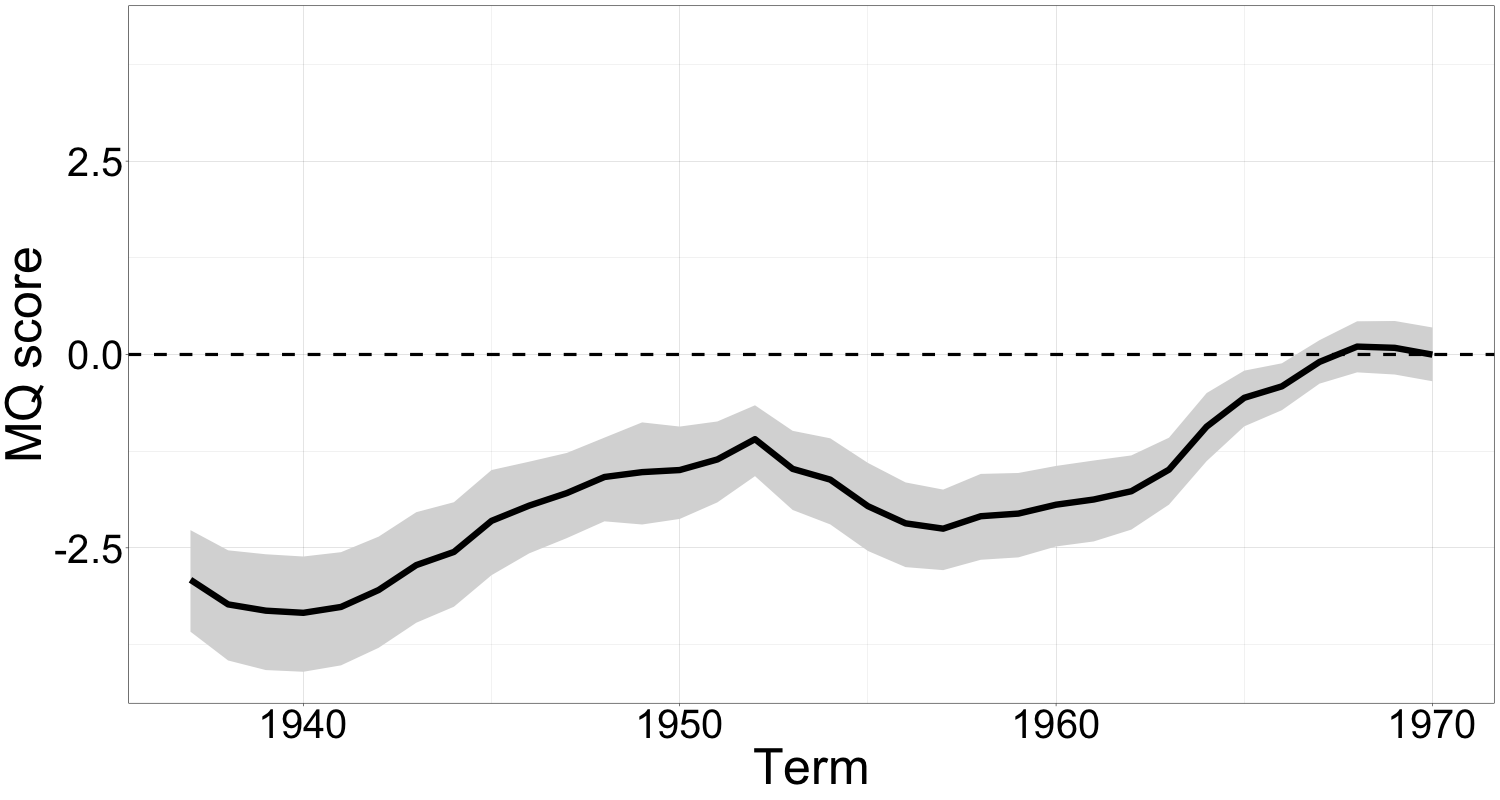}
    \subcaption{Hugo Black}
\end{subfigure}  
\begin{subfigure}{\textwidth}
    \centering
    \includegraphics[width = 0.37\textwidth]{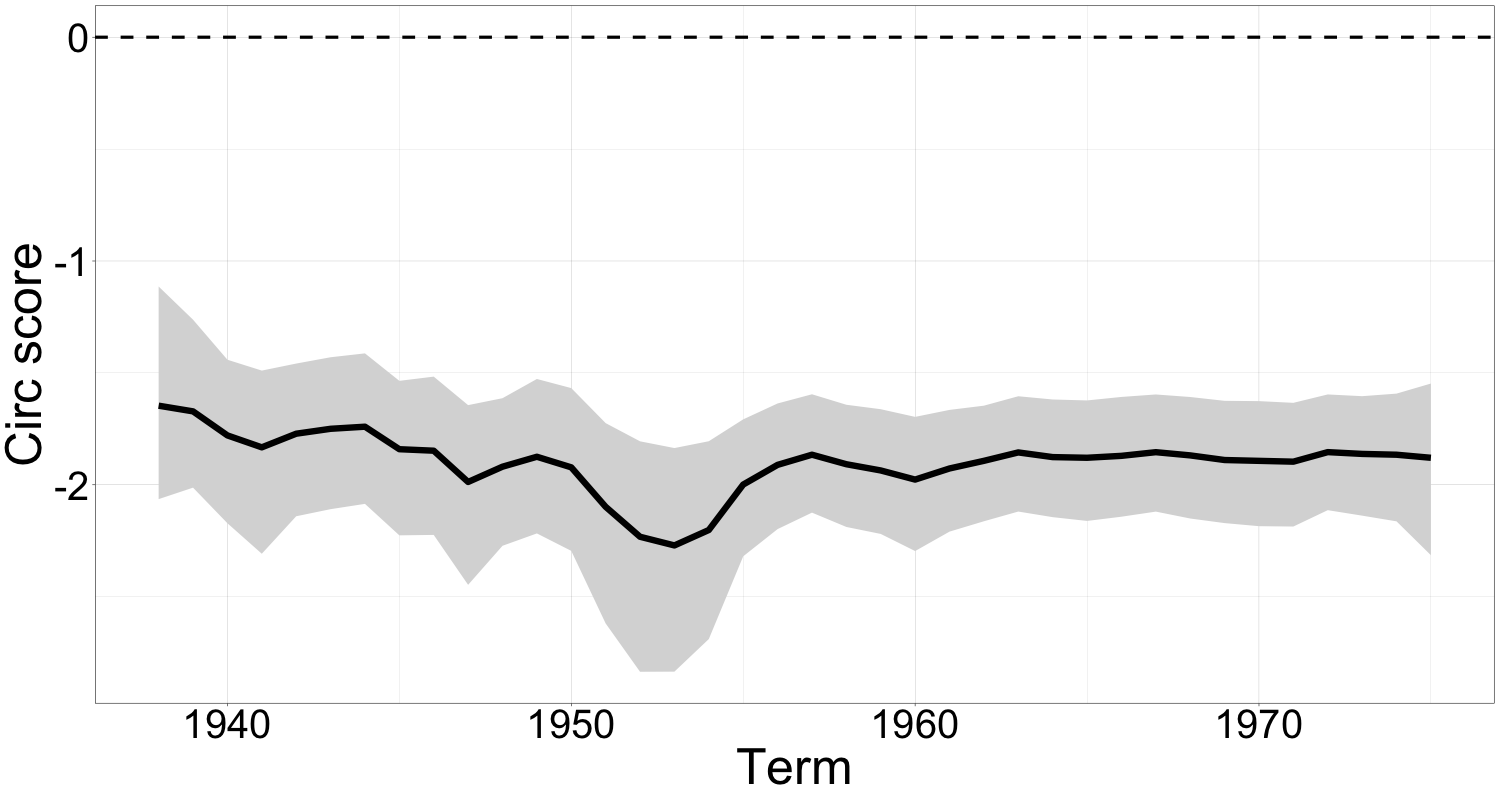}
    \includegraphics[width = 0.37\textwidth]{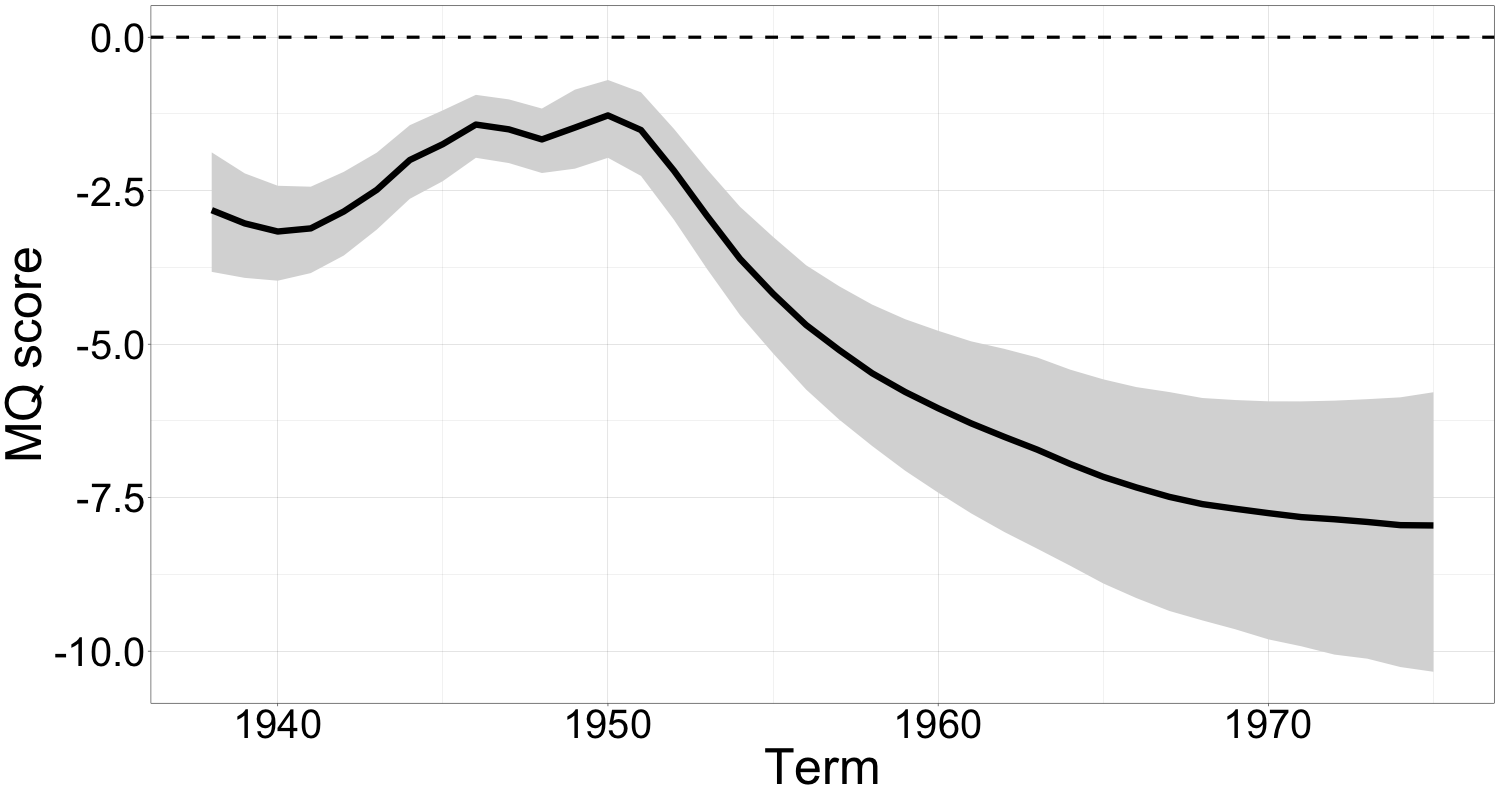}
    \subcaption{William Douglas}
\end{subfigure}
\begin{subfigure}{\textwidth}
    \centering
    \includegraphics[width = 0.37\textwidth]{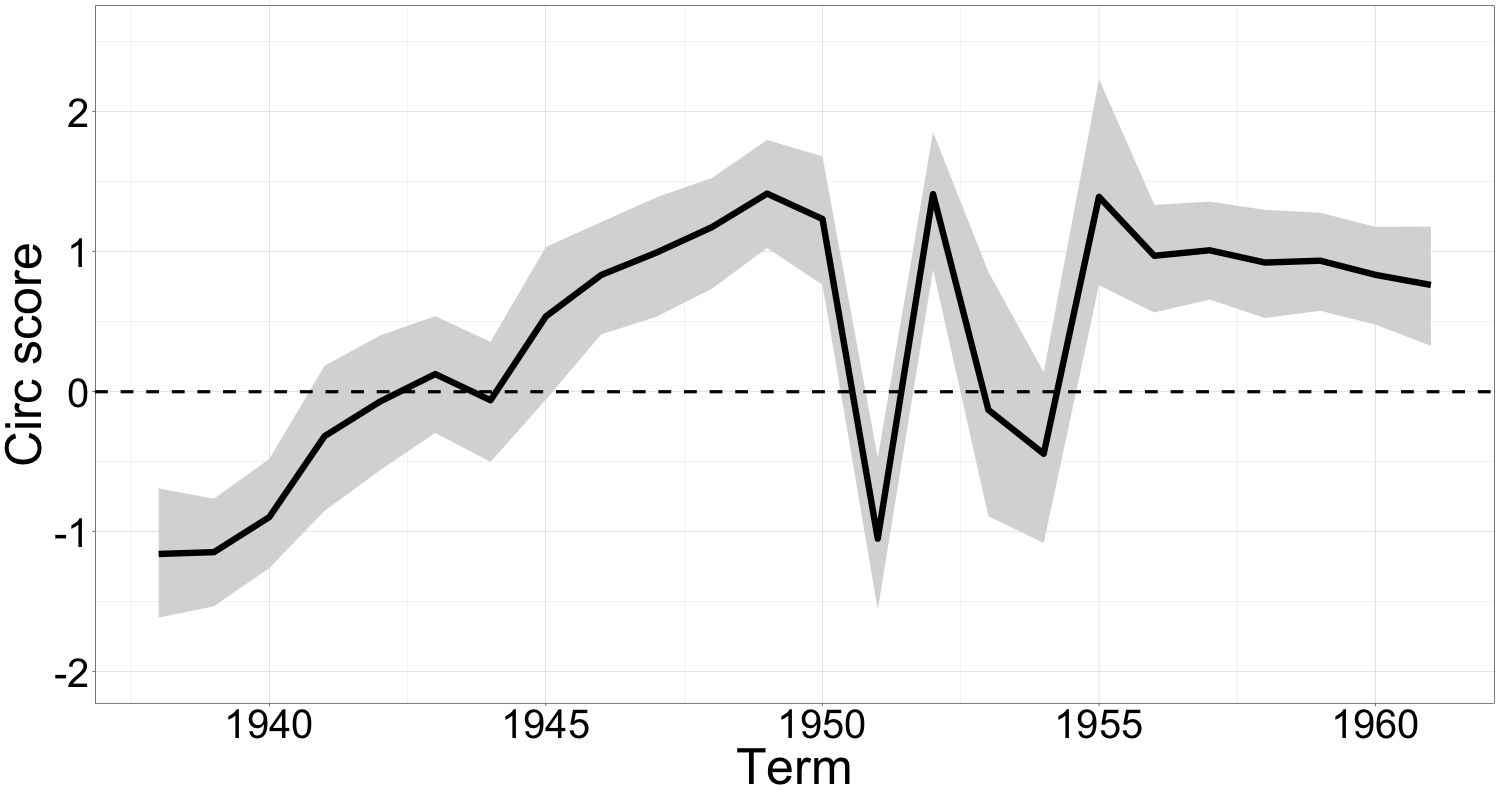}
    \includegraphics[width = 0.37\textwidth]{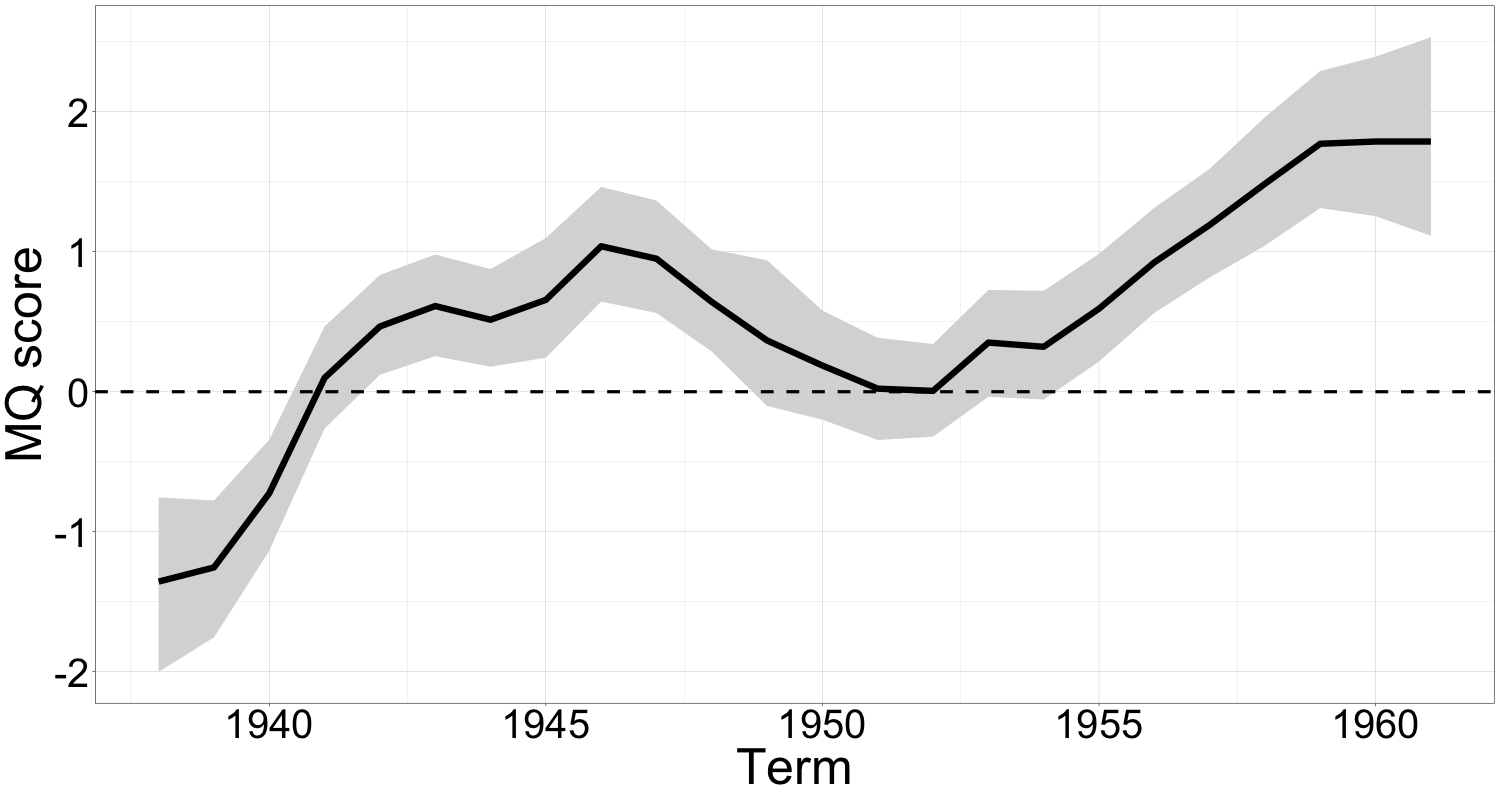}
    \subcaption{Felix Frankfurter}
\end{subfigure} 
\caption{Posterior mean (solid line) and 95\% credible intervals for the trajectories of the ideal points of Justices William Douglas, Hugo Black and Felix Frankfurter. The left column displays estimates based on our circular model, whereas the right column displays the MQ scores. Zero is indicated by a dashed line and $\pm\pi$ by a dotted line.  Circular scores are presented using Euclidean coordinates to facilitate comparisons.}
\label{fig:trajectories2}
\end{figure}

In order to provide additional insight into our circular model, Figures \ref{fig:trajectories1} and \ref{fig:trajectories2} show the estimated trajectories of the ideal points of eight justices with relatively long tenures under our circular (left column) and the MQ (right column) models.  The trajectories for the five justices included in Figure \ref{fig:trajectories1} are very similar under both models.  These similarities are representative of the results for most (but not all!) of the 48 justices in the dataset.  In contrast, Figure \ref{fig:trajectories2} shows three examples of judges for which the estimated trajectories of the ideal points arising from our circular model are qualitatively different from those that arise under the MQ model.  The most striking difference is for Justice Hugo Black. While the MQ model suggests that Black started as a liberal who tended to moderate his positions over time to become a centrist, the circular model suggests a much more stable behavior, with ideal points that are mostly close to the $\pm \pi$ boundary.  Similarly, the MQ model has Justice William Douglas becoming increasingly more liberal starting in 1960, while our circular model yields a stable ideological profile.  Finally, while the overall shape of Justice Felix Frankfurter's trajectory is similar across both models and reflects increasing conservative leanings in his voting patters, the circular model identifies a couple of spikes in the 1951 and 1953-1954 terms that are not present in the MQ scores.

\begin{figure}[t]
 \centering
   \begin{subfigure}{0.32\textwidth}
   \centering
     \includegraphics[width = 0.8\textwidth]{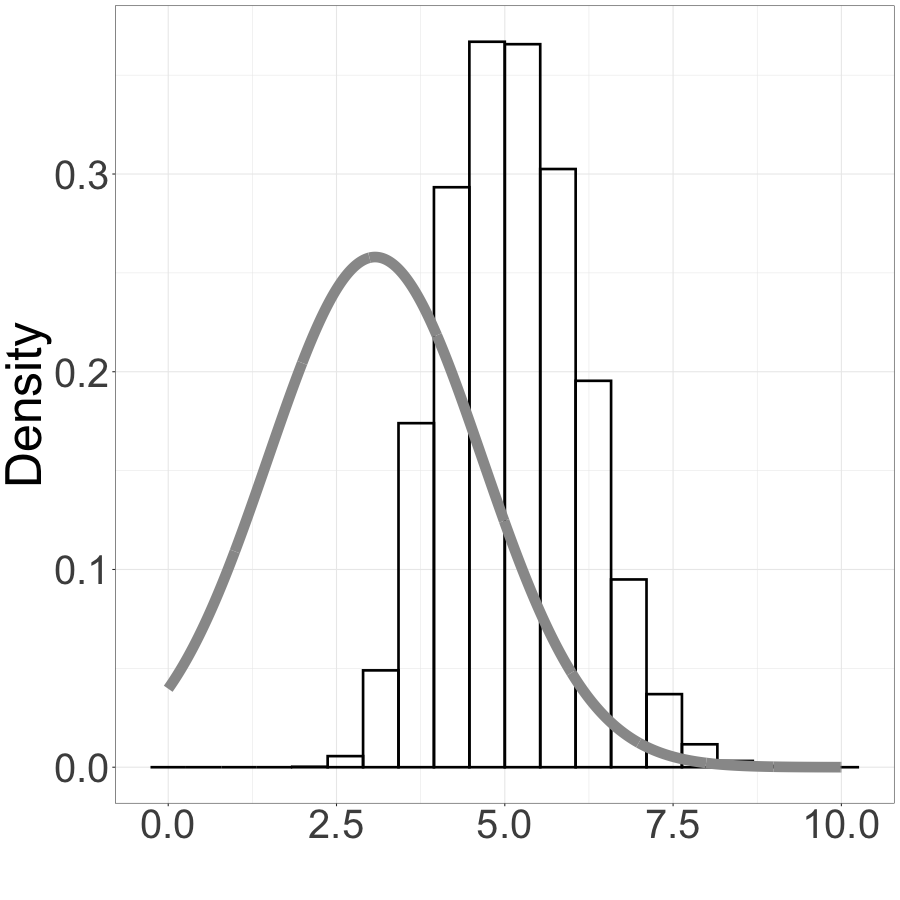}
     \subcaption{$\mu$}
   \end{subfigure}  
   \begin{subfigure}{0.32\textwidth}
   \centering
     \includegraphics[width = 0.8\textwidth]{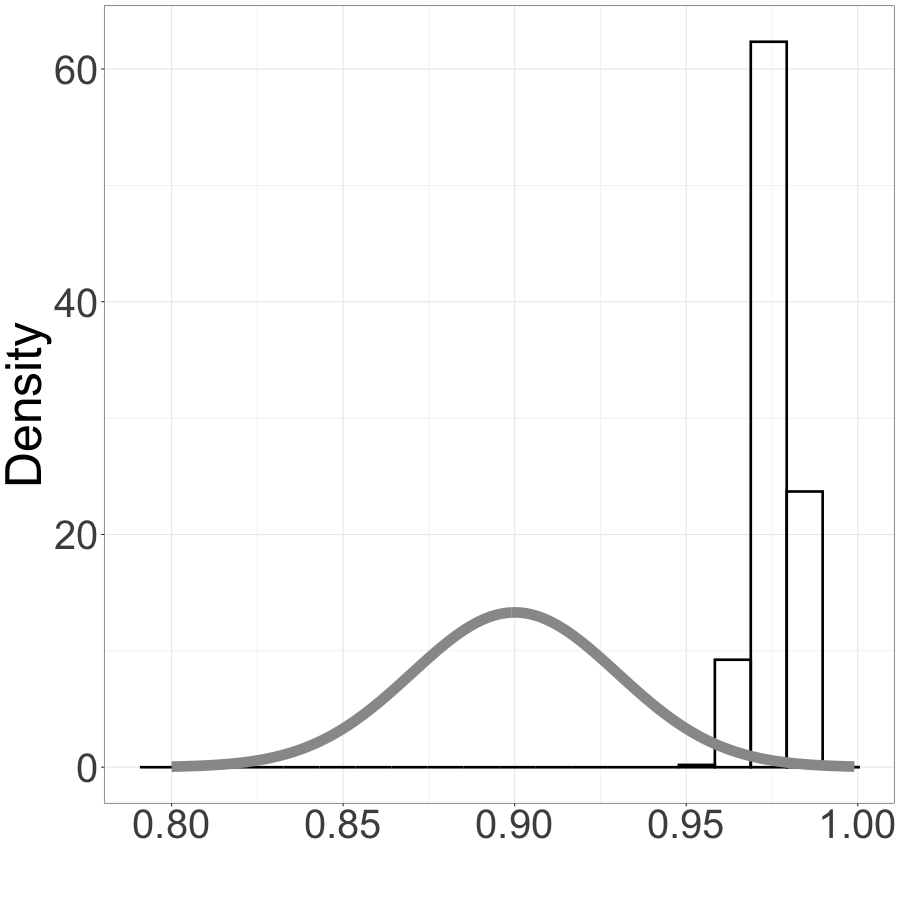}
     \subcaption{$\rho$}
   \end{subfigure}  
   \begin{subfigure}{0.32\textwidth}
   \centering
     \includegraphics[width = 0.8\textwidth]{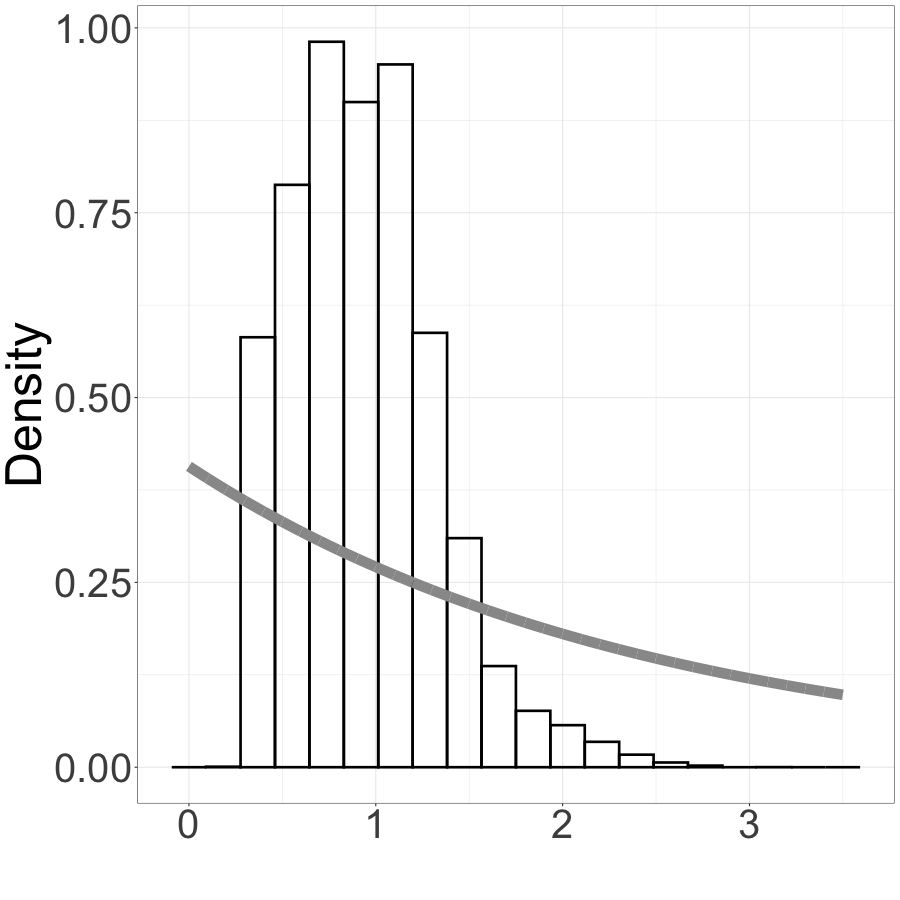}
     \subcaption{$\tau^2$}
   \end{subfigure}  \\
   \begin{subfigure}{0.32\textwidth}
   \centering
     \includegraphics[width = 0.8\textwidth]{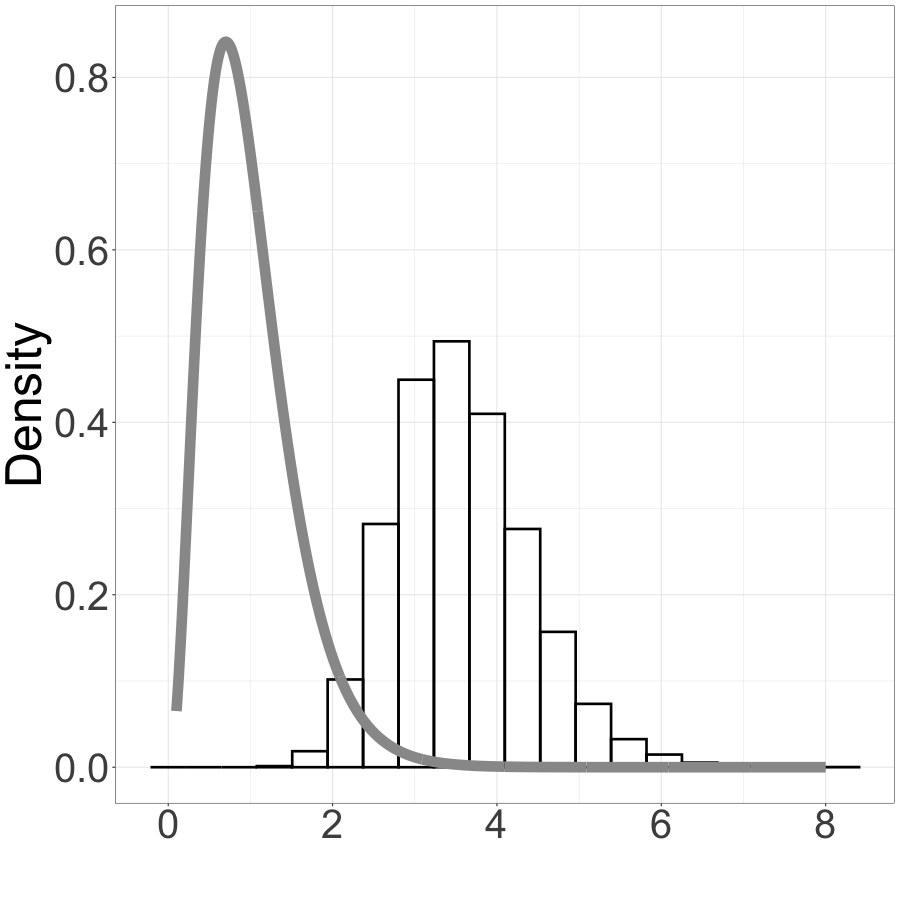}
     \subcaption{$\varsigma$}
   \end{subfigure}  
   \begin{subfigure}{0.32\textwidth}
   \centering
     \includegraphics[width = 0.8\textwidth]{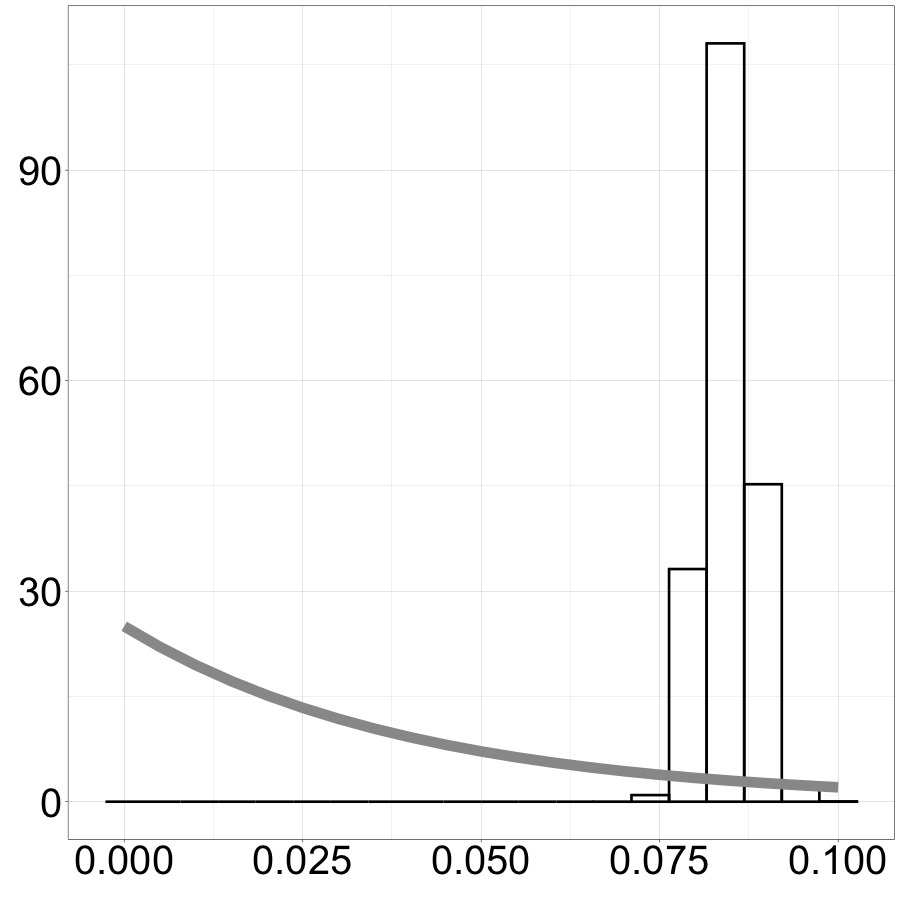}
     \subcaption{$\lambda$}
   \end{subfigure}  
   \begin{subfigure}{0.32\textwidth}
   \centering
    \includegraphics[width = 0.67\textwidth]{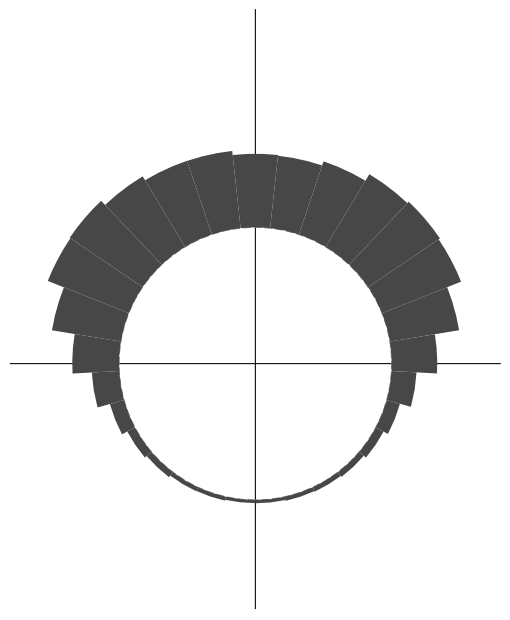}
    \subcaption{$\beta_{i, j}$} \label{fig:posteiorbeta}
   \end{subfigure}  

    \caption{Histogram of the posterior distribution of various hyperparameters and the implied  predictive distribution for the ideal point $\beta_{i, j}$ of new Justices. Solid lines correspond to the respective prior distributions.}
    \label{fig:posteriorshyperparameters}
\end{figure}

Finally, Figure \ref{fig:posteriorshyperparameters} displays histograms of posterior samples of the hyperparameters $\mu$, $\rho$, $\tau^2$, $\varsigma$ and $\lambda$, as well as the implied distribution on $\beta_{i,j}$ induced by these posteriors. Notice that the posteriors for all the parameters are unimodal with potentially some skewness to the right. With the exception of $\tau^2$, the posteriors are shifted to the right of the prior distribution. For instance, the mean of $\rho$ changes from $0.9$ in the prior to $0.976$ in the posterior. Furthermore, the posteriors of $\lambda$, $\rho$ and $\tau^2$ are particularly concentrated compared to their respective prior distribution. Meanwhile, the posterior of $\mu$ is about as concentrated as its prior, whereas the posterior of $\varsigma$ is significantly less concentrated.  It is also worthwhile noting that the implied posterior distribution for $\beta_{i,j}$ is concentrated on the $[-\pi/2,\pi/2]$ interval but, unlike the prior, it seems to be bimodal.  This feature illustrates one of the key advantages of moving to a projected Gaussian process to model the ideal points:  this prior allows for bimodality (a very common feature when estimating preferences from voting data), while the von Mises distributions does not.

\subsection{The U.S.\ Supreme Court, 1949--1952}\label{se:SCOTUS1949-1952}

The early and mid 1930s were a period of increasing conflict between the executive and judicial branches of the U.S.\ Federal Government. Put briefly, this conflict was driven by repeated instances of a (``conservative'') Supreme Court bent on protecting property rights striking down ``New Deal'' legislation put forward by the (``liberal'') administration of Franklin D.\ Roosevelt to address the causes and consequences of the Great Recession.  By 1941, however, the conflict had been resolved by the fact that Roosevelt had been able to appoint all nine members of SCOTUS.  By the 1949 term, the Court composition had changed again.  Five Roosevelt appointees remained in the court (Justices Felix Frankfurter, Robert Jackson, Hugo Black, Willian O.\ Douglas and Stanley F.\ Reed) and were joined by four recent Truman appointees (Justices Harold H.\ Burton, Fred M.\ Vinson, Tom C.\ Clark and Sherman Minton, with Clark and Minton in their first term).  

While Frankfurter, Jackson, Black and Douglas started their appointments in the Court wih similar liberal political inclinations and were committed New Dealers, once on the bench, they developed radically different judicial philosophies and strained interpersonal relationships that often put them at odds with each other. Before joining the court, Frankfurter  was a staunch supporter of liberal political ideas and one of the founders of the American Civil Liberties Union.  However, during his tenure on the Court, he became the Court's most outspoken advocate of judicial restraint, the view that courts should not interpret the Constitution in such a way as to impose sharp limits upon the authority of the legislative and executive branches \citep{irons2006people}.  He also had a pompous style that strained his personal relationship with other justices.  Many historians have come to see Frankfurter as the eventual leader of the conservative faction of the Supreme Court (e.g., see \citealp{eisler1993justice}), to which Jackson (a frequent ally of Frankfurter, specially in his animosity towards Justices Douglas and Black) also belonged. 
Black was also an adherent of judicial restraint (specially during the early part of his tenure on the Court), but he was also a committed literalist and absolutist who disliked and repeatedly clashed with Frankfurter and Jackson \citep{ball1996hugo,magee1980mr}.  Finally, Douglas espoused a pragmatic judicial philosophy that did not highly value judicial consistency when deciding cases and relied on philosophical insights and observations about current politics as much as more conventional judicial sources \citep{tomlins2005united}.  

\begin{figure}[!t]
    \centering
    \begin{subfigure}[t]{0.49\textwidth}
    \centering
    \includegraphics[width = .95\textwidth]{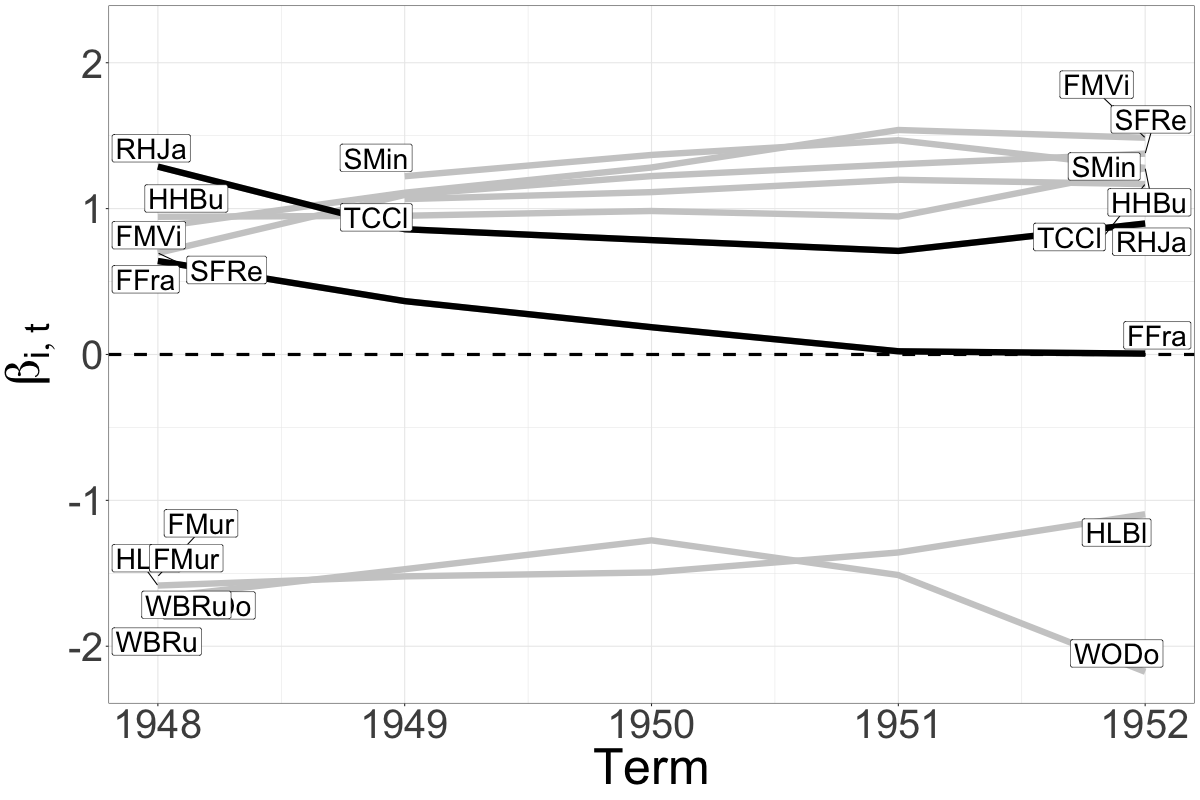}
    \caption{Martin-Quinn scores}\label{fig:scores1949to1952MQ}
    \end{subfigure}
    \begin{subfigure}[t]{0.49\textwidth}
    \centering
    \includegraphics[width = .8\textwidth]{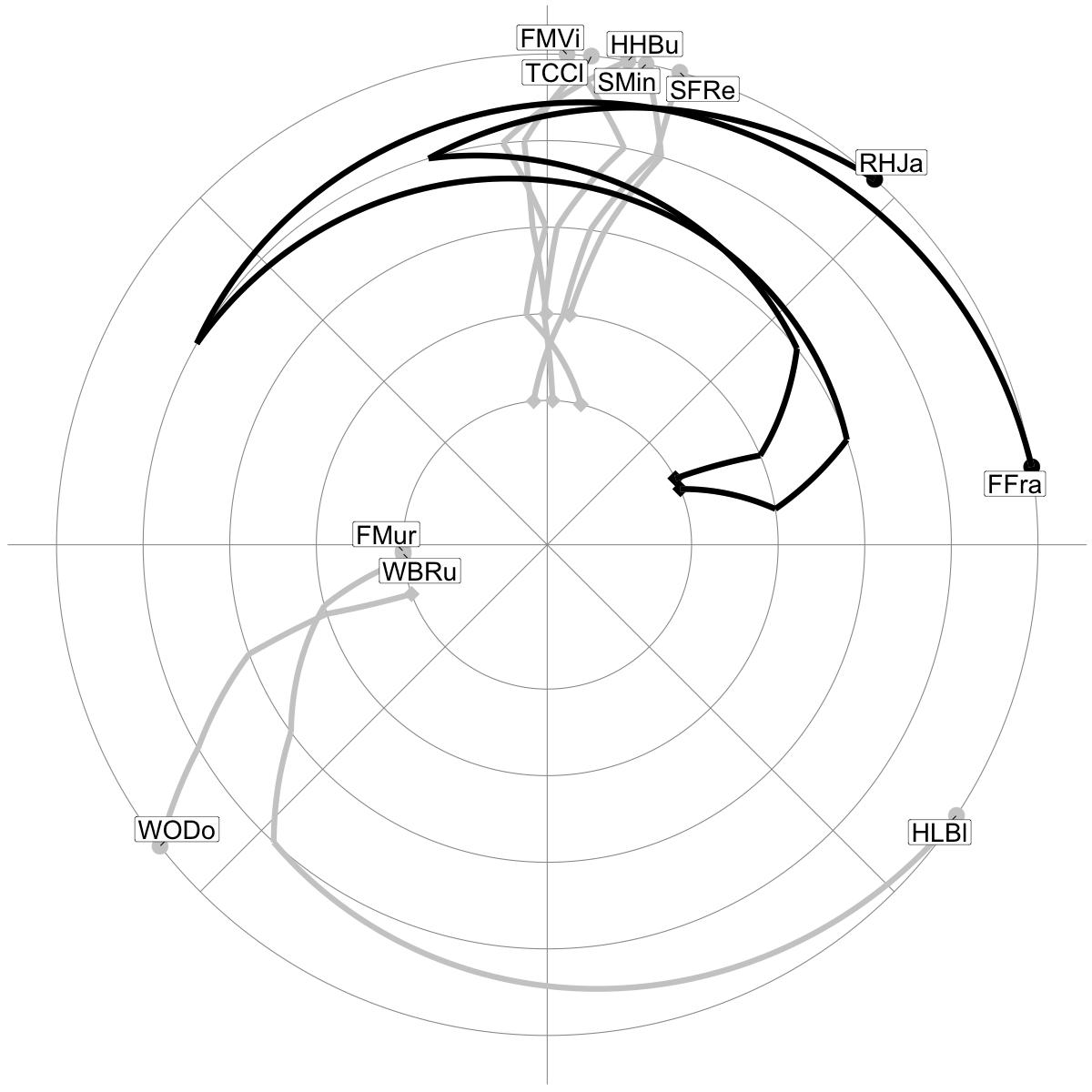}
    \caption{Circular scores}\label{fig:scores1949to1952circcirc}
\end{subfigure}
    \caption{Martin-Quinn (left panel) and circular scores (right panel) of the justices on the Supreme Court from 1948 to 1952.  In the right panel, each circle represents a SCOTUS term, with the innermost one representing 1948 and the outermost one representing 1952.  The trajectories of Justices Felix Frankfurter and Robert Jackson are highlighted.} \label{fig:scores1948to1952}
\end{figure}

With this background in mind, we now focus on the trajectories of the ideal points of the nine SCOTUS justices from 1949 to 1952 (see Figure \ref{fig:scores1948to1952}).  We can see in Figure \ref{fig:scores1949to1952MQ} that the MQ model places both Frankfurter and Jackson as centrists.  In fact, the MQ model makes them seem like the two judges whose ideal points are closest to those of Douglas and Black.  This is clearly at odds with the consensus historical analysis that we described before, especially as it relates to Frankfurter's ideology.  In contrast, we can see from Figure \ref{fig:scores1949to1952circcirc} that for the 1948 to 1950 terms, it is more reasonable to interpret the circular model as suggesting the presence of three groups of justices:  one made of all Truman appointees plus Justice Reed, another made of Douglas and Black, and a third made of Frankfurter and Jackson, roughly lined up as the vertices of a triangle.  Outcomes on specific cases are driven by ad-hoc alliances among justices across these three groups.  Such three-ways alliances, which have low-probability under the MQ model, are much more likely under the circular geometry, explaining the much better fit our circular model.  These observations are consistent with the results of the clustering procedure presented in Figure \ref{fig:means_over_time} (which indicate three clusters for the 1948 to 1950 terms) and with the historical narrative, which portrays Frankfurter and Jackson as members of a faction opposed to a second alliance made of Douglas and Black.

At a substantive level, we believe that the dominance of the circular model during this  period of the court can be explained in part by shifting views on the doctrine of judicial restraint.  As discussed in  \cite{FeldmanScorpionsBattlesTriumphs2010}, judicial restraint had been favored judicial philosophy among liberal lawyers and justices during the 1920s and the 1930s, as it served as check on what they saw as an overly conservative SCOTUS.  With the changes to the Court composition occurring during the Roosevelt years, judicial restraint acted as a brake on the ambitions of liberal justices, most of whom had come to SCOTUS with clear political ambitions and after serving in the legislative and executive branches rather than in academia or the courts.  As they settled on their positions in the Court, several of Roosevelt's appointees either ditched judicial restraint completely, or applied it selectively depending on whether the consequences on a particular case agreed with their political goals. This explains how Justice Felix Frankfurter, by holding on to judicial restraint as the guiding principle of his jurisprudence, became known as the leader of the ``conservative'' wing of the Court, and Douglas and Black as defendants of ``liberalism'', while still being able to vote together on a number nonunanimous decisions.

\subsection{The U.S.\ Supreme Court, 1967--1970}\label{se:SCOTUS1967-1970}

The 1967--1970 period covers the last two terms of the Warren Court and the first two of the Burger Court. These terms are considered by historians as strongly liberal.  The Warren court in particular is famous for expanding civil rights, civil liberties, and judicial power in dramatic ways.  Six justices (Hugo Black, William Douglas, John Harlan, William Brennan, Steward Potter, and Byron White) were active over this whole period.  As for the remaining two years, Justice Thurgood Marshall replaced Justice Tom Clark in 1967, Justice Harry Blackmum replaced Justice Abe Fortas in 1970, and Warren Burger replaced Earl Warren as Chief Justice in 1970.

\begin{figure}[!t]
    \centering

    \begin{subfigure}[t]{0.49\textwidth}
        \centering
        \includegraphics[width = .95\textwidth]{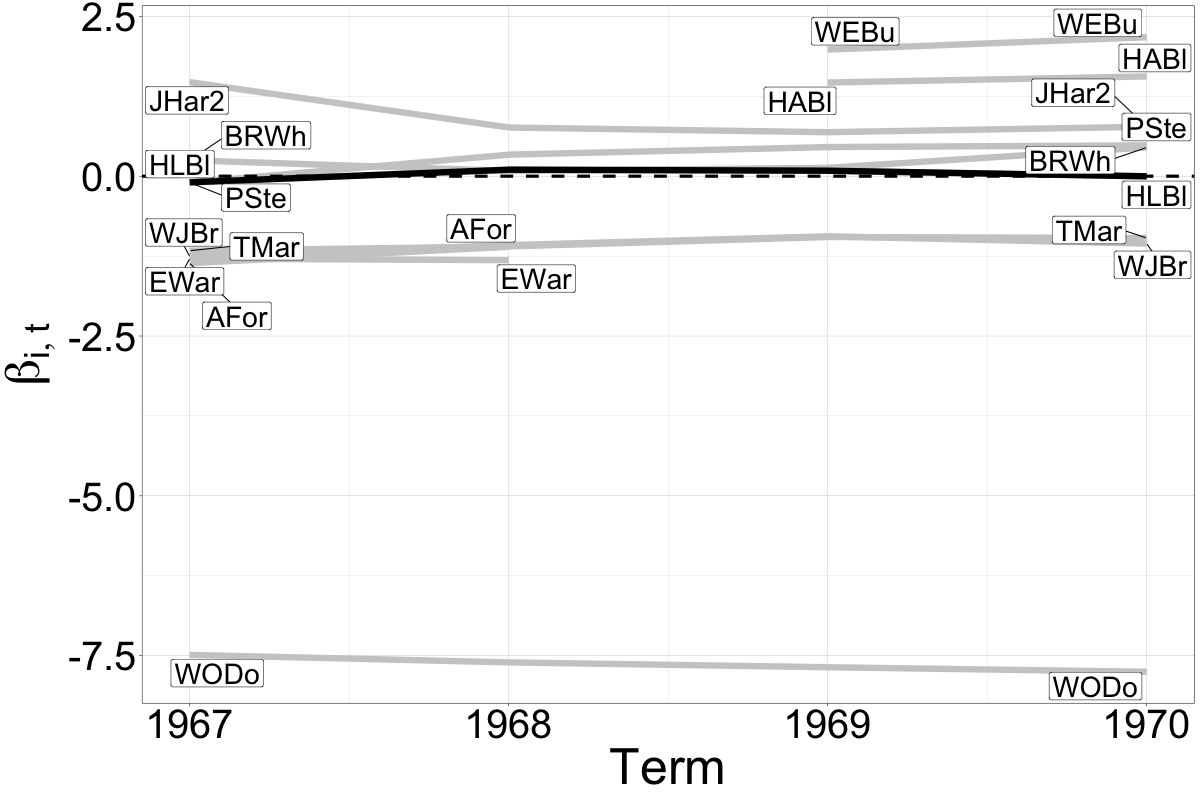}
    \caption{Euclidean coordinates}\label{fig:scores1969to1972MQ}
    \end{subfigure}
    \begin{subfigure}[t]{0.49\textwidth}
        \centering
        \includegraphics[width = .8\textwidth]{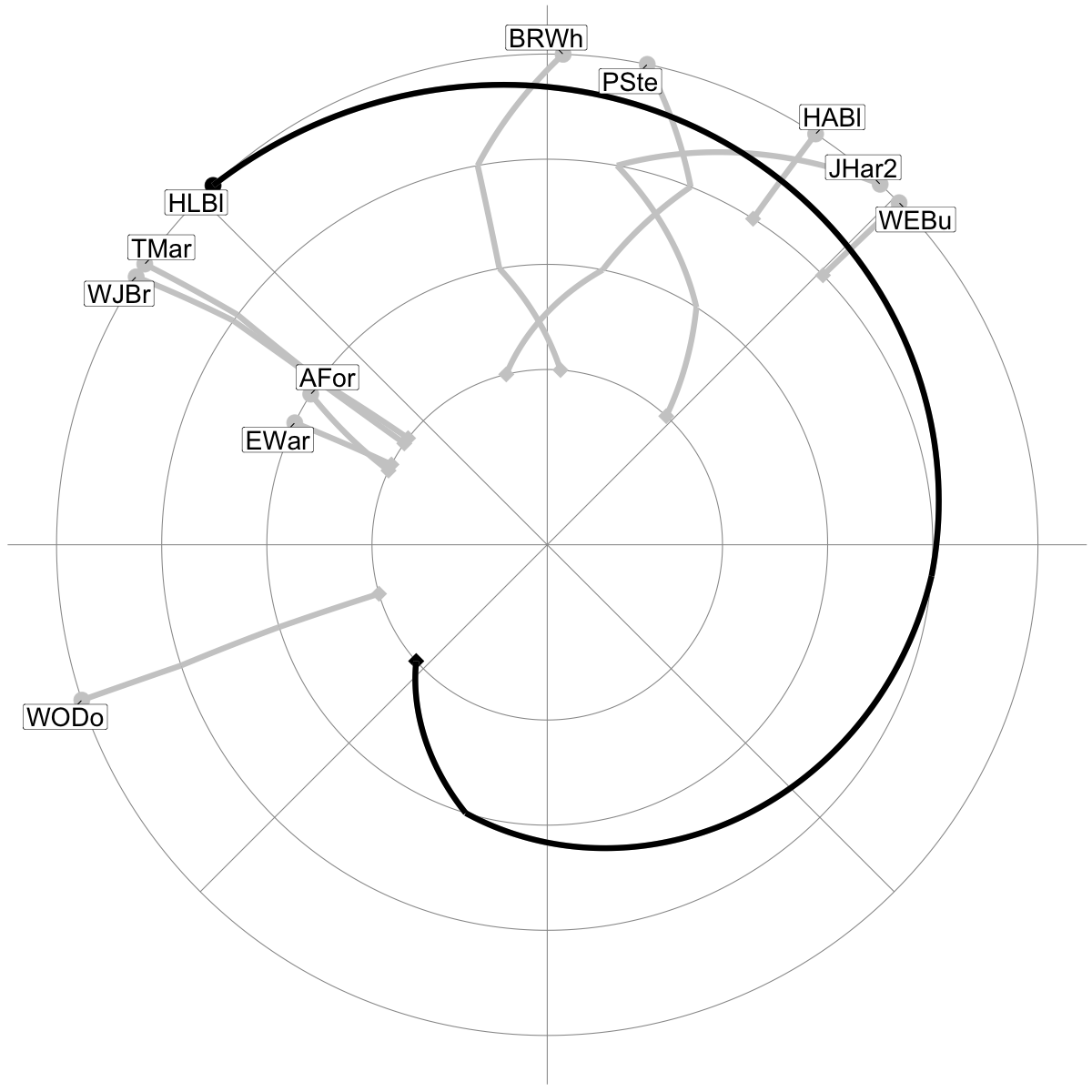}
        \caption{Polar coordinates}\label{fig:scores1967to1970circcirc}
    \end{subfigure} 
\caption{Martin-Quinn (left panel) and circular scores (right panel) of the justices on the Supreme Court from 1967 to 1970.  In the right panel, each circle represents a SCOTUS term, with the innermost one representing 1967 and the outermost one representing 1970.  The trajectory of Justice Hugo Black is highlighted on both graphs.} \label{fig:scores1968to1970}
\end{figure}

Figure \ref{fig:scores1969to1972MQ} shows the trajectories of the Martin-Quinn scores of all the justices that served in the Court during this period.  Under Martin-Quinn's model, the justices'  preferences seem to be very stable over time.  In particular, Black appears to be a relative moderate during this period.  This is at odds with the opinions of various experts.  Indeed, while Black was considered a liberal during the early part of his tenure in the Court, towards the end of his career, his ``originalism was turning him into a judicial split personality: now liberal, now conservative'' \citep{FeldmanScorpionsBattlesTriumphs2010}.  Black opposed the Warren Court's expansion of civil liberties, as he did not believe that the original text of the Constitution supported this expansion, except in the case of free speech.

In contrast, Figure \ref{fig:scores1967to1970circcirc} shows the estimates the ideal points under our circular model.  These show two key differences with respect to the Martin-Quinn scores.  First, the preferences of Justice Douglas now seem less extreme.  More importantly, we can see a dramatic change in the preferences of Justice Black between 1968 and 1969, precisely at the time of the transition between Warren and Burger as Chief Justices. This pattern of behavior agrees much more closely the narrative around Justice Black's behavior that we discussed above.

\subsection{Sensitivity Analysis}\label{sec:sensitivity}

To investigate the sensitivity of the results to the choice of priors, we repeated our analyses using an alternative set of priors in which $\rho$ is assigned a Gaussian distribution with mean 0.9 and standard deviation 0.04 (i.e. $\Pr(0.85 \le \rho \le 1) \approx 0.9$), truncated to the $[0,1]$ interval, and $\mu$ follows a normal distribution with mean $0$ and standard deviation $1.4$, truncated to the positive numbers, $\tau^2$ follows an exponential distribution with mean 0.1, and $\varsigma$ follows a Gamma distribution with mean 1 and variance 0.5. The implied prior on $\beta_{i,t}$ is more concentrated than the one we used for our original analysis, while the implied prior on $\theta_{i,j,t}$ is still trimodal and similar to the one used in this paper. Please see Section 2 of the supplementary materials for a visual representation of these two implied priors.

The results appear to be reasonably robust to our choice of priors, although there are some small differences.  In particular, while the posterior distribution of the hyperparameters $\rho$, $\varsigma$ and $\lambda$ remain the same, the posterior distributions for $\mu$ and $\tau^2$ seem to favor slightly lower values under our alternative priors.  On the other hand, the trajectories of a couple of the justices' ideologies seem to be slightly smoother under our alternative prior.  However, the key conclusions from our previous analyses do not seem to be affected by the new priors.  Please see the supplementary materials for additional details.

\section{Discussion}\label{se:discussion}

This paper introduces a novel dynamic factor model for binary data that relies on a circular latent space. 
The model is motivated by the problem of assessing the evolution of the ideological preferences of U.S.\ Supreme Court justices.  We show that the circular model replicates the results of more traditional Euclidean voting models when the data supports such models, while providing a richer and more accurate description of the justices' behavior at key historical periods in which the results from traditional Euclidean models contradict well-established narratives around judges' ideology.


Our circular model suffers from many of the same shortcomings as other factor models used for scaling preferences, including the one proposed by \cite{MartinQuinnDynamicIdealPoint2002a}.  In particular, a key assumption that is rarely explicitly acknowledged is that the distribution of the decision-specific parameters $(\psi_{i,t}, \zeta_{i,t})$ remains constant over time.  Because Supreme Court justices can select the cases they will consider in a given term, this assumption is suspect in our application.  In other words, it is possible that some of the fluctuations in ideal points capture by these models might be due to changes in the nature of the cases in the SCOTUS' docket rather than a reflection of changing ideology among justices.  In our application, this shortcoming might be particularly impactful starting in the mid-90s, when the number of SCOTUS decisions drops dramatically.

During the review process, one referee wondered about the possibility of using other similarity metrics different from the geodesic distance, and suggested the possibility of relying on the cosine dissimilarity, $d_{C}(x_1,x_2) = 1 - \cos(x_1-x_2)$.  While the use of distances different from the geodesic is possible with minimal modifications to our framework, the theoretical motivation of the model requires that the similarity used to construct the utility functions corresponds to a well-defined distance on the circle.  Indeed, as we mentioned in the introduction, the theoretical underpinning of spatial voting models is rational choice theory.  One consequence of its axioms is that, if preferences can be represented through utility functions, those utility functions must satisfy the triangle inequality (e.g., see \citealp{de1983mixed}).  Hence, since the cosine dissimilarity does not satisfy the triangle inequality, it follows that a model that uses it cannot represent a well-defined set of underlying preferences.  Of course, rational choice theory can (and has) been criticized, and generalized utility theories have been proposed (e.g., see \citealp{machina1982expected} and \citealp{quiggin2012generalized}).  Studying the connection between these and various possible similarity metrics is an interesting direction for future reach, but it is outside of the scope of this paper.


Future extensions of the model presented in this paper that we would like to consider include dynamic models that rely on higher-dimensional spherical policy spaces, as well as models based on more general latent geometries.  We are also interested in extending the model by replacing the single projected autoregressive process priors on the vectors of ideal points with a mixture of autoregressive process, in the spirit of \cite{handcock2007model}.  This would avoid the need for ad-hoc clustering of the ideal points.  A challenge to this kind of extension, however, is the fact that projected normal distributions can be bimodal (recall Figure \ref{fig:posteiorbeta}), which might lead to identifiability issues in the case of mixture priors. A final area of clear interest is the comparison of our circular models with (yet-to-be-developed) dynamic versions of the generalized unfolding model of \cite{duck2022ends}.  As we discussed in the introduction, the implementation of such methods involves substantial modeling and computational challenges.  Furthermore, our own (unpublished) investigation of the static versions of these models in the context of the U.S.\ House of Representatives suggests that models based on circular voting spaces provide a better complexity-adjusted fit (as measured by WAIC) than generalized unfolding model.  We expect to report on these comparisons elsewhere.

\section*{Acknowledgements}
We would like to thank Kevin Quinn for his help with the \texttt{R} package \texttt{mcmcPack}. This research was partially funded by grants NSF-DMS 2114727 and NSF-CIS 2023495.

\bibliographystyle{rss} 
\bibliography{my_refs,newref}

\end{document}




\title{Supplementary Material for ``Dynamic Factor Models for Binary Data in Circular Spaces: An Application to the U.S.\ Supreme Court''}
\author{Rayleigh Lei and Abel Rodriguez}
\date{}
\maketitle

\section{Interpretation of the parameters of the circular autoregressive process}

\begin{figure}[!bt]
\centering
\captionsetup{justification=centering}
\begin{subfigure}[t]{0.49\textwidth}
    \includegraphics[width = 0.75\textwidth]{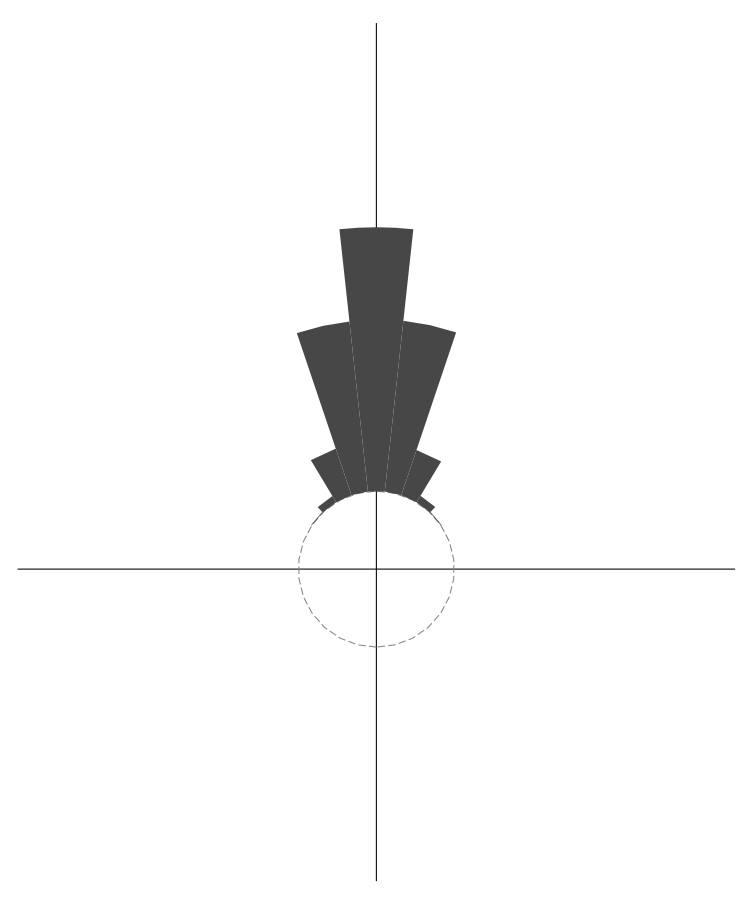}
    \caption{$\mu_1 = 10$, $\mu_2 = 0$, $\rho = 0.95$, $\tau_1 = 0.5$, $\tau_2 = 0.5$}\label{fig:prior_ref}
\end{subfigure}
\begin{subfigure}[t]{0.49\textwidth}
    \includegraphics[width = 0.75\textwidth]{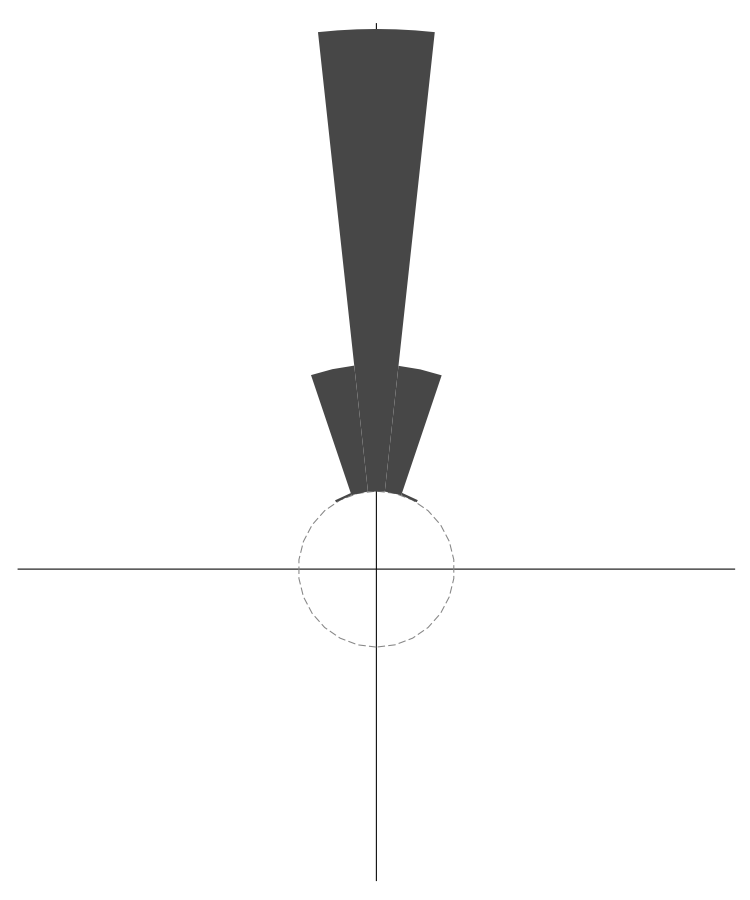}
    \caption{$\mu_1 =10$, $\mu_2 = 0$, $\rho = 0.8$, $\tau_1 = 0.5$, $\tau_2 = 0.5$}\label{fig:prior_rho}
\end{subfigure} \\
\begin{subfigure}[t]{0.49\textwidth}
    \includegraphics[width = 0.75\textwidth]{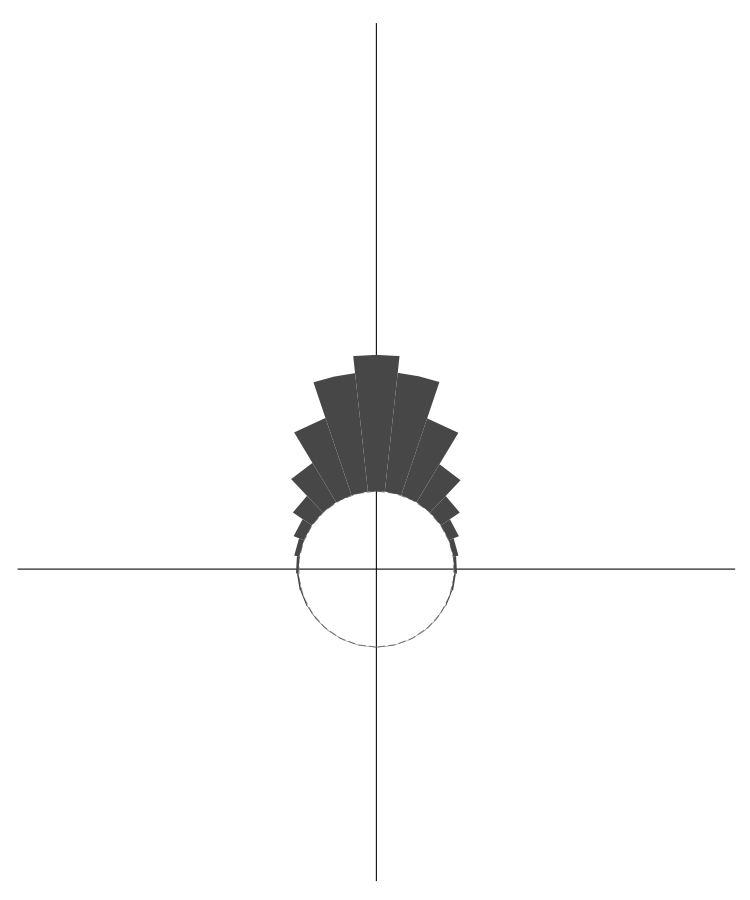}
    \caption{$\mu_1 = 5$, $\mu_2 = 0$, $\rho = 0.95$, $\tau_1 = 0.5$, $\tau_2 = 0.5$}\label{fig:prior_mu}
\end{subfigure}
\begin{subfigure}[t]{0.49\textwidth}
    \includegraphics[width = 0.75\textwidth]{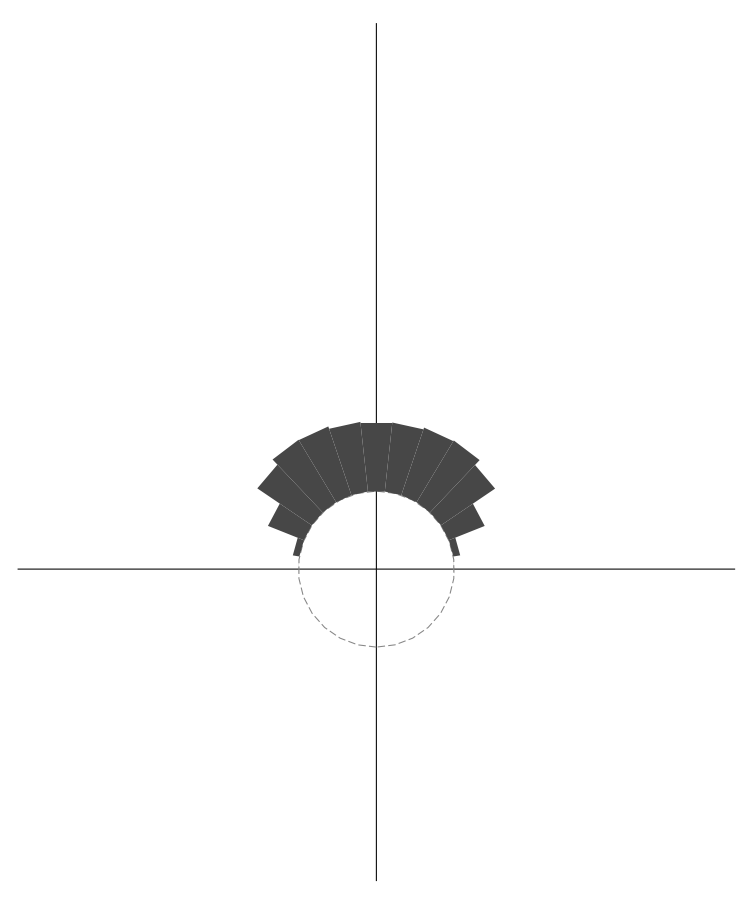}
    \caption{$\mu_1 =10$, $\mu_2 = 0$, $\rho = 0.95$, $\tau_1 = 0.5$, $\tau_2 = 8$}\label{fig:prior_tau_v1}
\end{subfigure} \\
\begin{subfigure}[t]{0.49\textwidth}
    \includegraphics[width = 0.75\textwidth]{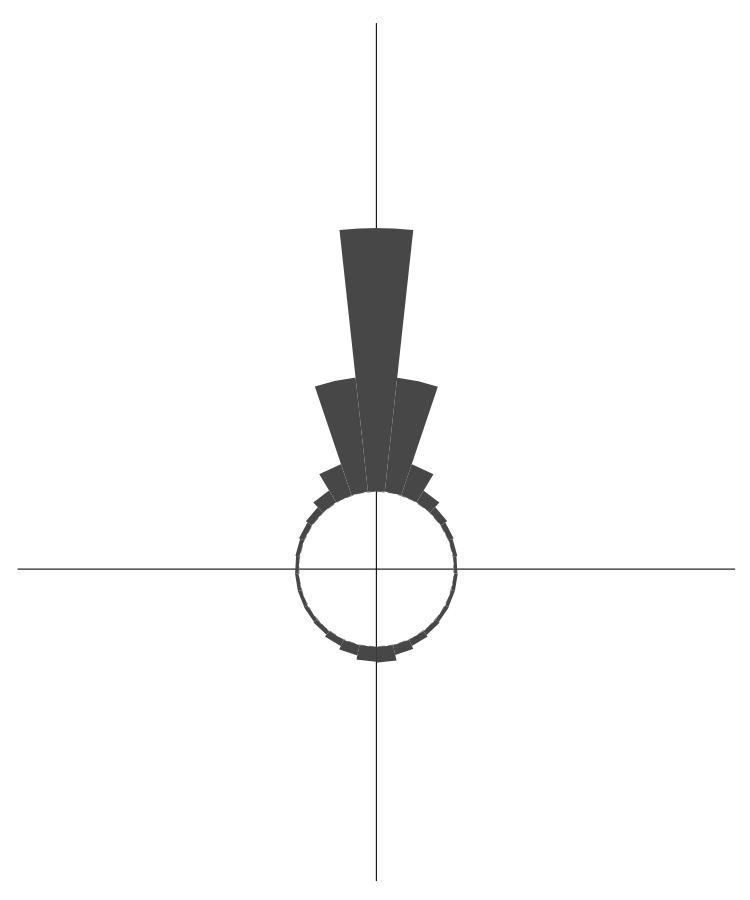}
    \caption{$\mu_1 = 10$, $\mu_2 = 0$, $\rho = 0.95$, $\tau_1 = 8$, $\tau_2 = 0.5$}\label{fig:prior_tau_v2}
\end{subfigure}
\begin{subfigure}[t]{0.49\textwidth}
    \includegraphics[width = 0.75\textwidth]{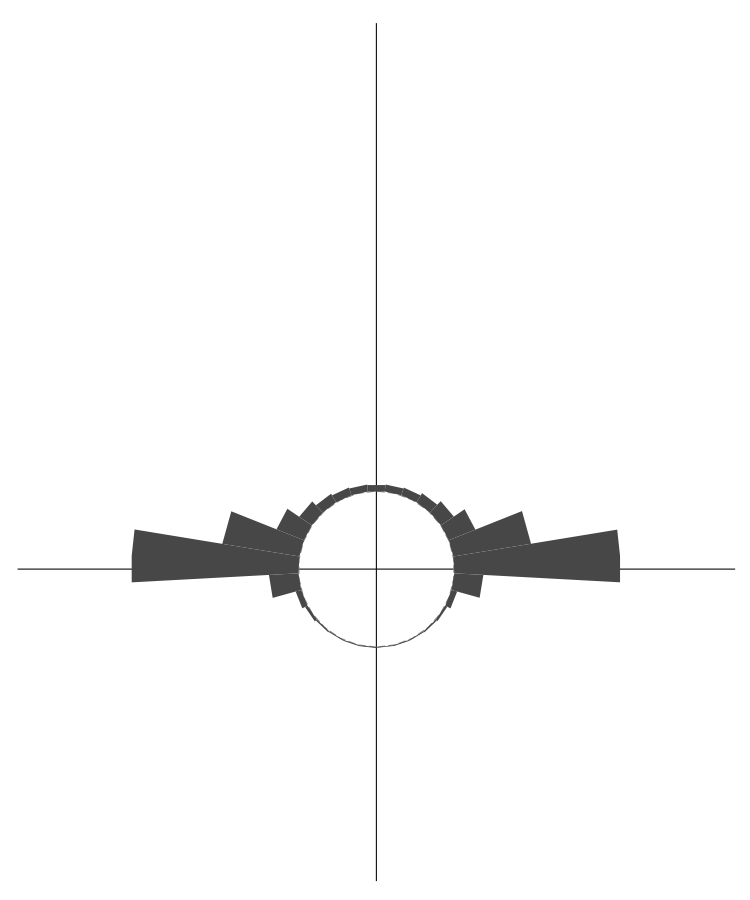}
    \caption{$\mu_1 = 3$, $\mu_2 = 0$, $\rho = 0.99$, $\tau_1 = 0.25$, $\tau_2 = 20$}\label{fig:prior_tau_v3}
\end{subfigure}
\caption{Plots showing 10,000 draws from the marginal prior distribution for $\judgeI$ according to various hyperpriors.}
\label{fig:betashapes}
\end{figure}

Figure \ref{fig:betashapes} presents histograms based on 10,000 samples from the marginal distribution of a circular autoregressive process defined in Section 3.1 of the main manuscript for various combinations of (fixed) hyperparameters. Because $\mu_2 = 0$ in all cases, the mean of all the distributions is zero.
From these plots, it can be seen that lower values of $\rho$ lead reduce the variance of the stationary distribution (see Figure \ref{fig:prior_rho}). On the other hand, either lower values of $\mu_1$ or higher values of $\tau_1$ or $\tau_2$ lead to distributions with higher marginal variance. Note however, that while $\mu_1$ seems to only affect the variance (Figure \ref{fig:prior_mu}), $\tau_1$ or $\tau_2$ also affect the shape of the marginal distribution in other ways.  In particular, when the ratio $\tau_1/\tau_2$ becomes either much larger or much smaller than one, the distribution becomes first platykurtic (as in Figure \ref{fig:prior_tau_v1}) and eventually bimodal (as in Figure \ref{fig:prior_tau_v2} and \ref{fig:prior_tau_v3}).  Note that when $\tau_1/\tau_2$ is much larger than one, the two modes of the distribution of angles are located at $\beta_{i,t} = 0$ and $\beta_{i,t} = \pi$.  On the other hand, when $\tau_1/\tau_2$ is much smaller than 1, the two modes are symmetric around $\beta_{i,t} = 0$, but clearly away from it. 

\section{Sensitivity analysis}

In this Section we provide a more detailed comparison of the results under the original priors described in Section 3.2 of the main manuscript and those under the alternative prior described in Section 5.3.



\begin{figure}[!t]
  \centering
  \begin{subfigure}[t]{0.45\textwidth}
  \centering
  \includegraphics[width = 0.8\textwidth]{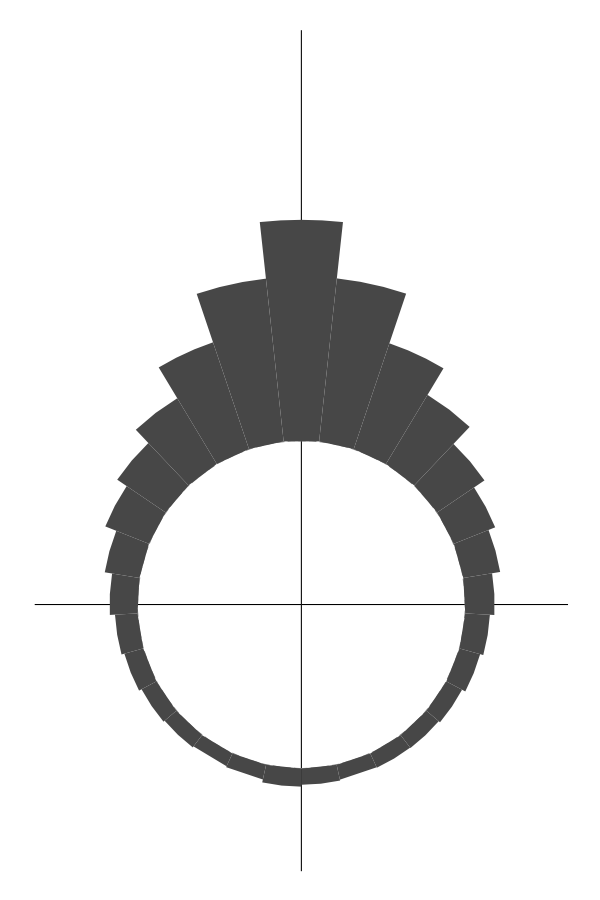}
  \subcaption{Induced prior on $\beta_{i,t}$ (original)}
  \end{subfigure}
  \begin{subfigure}[t]{0.45\textwidth}
  \centering
  \includegraphics[width = 0.8\textwidth]{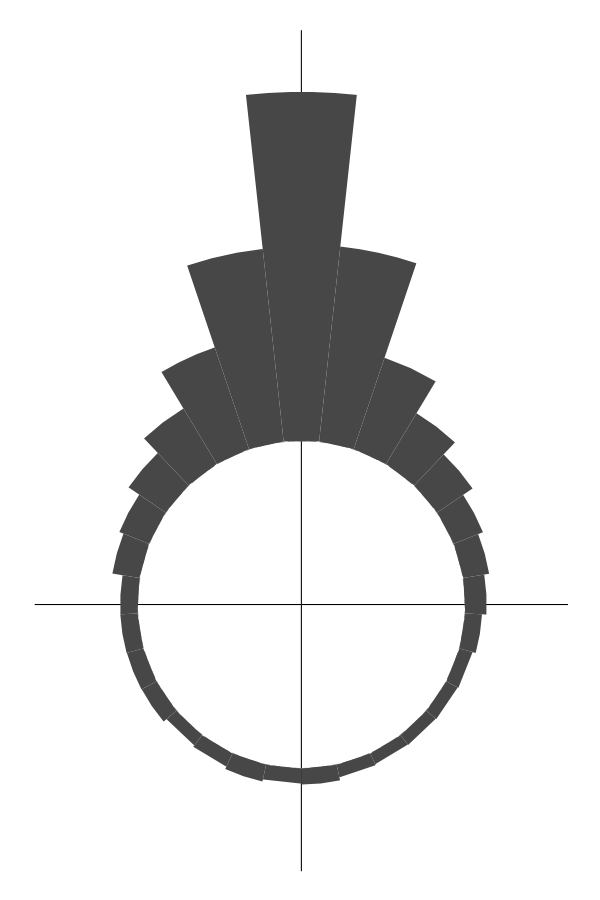}
  \subcaption{Induced prior on $\beta_{i,t}$ (alternative)}
  \end{subfigure}\\
  \begin{subfigure}[t]{0.45\textwidth}
  \centering
  \includegraphics[width = \textwidth]{img/circ_hyperparameters/marginal_theta_prior_7.png}
  \subcaption{Induced prior on $\theta_{i,j,t}$ (original)}
  \end{subfigure}
  \begin{subfigure}[t]{0.45\textwidth}
  \centering
  \includegraphics[width = \textwidth]{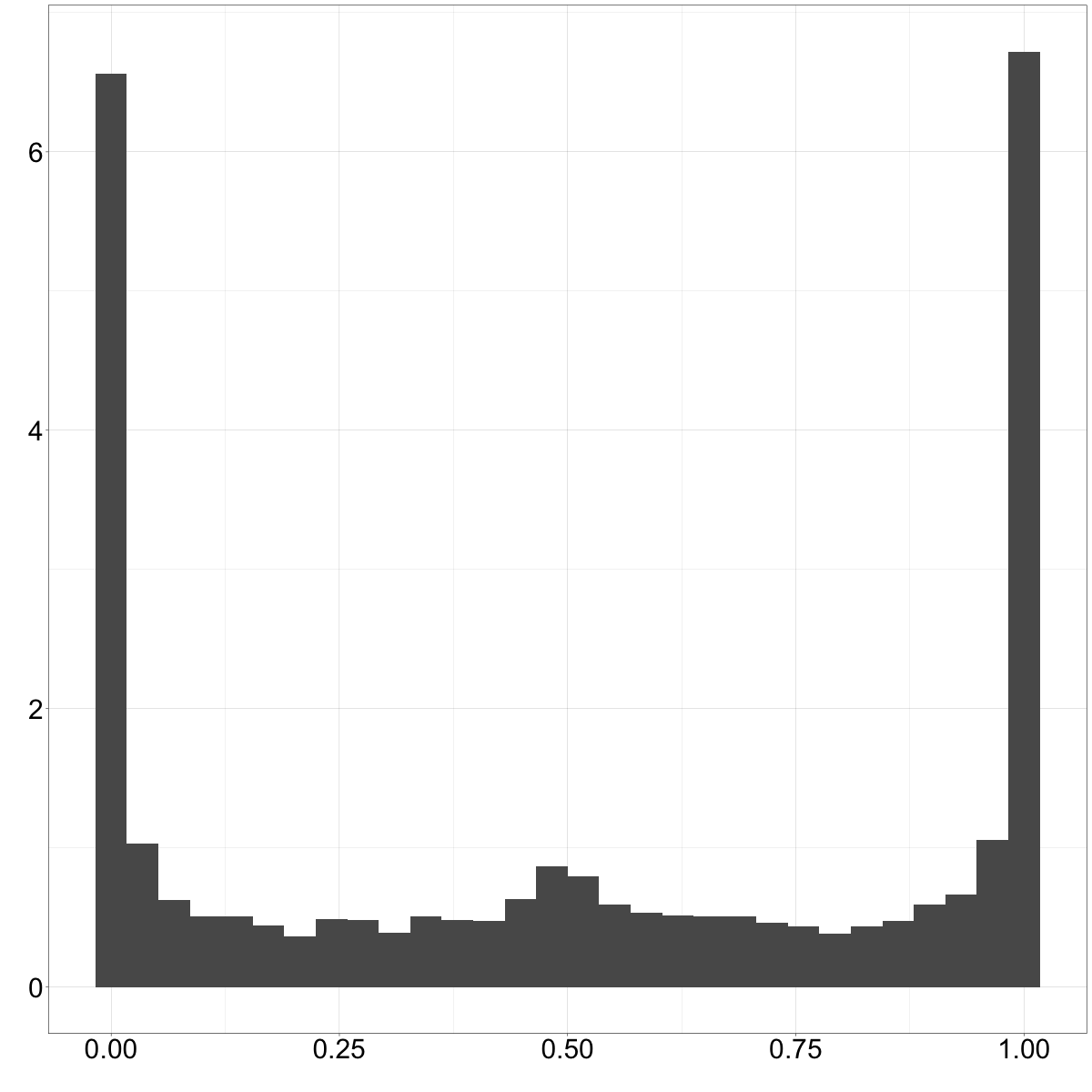}
  \subcaption{Induced prior on $\theta_{i,t}$ (alternative)}
  \end{subfigure}
  %
  \caption{Histograms showing 10,000 draws from the induced marginal distributions of $\beta_{i,t}$ (top row) and $\theta_{i,j,t}$ (bottom row) under the main prior in Section 3.2 (left column) paper and the alternative alternative prior in Section 5.3 (right column). Note that the scale for the circular histogram is different than the one used in Figure \ref{fig:betashapes}.}
  \label{fig:hyperparam_comp}
\end{figure}

Figure \ref{fig:hyperparam_comp} compares the marginal prior on $\beta_{i,t}$ (the ideal point of justice $i$ at time $t$) and $\theta_{i,j,t}$ (the probability that a justice will vote to reverse a lower court decision) induced by both sets of hyperparameters.  For $\beta_{i,t}$, the shapes are very similar, but we can see that the prior is slightly more concentrated around zero under the alternative set of hyperparameters.  Correspondingly, in the case of $\theta_{i,j,t}$, the alternative prior places slightly less mass on values of closer to either 0 or 1.

\begin{figure}[t]
 \centering
   \begin{subfigure}{0.9\textwidth}
    \centering
    \includegraphics[width = 0.3\textwidth]{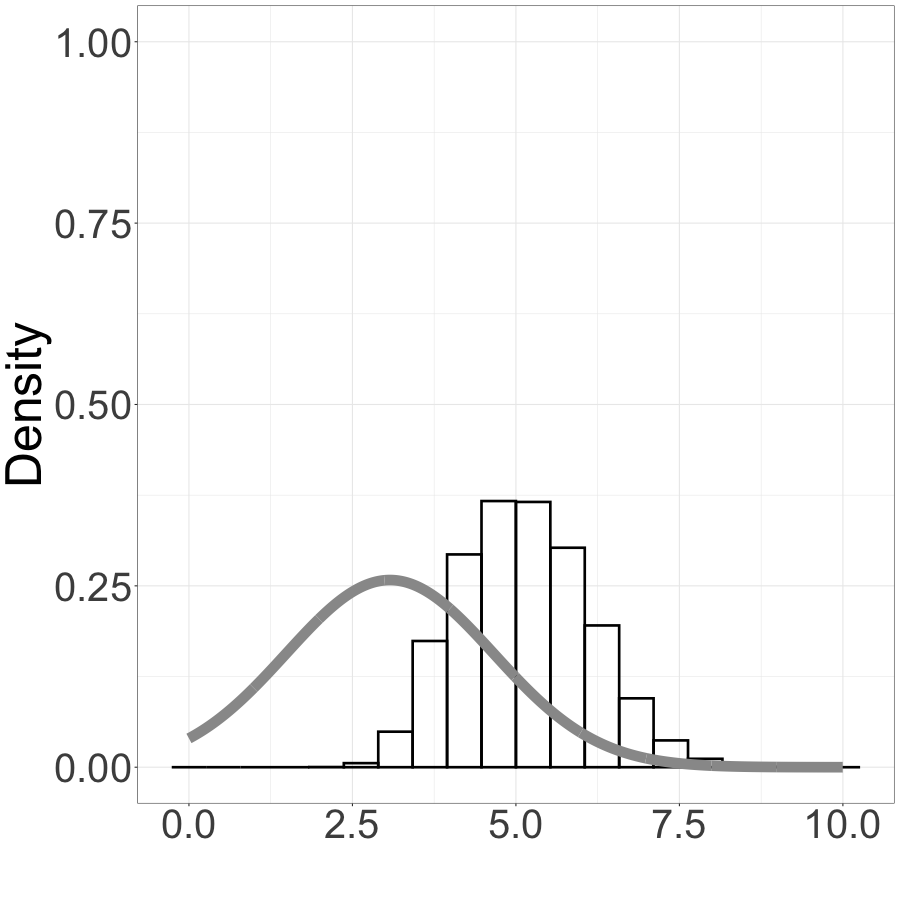}
    \qquad
    \includegraphics[width = 0.3\textwidth]{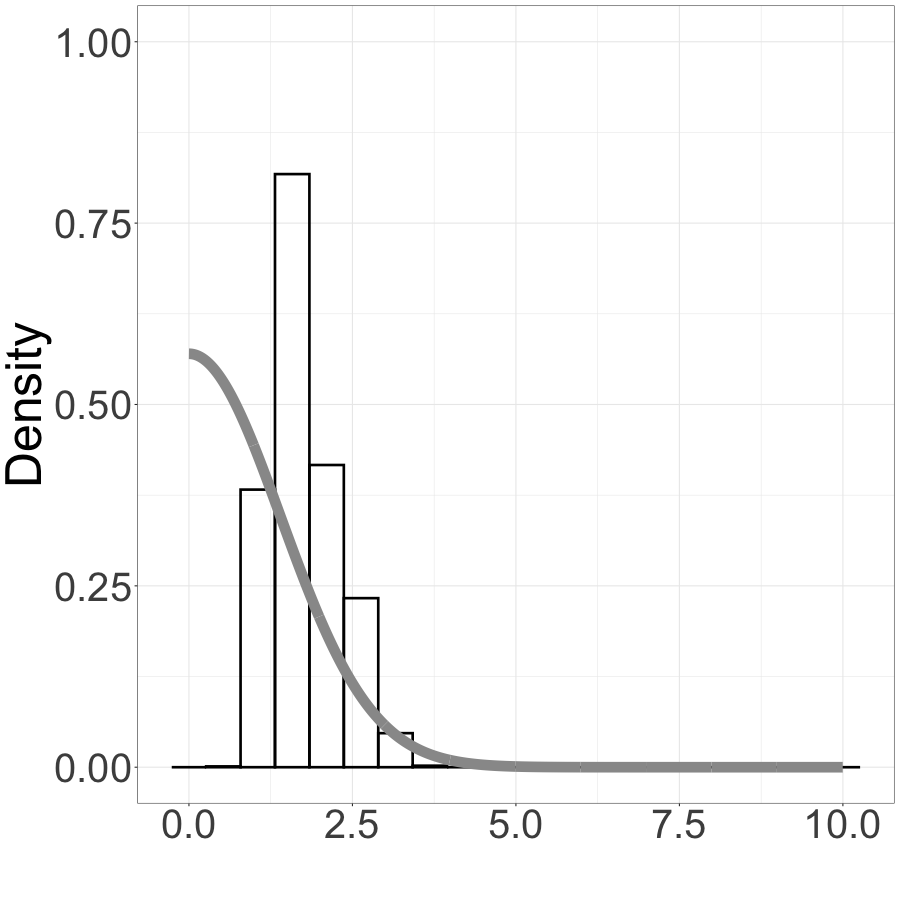}
    \subcaption{$\mu$}
   \end{subfigure}  
   \begin{subfigure}{0.9\textwidth}
   \centering
    \includegraphics[width = 0.3\textwidth]{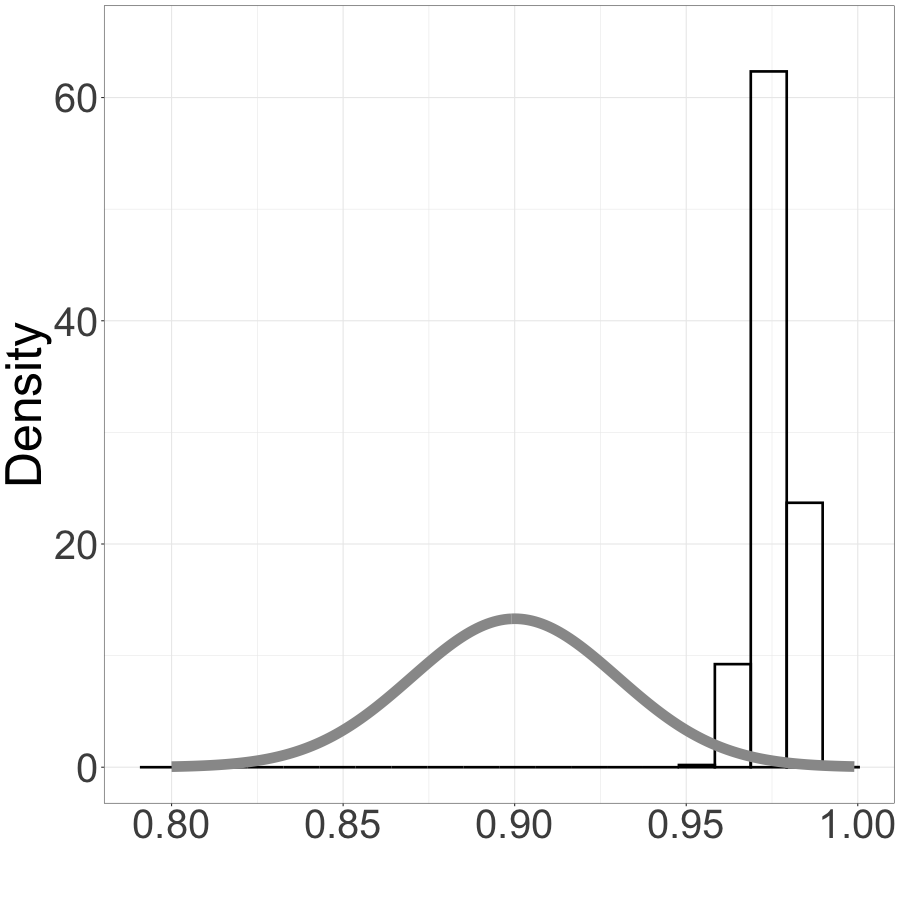}
    \qquad
    \includegraphics[width = 0.3\textwidth]{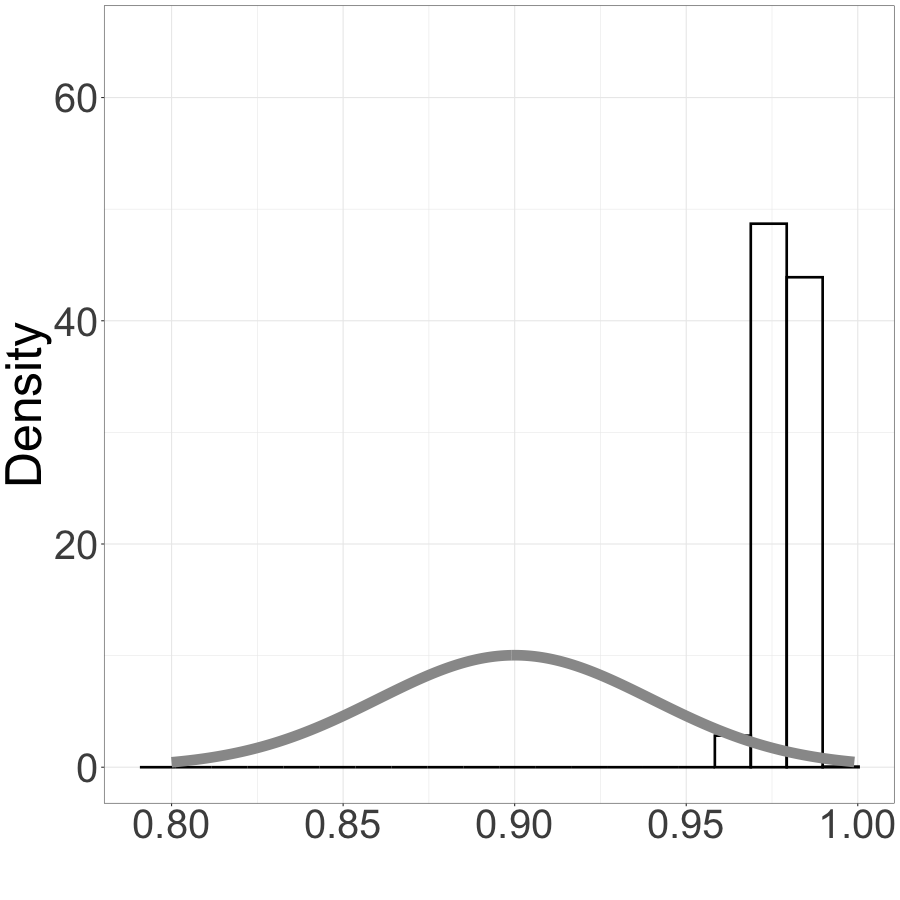}
     \subcaption{$\rho$}
   \end{subfigure}  
   \begin{subfigure}{0.9\textwidth}
   \centering
     \includegraphics[width = 0.3\textwidth]{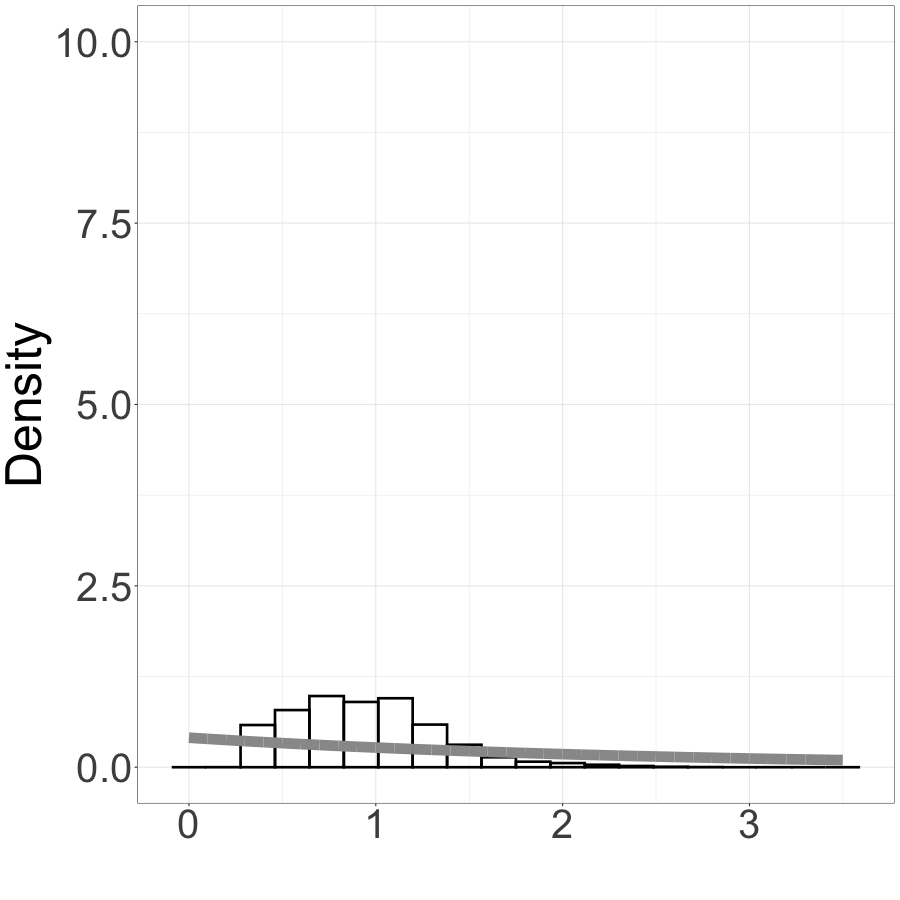}
     \qquad
    \includegraphics[width = 0.3\textwidth]{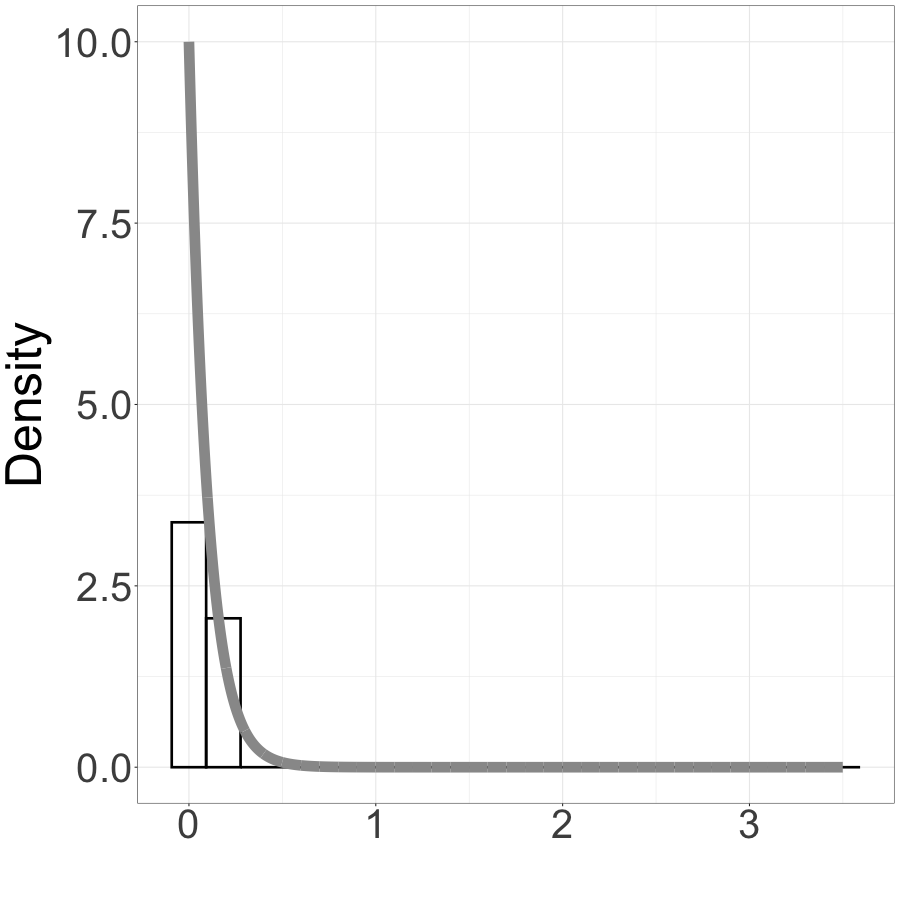}
     \subcaption{$\tau^2$}
   \end{subfigure}  \\
   \begin{subfigure}{0.9\textwidth}
   \centering
     \includegraphics[width = 0.3\textwidth]{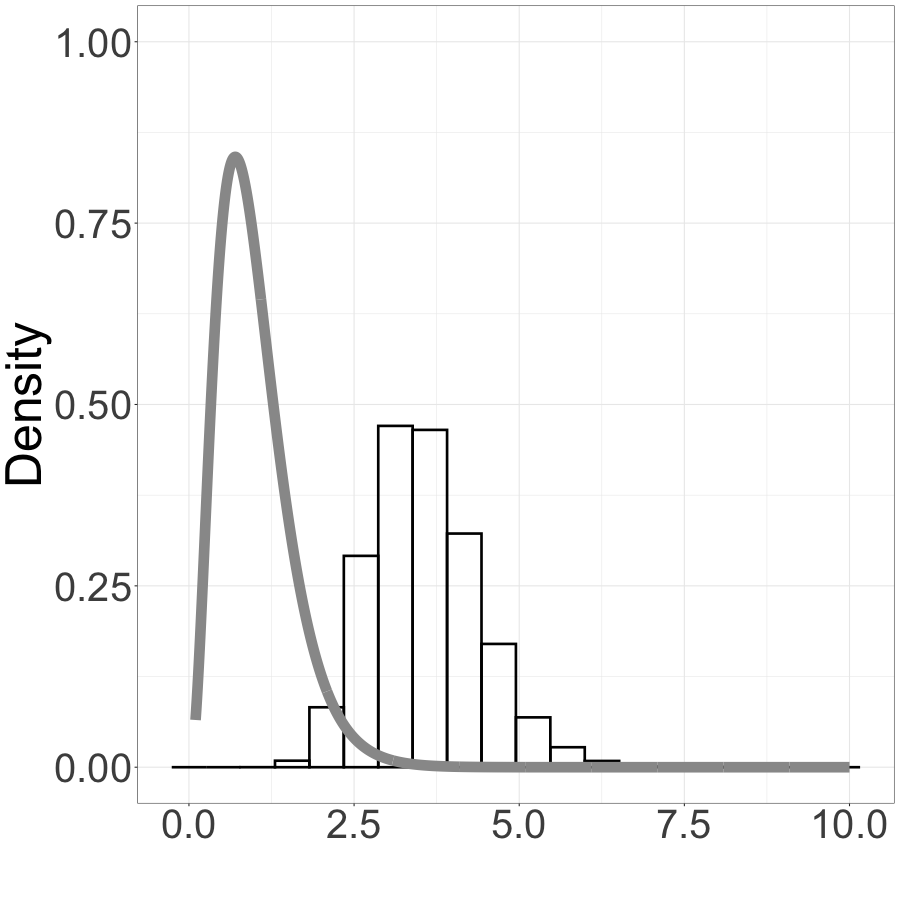}
     \qquad
    \includegraphics[width = 0.3\textwidth]{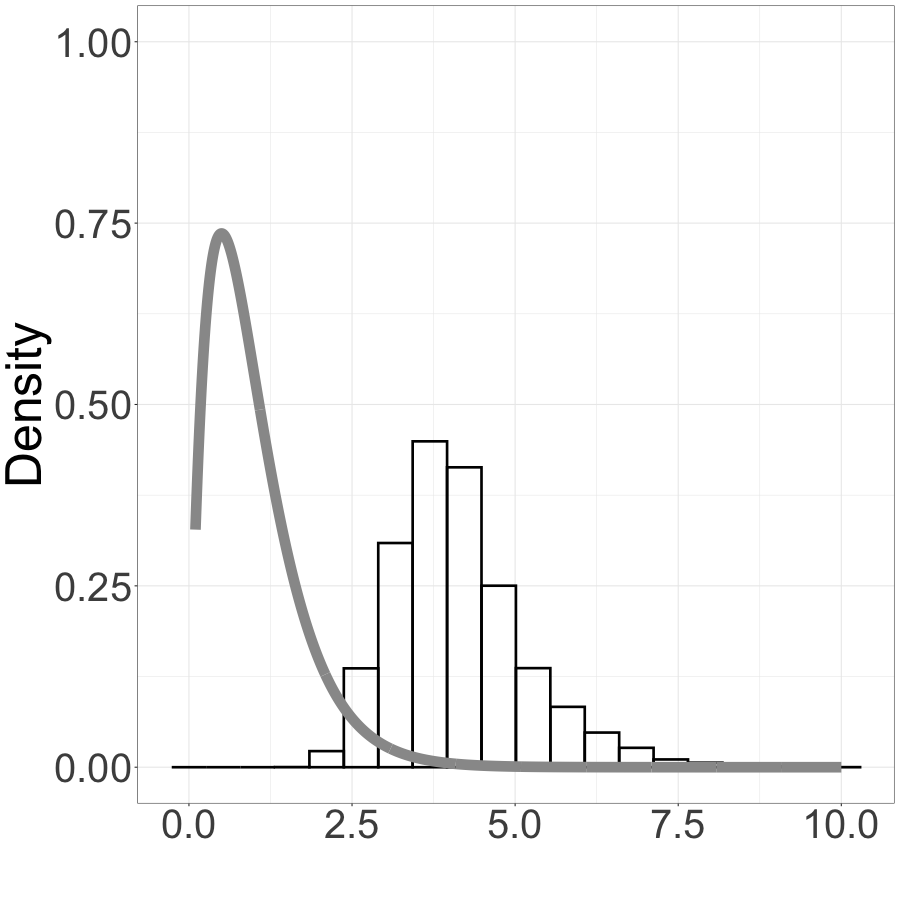}
     \subcaption{$\varsigma$}
   \end{subfigure}  
   \begin{subfigure}{0.9\textwidth}
   \centering
     \includegraphics[width = 0.3\textwidth]{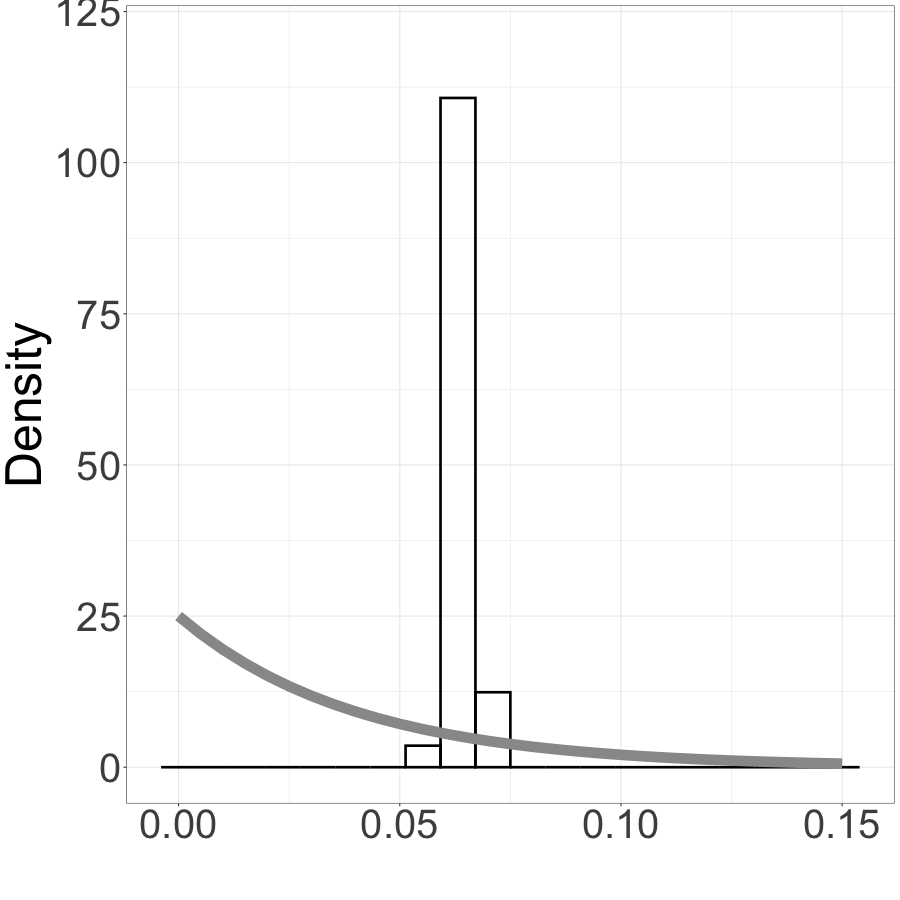}
     \qquad
    \includegraphics[width = 0.3\textwidth]{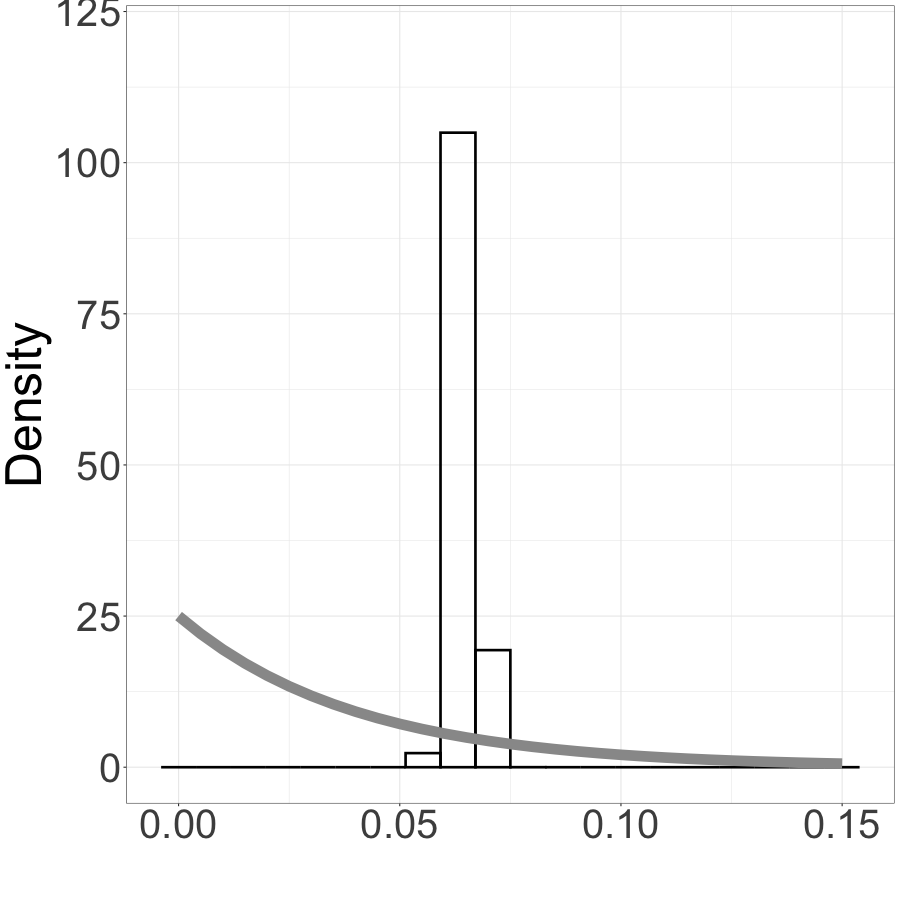}
     \subcaption{$\lambda$}
   \end{subfigure}  
    \caption{Histograms of the posterior distributions of various hyperparameters under the main prior (left column) and the alternative prior (right column). Solid lines correspond to the respective prior distributions.}
    \label{fig:posteriorshyperparameters}
\end{figure}

\begin{figure}[!t]
  \centering
  \begin{subfigure}[t]{0.45\textwidth}
  \centering
  \includegraphics[width = \textwidth]{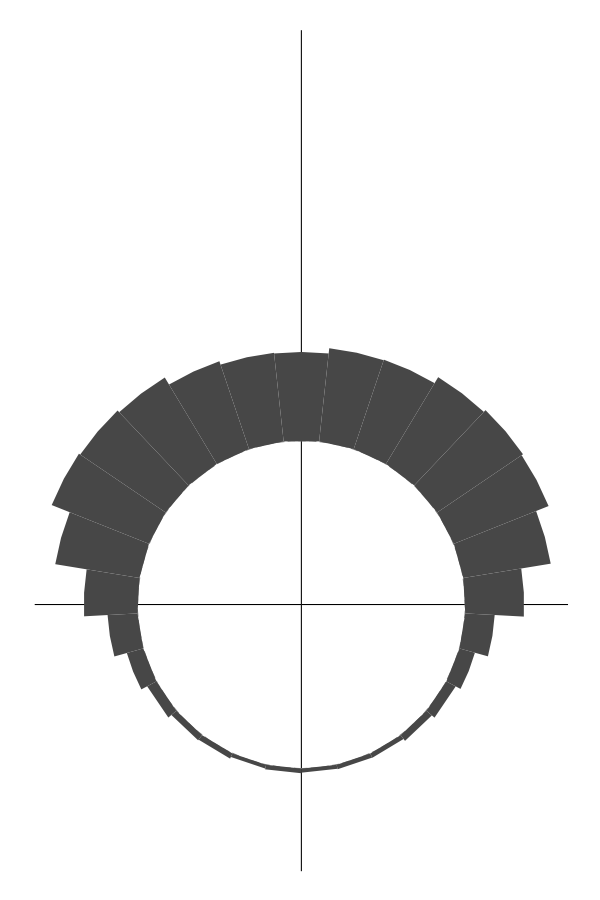}
  \subcaption{Original prior}
  \end{subfigure}
  \begin{subfigure}[t]{0.45\textwidth}
  \centering
  \includegraphics[width = \textwidth]{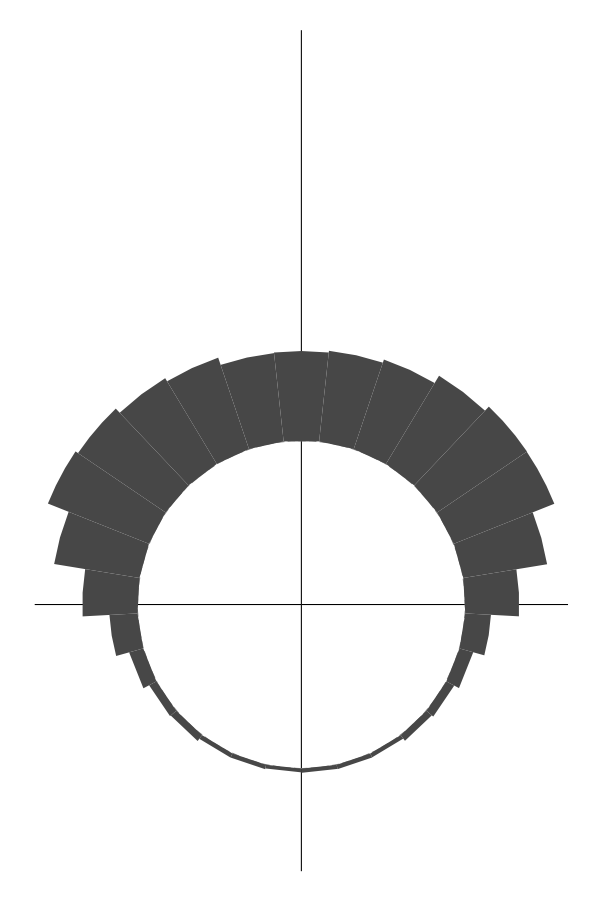}
  \subcaption{Alternative prior}
  \end{subfigure}
  \caption{Marginal posterior distribution on $\beta_{i,t}$ under our original and alternative priors. Note that the scale is the same one that is used in Figure \ref{fig:hyperparam_comp}.}\label{fig:implied_beta_comparison}
\end{figure}

Figure \ref{fig:posteriorshyperparameters} presents histograms of the samples from the posterior distribution of the hyperparameters $\mu$, $\rho$, $\tau^2$, $\varsigma$ and $\lambda$ under both hyperpriors.  As we discussed in Section 5.2 of the main manuscript, the posteriors on  $\varsigma$, $\lambda$, $\rho$ and $\mu$ seem to be reasonably robust to moderate changes in he prior distribution.  The biggest effect can be seen on $\tau^2$ where the prior favors much smaller values (and, therefore, smoother trajectories for the ideal points).  Accordingly, we see a more marked effect on the posterior distribution.
To complement this comparison, we present in Figure \ref{fig:implied_beta_comparison} the posterior distribution of $\beta_{i,t}$ under both sets of hyperparameters.  In spite of the difference in the posteriors for the underlying parameters, the implied marginal posteriors for $\beta_{i,t}$ are essentially identical.

\begin{figure}[!ht]
\centering
    \includegraphics[width = .7\textwidth]{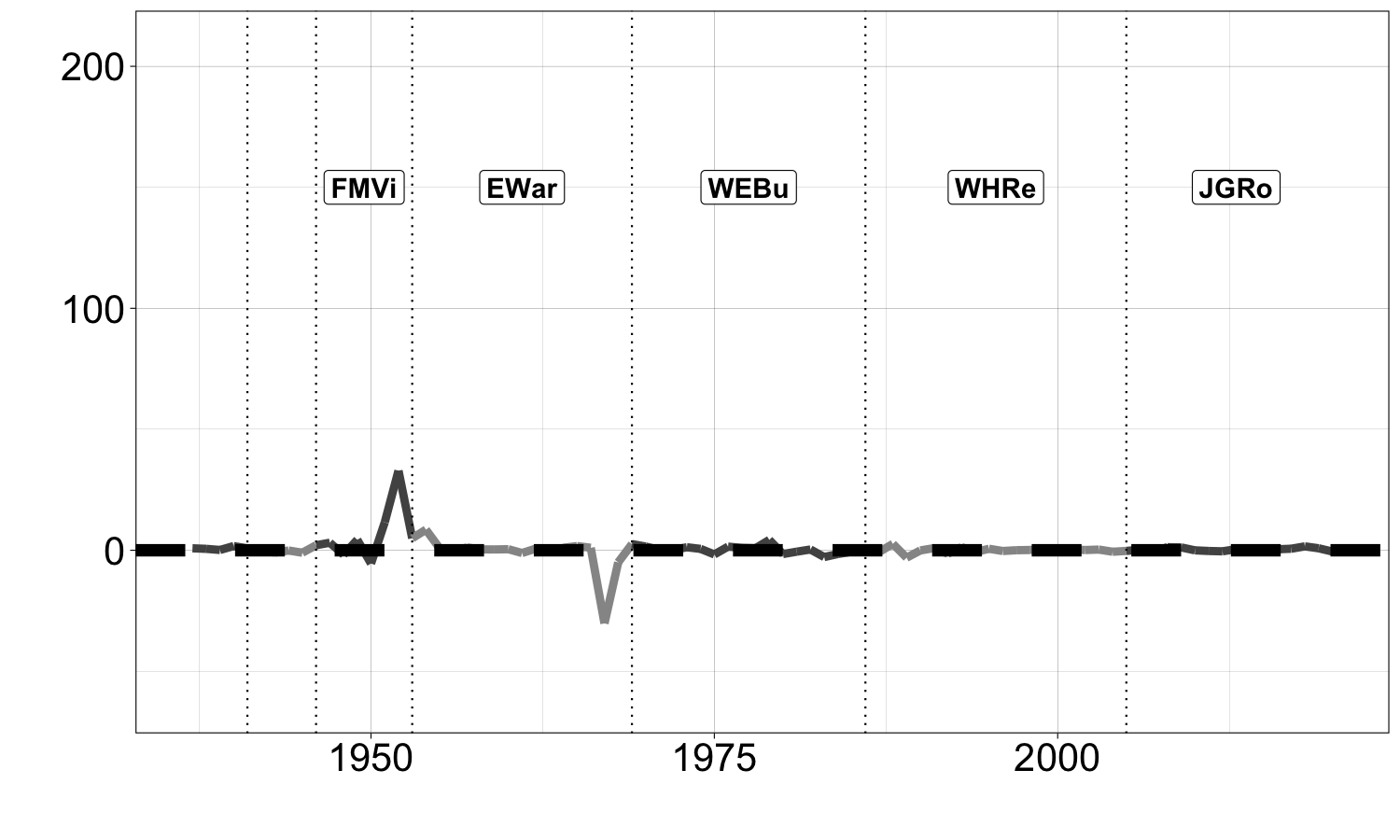}
\caption{Difference in WAIC scores under the original and alternative priors and other priors and models. Here, a positive difference indicates that the original prior is preferred. 
}
\label{fig:rank_WAIC_comparison_alt}
\end{figure}

Next, we present in Figure \ref{fig:rank_WAIC_comparison_alt} the difference in WAIC scores between the original and the alternative priors.  Note that, in order to facilitate comparisons, the scale of the graph was selected to match that of Figure 6 in the main manuscript. We can see that for all but two terms, the difference in WAIC is negligible and attributable to Monte Carlo error.  The two terms for which there is a substantial difference in WAIC scores are 1951 and 1967, with the original set of hyperparameters leading to a better complexity-adjusted fit in 1951, but a worse one in 1967.  Notably, the absolute value of the difference in WAIC seems to be same in both terms, so that  the aggregate WAIC over the 1937 - 2021 period under study is essentially the same for both priors.

\begin{figure}[t]
 \centering
   \centering
  \begin{subfigure}[t]{0.45\textwidth}
  \centering
  \includegraphics[width = 0.8\textwidth]{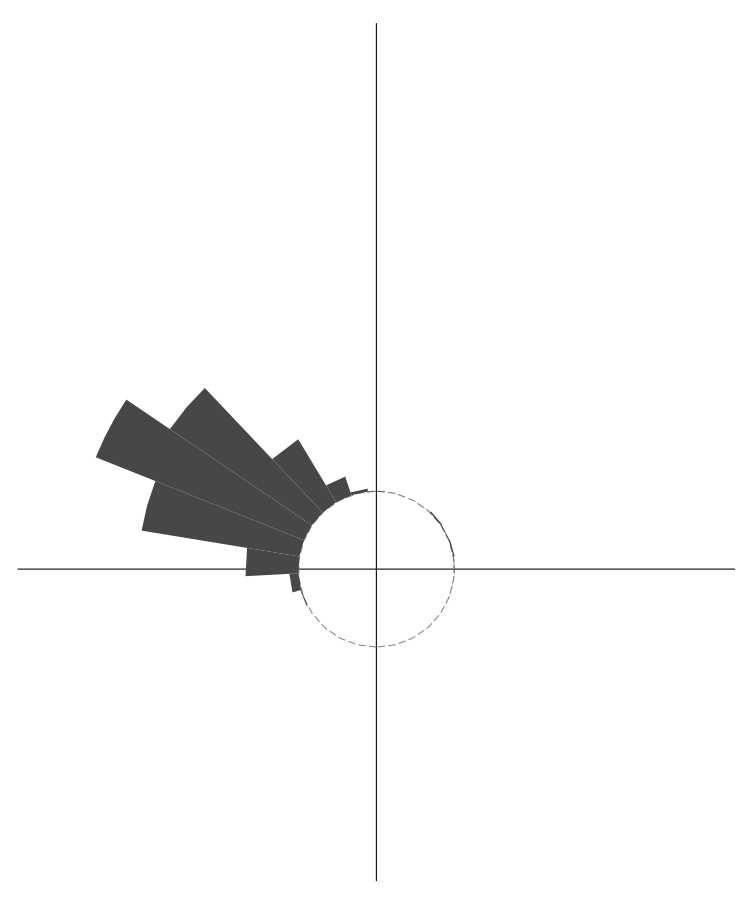}
  \subcaption{Justice Frankfurter 1951 (original prior)}
  \end{subfigure}
  \begin{subfigure}[t]{0.45\textwidth}
  \centering
  \includegraphics[width = 0.8\textwidth]{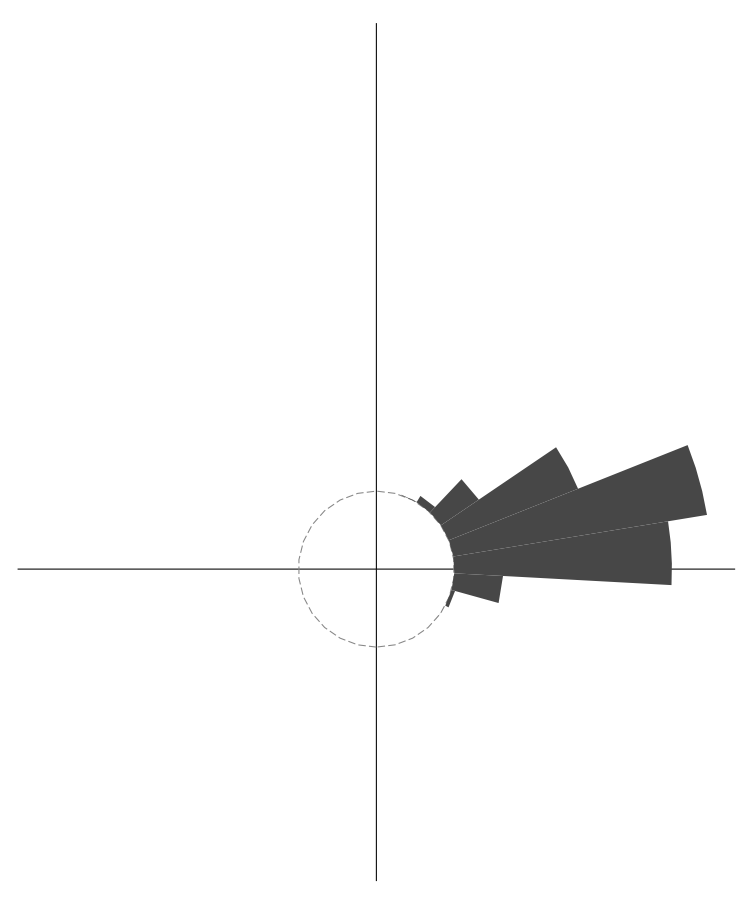}
  \subcaption{Justice Frankfurter 1951 (alternative prior)}
  \end{subfigure}\\
  \begin{subfigure}[t]{0.45\textwidth}
  \centering
  \includegraphics[width = \textwidth]{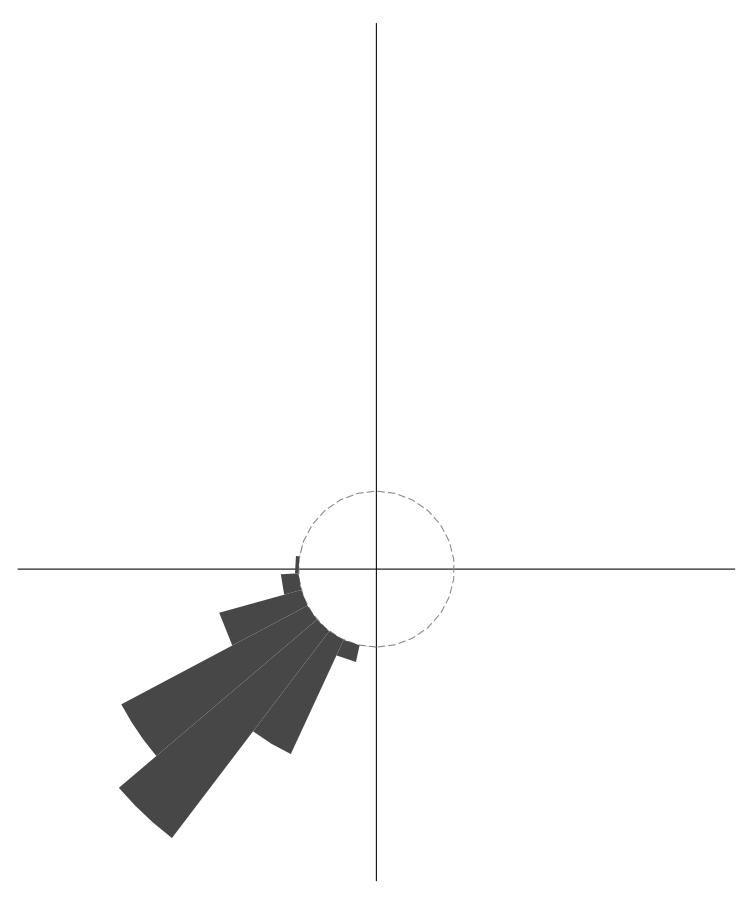}
  \subcaption{Justice Black 1967 (original prior)}
  \end{subfigure}
  \begin{subfigure}[t]{0.45\textwidth}
  \centering
  \includegraphics[width = \textwidth]{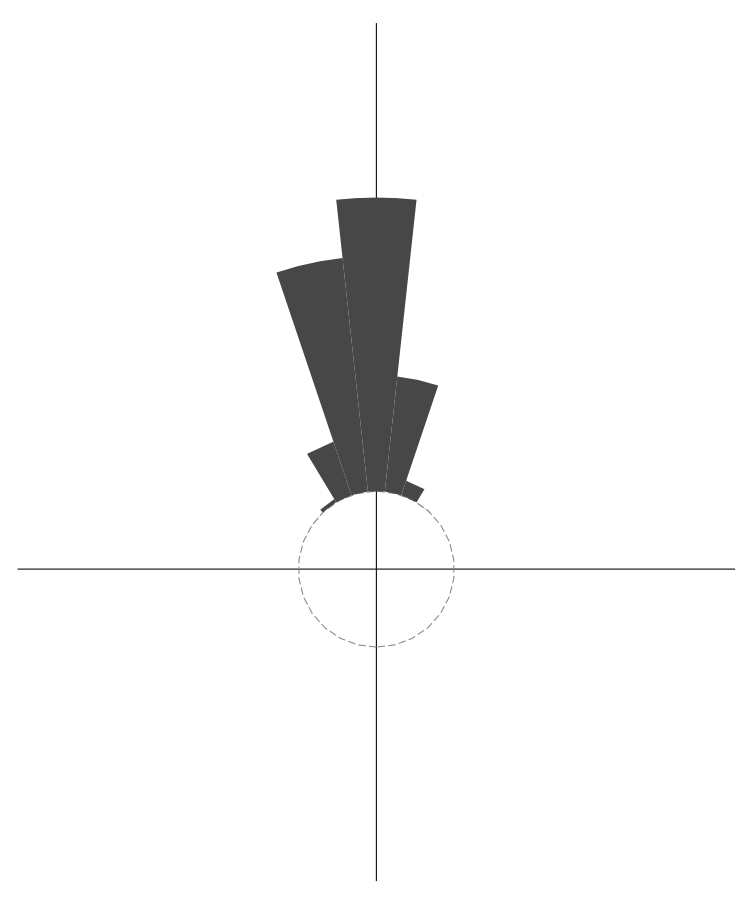}
  \subcaption{Justice Black 1967 (alternative prior)}
  \end{subfigure}
    \caption{Circular histograms for the samples of the posterior distributions of the ideal points of Justice Felix Frankfurter in 1951 (top row) and Justice Hugo Black in 1967 (bottom row) under the original (left column) and alternative (right column) priors. Note that the scale is the same one that is used in Figure \ref{fig:betashapes}.}\label{fig:frank1951_black1067_hist}
\end{figure}

\begin{figure}[!ht]
\centering

\begin{subfigure}{\textwidth}
    \centering
    \includegraphics[width = 0.38\textwidth]{img/circ_results/circular_frankfurter_trajectory.png}
    \includegraphics[width = 0.38\textwidth]{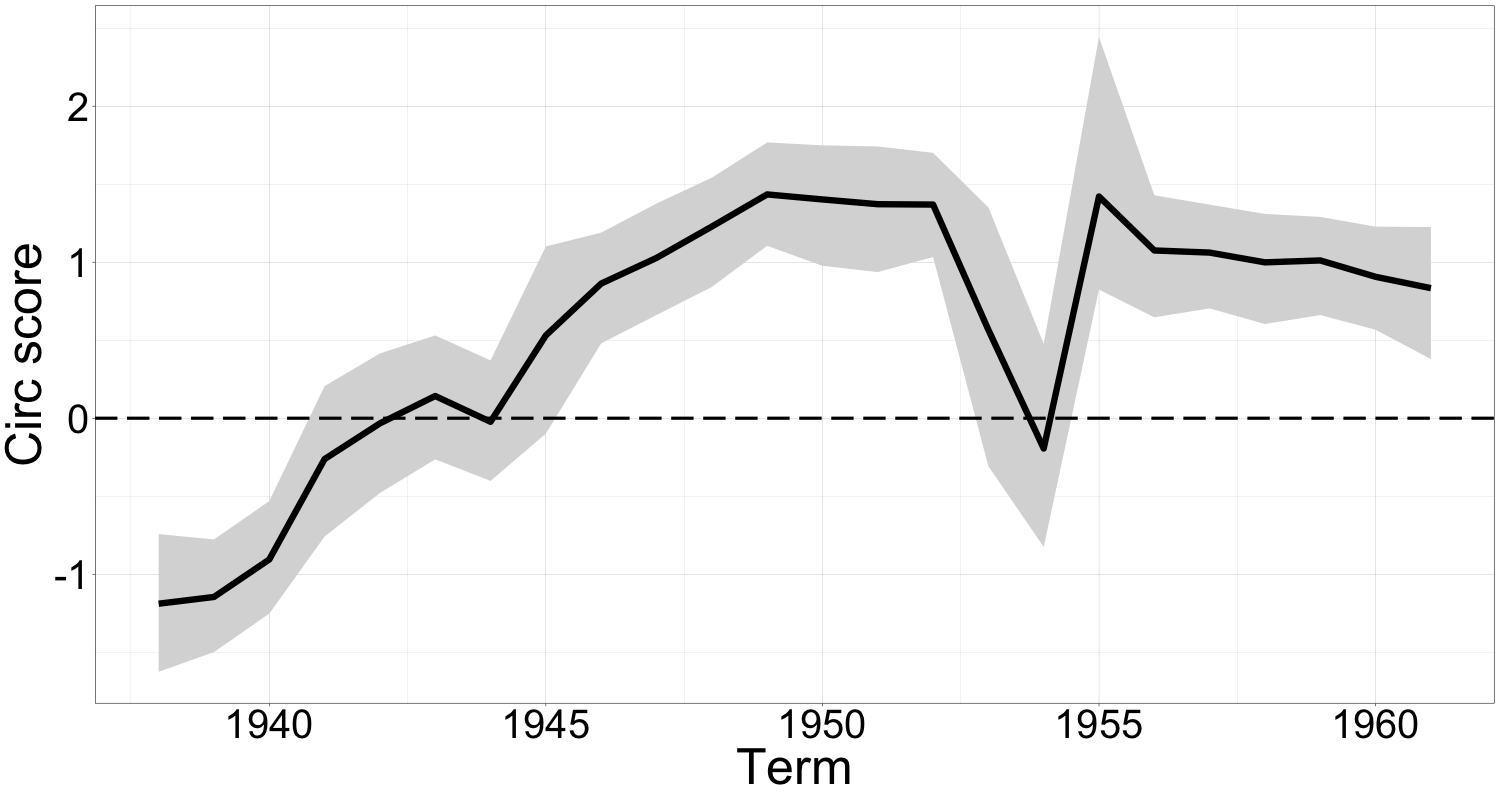}
    \subcaption{Felix Frankfurter}
\end{subfigure} 

\begin{subfigure}{\textwidth}
    \centering
    \includegraphics[width = 0.38\textwidth]{img/circ_results/circular_black_trajectory.png}
    \includegraphics[width = 0.38\textwidth]{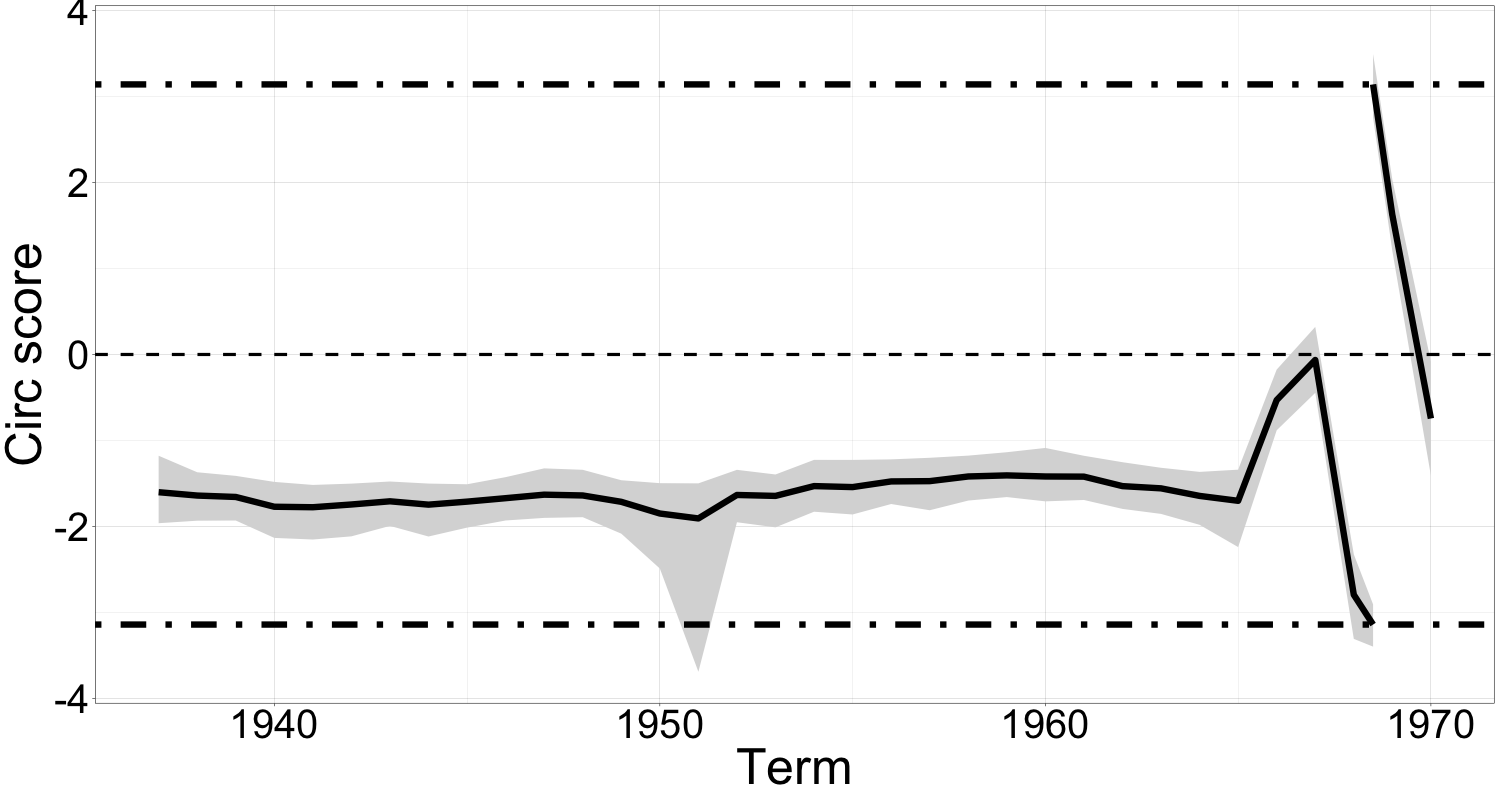}
    \subcaption{Hugo Black}
\end{subfigure}  
\caption{Posterior mean (solid line) and 95\% credible intervals for the ideal points of Justices Felix Frankfurter (top row) and Hugo Black (bottom row). The left column displays estimates based on the original prior, whereas the right column displays the estimates based on the alternative prior. Zero is indicated by a dashed line and $\pm\pi$ by a dotted line.  Euclidean coordinates are used to facilitate comparisons against the main manuscript and the MQ model.}
\label{fig:trajectories2_comp}
\end{figure}

The differences in WAIC scores that we just discussed seem to be driven by differences in the estimates of the ideal points of Justice Felix Frankfurter in 1951, and that of Justice Hugo Black in 1967 (see Figure \ref{fig:frank1951_black1067_hist}).  Indeed, the original prior favors a negative circular score for Justice Frankfurter in 1951, while the alternative prior favors a positive one.  In the case of Justice Black, the original prior again clearly favors a negative score in 1967, while the alternative prior favors values close to zero.  To provide additional context, Figure \ref{fig:trajectories2_comp} shows the estimated trajectories of the ideal points of both justices over the full tenures.  In the case of Justice Frankfurter, we can immediately see that the key difference is the level of smoothness in the trajectory.  In particular, the original prior allows for a sharp spike in the trajectory for 1951, something that the alternative prior (which, as we discussed before, favors very small values of $\tau^2$) does not allow.  Similarly, the trajectory of Justice Black is smoother under the alternative prior.

